\documentclass[review]{elsarticle}

\usepackage{lineno,hyperref}
\usepackage{amssymb}
\usepackage{amsmath}
\usepackage{mathrsfs}
\modulolinenumbers[1]

\journal{ }
\begin{document}

\begin{frontmatter}

\title{Microscopic theory of capillary pressure hysteresis based on pore-space accessivity and radius-resolved saturation}
\author[cheme]{Zongyu~Gu}
\ead{zygu@mit.edu}
\author[cheme,math]{Martin~Z.~Bazant\corref{cor1}}
\ead{bazant@mit.edu}
\cortext[cor1]{Corresponding author}
\address[cheme]{Department of Chemical Engineering, Massachusetts Institute of Technology,
Cambridge, Massachusetts 02139, USA}
\address[math]{Department of Mathematics, Massachusetts Institute of Technology,
Cambridge, Massachusetts 02139, USA}

\begin{abstract}
Continuum models of porous media use macroscopic parameters and state variables to capture essential features of pore-scale physics. We propose a macroscopic property ``accessivity" ($\alpha$) to characterize the network connectivity of different sized pores in a porous medium, and macroscopic state descriptors ``radius-resolved saturations'' ($\psi_w(F),\psi_n(F)$) to characterize the distribution of fluid phases within. Small accessivity ($\alpha\to0$) implies serial connections between different sized pores, while large accessivity ($\alpha\to1$) corresponds to more parallel arrangements, as the classical capillary bundle model implicitly assumes. Based on these concepts, we develop a statistical theory for quasistatic immiscible drainage-imbibition in arbitrary cycles, and arrive at simple algebraic formulae for updating $\psi_n(F)$ that naturally capture capillary pressure hysteresis, with $\alpha$ controlling the amount of hysteresis. These concepts may be used to interpret hysteretic data, upscale pore-scale observations, and formulate new constitutive laws by providing a simple conceptual framework for quantifying connectivity effects, and may have broader utility in continuum modeling of transport, reactions, and phase transformations in porous media.
\end{abstract}


\end{frontmatter}


\section{Introduction}

From rocks and wood to concrete and catalysts, porous materials vary widely in origin, properties, and applications. Despite their macroscopic appearance as solid objects, porous media are distinguished by their ability to contain fluids internally, owing to their heterogeneous microstructure -- the solid matrix occupies only a portion of the macroscopic domain, while the complementary pore space is able to accommodate one or more fluid phases \cite{bear1972}.

In the typical case of well-connected pores, diverse physical phenomena in porous media, such as fluid flow, heat and mass transfer, gas adsorption, and phase transformations, may be amenable to homogenized macroscopic descriptions \cite{torquato2013}, although the exact connection with microscopic details of the porous medium is not always clear. Remarkably, any continuum model implicitly assumes that the overall effects of the often nontrivial pore-space morphologies \cite{sahimi2011} can be encapsulated in a small number of parameters, e.g., porosity, $\phi$, tortuosity, $\tau$, intrinsic permeability, $k_s$, effective thermal conductivity, $k_e$, etc., and the state of any fluid phase can be described by a small number of distributed state variables, e.g., pressure, $p$, saturation, $s$, etc.. For certain simple physical processes, especially those involving a single fluid phase, simple continuum formulations generally work well, and a rigorous connection between the pore-scale and continuum-scale governing equations can be sought -- examples include single-phase flow \cite{allaire1989,hornung2012} and heat transfer \cite{loeb1954,carson2005} -- although estimating the transport coefficients from microscopic features of the porous medium is still an open research problem \cite{ranut2016,pabst2017,neithalath2010,chareyre2012,mostaghimi2013}.

In comparison, it is considerably more challenging to develop continuum models for immiscible multiphase flow in porous media (including unsaturated flow and condensate transport) \cite{bear1972,scheidegger1974,richards1931,buckley1942,childs1950,klute1952,tamon1981,lee1986,jaguste1995,do2001} based on microscopic physics. While it is possible to upscale pore-scale equations by careful averaging \cite{durlofsky1998,arbogast2000,cushman2002,wood2009,li2006}, the resulting model varies depending on the macroscopic state variables selected and the scaling laws assumed for the application considered, not to mention that the mechanisms for pore-scale fluid motions are highly complex and are still actively researched \cite{lenormand1983,pak2015,holtzman2015,zhao2016}. Thus, in formulating continuum models of multiphase flow in porous media, there exists a trade-off between mathematical simplicity and consideration of pore-scale physics. In conventional models that are widely accepted in practice, saturation is the primary state variable; it is used to compute capillary pressure \cite{brooks1964,van_genuchten1980} and relative permeabilities \cite{corey1954,irmay1954,brooks1964} via empirical constitutive relationships, which are able to fit typical experimental measurements by virtue of having several adjustable parameters \cite{pinder2008}. By recognizing that pore-scale phenomena like viscous flow and capillary equilibrium must first and foremost depend on the microscopic dimensions of pores, some of the most popular continuum constitutive relationships in the literature have also incorporated the concept of a pore-size distribution \cite{thomson1872,washburn1921,burdine1953,mualem1976}, conceptualizing the pore space as interconnected pores of various sizes, either implicitly or explicitly \cite{van_brakel1975,quiblier1984}, such as in the well-known ``capillary bundle'' model.

However, these conventional models suffer from hysteresis, meaning that the relationships between state variables are non-unique and history-dependent. This suggests that saturation alone cannot fully describe the microscopic state of fluid phases in real porous media, and additional macroscopic state variables are required to capture hysteresis. Many authors have considered extending the conventional constitutive relationship between capillary pressure and saturation, $p_c(s_w)$, by assuming that $p_c$ may also depends on the ``rate of saturation'', $\partial s_w / \partial t$ (or at least its sign) \cite{barenblatt1971,luckner1989,hassanizadeh1990,barenblatt2003,juanes2008}, thereby attributing hysteresis to nonequilibrium effects. Some authors have identified ``specific interfacial area'', $a_{wn}$, or the interfacial area between fluid phases $w$ and $n$ per unit volume of the porous medium, as a physically relevant state variable on thermodynamic grounds, and have advocated for its inclusion in continuum models to reduce capillary pressure hysteresis \cite{kalaydjian1987,hassanizadeh1990,hassanizadeh1993,beliaev2001,hassanizadeh2002}. This hypothesis seems to hold in many but not all cases, as revealed by micromodel experiments \cite{cheng2004,chen2007,pyrak-nolte2008,karadimitriou2014}, lattice-Boltzmann simulations \cite{porter2009}, pore-network simulations \cite{reeves1996,held2001,joekar-niasar2008,joekar-niasar2012-1}, and analysis \cite{helland2007}, while the exact form of any new constitutive relationships required may not be completely clear. More recently, Hilfer put forth a new class of continuum models for two-phase flow in porous media \cite{hilfer1998,hilfer2006,doster2010} involving four fluid saturation variables, $\{s_1,s_2,s_3,s_4\}$, as opposed the conventional two, $\{s_w,s_n\}$, for the two fluid phases (in either case, all saturation variables must sum up to unity); we have $s_w=s_1+s_2$ and $s_n=s_3+s_4$, where $s_1$ and $s_3$ correspond to ``percolating regions'' of the respective fluid phases, and $s_2$ and $s_4$ correspond to ``non-percolating regions''. By differentiating between the contributions of percolating and non-percolating fluid ``subphases'', Hilfer's model naturally predicts hysteresis as a result of the dynamics of the newly introduced state variables. Because the model is not derived from the principles of microscopic physics, phenomenological assumptions are still required to, say, model the ``mass transfer rates'' between $s_1$ and $s_2$, or $s_3$ and $s_4$ (namely, percolating fluid regions becoming non-percolating and vice versa), which result in model parameters that may be difficult to physically interpret, though potentially determinable from experiments.

It appears that the intrinsic complexity of multiphase flow in porous media would ensure that any continuum model to come in the foreseeable future will not match the performance of pore-scale methods (pore-network, Lattice-Boltzmann, phase-field, etc.; see this review \cite{meakin2009} for instance) in terms of either predictability or connection to first principles. On the one hand, conventional models are preferred for macroscopic simulations and used routinely in practical applications, where hysteresis is either described empirically or neglected altogether, although it is considered crucial for making certain types of predictions (e.g., \cite{essaid1993}). On the other hand, new continuum models, despite having rightly introduced new and physically meaningful state variables ($a_{wn}$, Hilfer's $s_1,\ldots,s_4$) so as to naturally predict hysteresis, deviate significantly from conventional models, and involve somewhat unintuitive constitutive laws with phenomenological constants that lack a clear connection to the pore-scale descriptions, possibly due to the emphasis placed on reproducing certain macroscopic observations.

In this work, we take the view that there is great value in identifying new physically meaningful concepts that are relevant to continuum modeling of multiphase flow in porous media. These concepts should be connected to essential aspects of pore-scale physics, yet intuitive enough for a wide range of continuum-scale applications -- the pore-size distribution would fall within this category. In the short term, these concepts may be incorporated into conventional continuum models for incremental improvements, while in the long term, they may be subject to pore-scale investigations, and ultimately play a role in future continuum models of multiphase flow.

The first concept we shall propose is the ``pore-space accessivity'', denoted by $\alpha$. Accessivity is a continuum-scale property of porous media that, in the simplest possible fashion, contrasts serial and parallel arrangements of different sized pores. This extends the capillary bundle model, which is known to be insufficient, but still routinely invoked because of its simplicity \cite{diamond2000,hunt2013}. The capillary bundle picture coincides with the $\alpha\to1$ limit in our framework.

The second concept we shall propose is the ``radius-resolved saturation'', $\psi(F)$, where $0 \le F(r) \le 1$ is the cumulative distribution function (CDF) of the pore-size distribution, where $r$ denotes the pore radius. Radius-resolved saturation would replace saturation as a better continuum-scale descriptor for the distribution of fluid phases in the pore space, where ``saturation'' $\psi$ is now defined for pores of each particular size given by $F$, hence ``radius-resolved''.

The paper is organized as follows. In Sections 2 and 3, we introduce the concepts of accessivity and radius-resolved saturation, and relate them to existing ideas in the literature. In Section 4, we present a simple statistical theory based on pore branching that leads to simple governing equations involving the proposed concepts. In Section 5, we present simple illustrative examples to highlight the usefulness and limitations of our theory. Finally, we discuss the broader utility of accessivity and radius-resolved saturation and identify outstanding questions and future research directions in Section 6, before concluding in Section 7.

\section{Characterization of pore-space morphology}

In this section, we consider macroscopic descriptors for the pore-space morphology of a porous medium. For that purpose, the pore space need not be filled with any particular fluid and can be left ``empty'' (or, alternatively and conceptually equivalently, filled uniformly with an inert fluid). We will briefly review some existing descriptors and introduce pore-space accessivity at the end of the section.

\subsection{Porosity}

\begin{figure}[!h]
\centering
\includegraphics{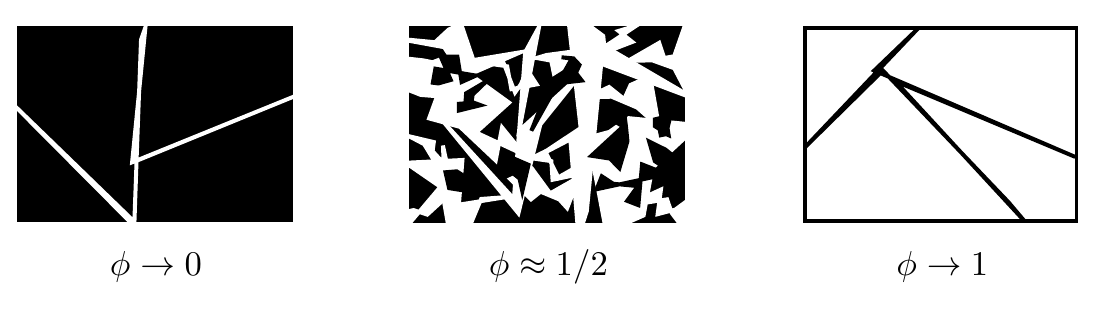}
\caption{Illustrations of a series of realizations of porous media with low, medium, and high porosities. Black and white regions correspond to the solid matrix and the pore space, respectively.}\label{figure:porosity_cartoon}
\end{figure}

Porosity, denoted by $\phi$, is an intuitive macroscopic property of a porous medium that reflects the volume fraction of the pore space in the domain of the medium. We may think of it as:
\begin{align}
    \phi = \frac{V_p}{V},\label{eq:porosity_definition}
\end{align}
where $V$ is the volume of a representative domain of the porous medium (called a representative elementary volume), and $V_p$ is the volume of the pore space within that domain \cite{bear1972,torquato2013}. The definition of porosity entails:
\begin{align}
    \phi \in \left( 0,1 \right).
\end{align}
The porous medium becomes plainly the solid in the limit of $\phi\to0$, and a homogeneous free space available for fluid occupation as $\phi\to1$. As $\phi$ increases, the medium acquires a greater capacity for fluids, and generally becomes less dense and more permeable. See Figure \ref{figure:porosity_cartoon}.

\subsection{Tortuosity}

Tortuosity (also referred to as the tortuosity factor by some authors \cite{epstein1989}), denoted by $\tau$, is another familiar continuum-scale parameter for porous media \cite{torquato2013}. It is sometimes intuitively taken as:
\begin{align}
    \tau = \left(\frac{L_e}{L}\right)^2,\label{eq:tortuosity_definition_geometric}
\end{align}
where $L$ is the straight-line distance along a particular direction of flow or diffusion in the porous medium, and $L_e$ is the typical arc length of a tortuous microscopic path spanning that distance that is followed by a tracer particle in the pore space \cite{epstein1989}, although the more general definition, based on the effective diffusivity through the medium:
\begin{align}
    D_{\text{eff}} = \frac{D\phi}{\tau},\label{eq:tortuosity_definition_transport}
\end{align}
may not have such a simple geometrical interpretation \cite{torquato2013,cooper2016}. Here, $D$ is the ``true'' diffusivity of the species in the pore space. We expect:
\begin{align}
    \tau \in \left[ 1,\infty \right).
\end{align}

\begin{figure}[!h]
\centering
\includegraphics{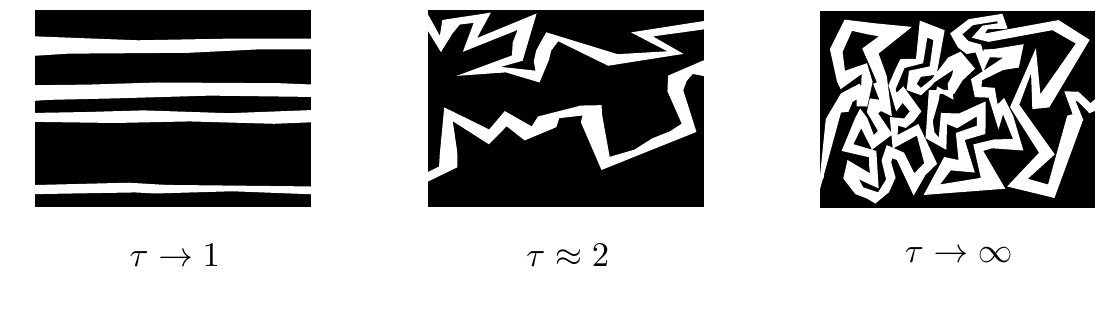}
\caption{Illustrations of a series of realizations of porous media with low, medium, and high tortuosities. Black and white regions correspond to the solid matrix and the pore space, respectively. Note that a porous medium with a large $\tau$ seldom looks like what is shown here, and the simple geometrical interpretation given by Eq. \eqref{eq:tortuosity_definition_geometric} is no longer useful or accurate.}\label{figure:tortuosity_cartoon}
\end{figure}

If we regard the pore space as a bundle of tortuous but non-intersecting capillaries, the limit of low tortuosity, $\tau\to1$, represents a bundle of straight capillaries that are perfectly aligned in the direction of flow or diffusion, in which case the effective value of the pertinent transport coefficient approaches its ``true'', pore-scale counterpart \cite{torquato2013}. As $\tau$ is raised, the typical residence time of a tracer particle grows longer, leading to increasingly slower transport \cite{epstein1989}, due to branched, non-percolating paths and series connections through the pore-network \cite{torquato2013}, which violate the capillary bundle model. See Figure \ref{figure:tortuosity_cartoon}.

\subsection{Pore-size distribution}

As we discussed in Section 1, the relevant length scale for pore-scale physics is often the local dimensions of the pore space. For example, in the simple case of a straight, cylindrical pore of radius $r$, viscous flow is described by the Hagen-Poiseuille equation \cite{deen2011}:
\begin{align}
    Q = \frac{\pi r^4}{8\mu}\left| \frac{\mathrm{d}\mathscr{P}}{\mathrm{d}z} \right|,\label{eq:poiseuille}
\end{align}
where $Q$ is the volumetric flow rate through the pore, $\mu$ is the viscosity of the fluid, and $\mathrm{d}\mathscr{P}/\mathrm{d}z$ is the magnitude of the dynamic pressure gradient along the pore axis; capillary equilibrium between two immiscible fluid phases $w$ and $n$ separated by a meniscus perpendicular to the pore axis is described by the Young-Laplace equation \cite{deen2011} (also known as the Washburn equation \cite{washburn1921}):
\begin{align}
    \left| \Delta p \right| = \left| \frac{2\gamma_{wn}\cos{\theta}}{r} \right|,\label{eq:Washburn}
\end{align}
where $\left| \Delta p \right|$ is the magnitude of the (microscopic) capillary pressure across the meniscus, $\gamma_{wn}$ is the surface tension between the two fluids, and $\theta$ is the contact angle. The pore radius $r$ appears in both equations, while the same principle also holds for pores with other geometries.

The pore-size distribution (PSD) describes the variability of pore radii throughout the medium at the continuum scale. We denote its probability density function and cumulative distribution function by $f\left(r\right)$ and $F\left(r\right)$, respectively; the two are related by:
\begin{align}
    F\left(r_0\right) = \int_{0}^{r_0}
    f\left(r\right)\mathrm{d}r.
\end{align}
Despite the lack of a universally accepted definition, the PSD is nevertheless an intuitive and useful concept \cite{hilfer1996,sahimi2011}. In this work, we will interpret $F\left(r_0\right)$ as the volume fraction of all ``pores'' whose ``radii'' are below $r_0$; the pores are taken to be cylindrical, or $r$ is to be interpreted as the effective radius of a cylindrical pore. It follows that $\mathrm{d}F\left(r_0\right) = f\left(r_0\right)\mathrm{d}r$ is the volume fraction of all pores whose radii fall within $\left[r_0,r_0+\mathrm{d}r\right)$.

Finally, we find it advantageous in the following discussions to refer to different pore sizes by $F$ instead of by $r$: we say a pore is ``of size $F_0$'' when it has a radius $r_0$ such that $F_0=F(r_0)$. There is no ambiguity because the cumulative function of the PSD is a bijective map from $r$ to $F$. Using $F$ as a surrogate for pore size is particularly useful when the pore scale events of interest are controlled only by the relative order of pore sizes (i.e., which pore is larger or smaller in size), such as in invasion percolation \cite{wilkinson1983}. This way, we can express general results that are independent from the PSD, as we shall see in subsequent analyses.

\subsection{Connectivity of different sized pores: accessivity}

Numerous continuum-scale processes in porous media are affected by not only the PSD, but also the way in which different sized pores are connected. The ``ink-bottle effect'' is a prominent example of connectivity effects, and is known to contribute to capillary pressure hysteresis, such as in mercury intrusion-extrusion porosimetry \cite{abell1999,diamond2000}: during intrusion, pores that are large enough to be invaded by mercury according to the Washburn equation will not become filled if they are only accessible through smaller pores; similarly, during extrusion, as $p$ is lowered quasistatically, smaller pores may not empty if they are preceded by larger ones.

How connectivity effects are handled \cite{van_brakel1975,quiblier1984} significant affects the resulting porous media model. Consider the capillary bundle model, which conceptualizes the pore space as a bundle of straight capillaries that are directly accessible from the surface of the sample. Here, connectivity effects -- and hence hysteresis (barring other explanations such as contact-angle hysteresis \cite{lowell2012}) -- are completely absent, as pores of different sizes are equally and consistently accessible. The simplistic nature of the capillary bundle model draws much criticism \cite{diamond2000,hunt2013}, while at the same time engendering widespread use in practice, e.g., in the standard laboratory interpretation of mercury intrusion porosimetry data and sorption isotherms. It is precisely the inadequacy of the simplistic capillary bundle model that led to the birth of pore-network modeling, as Fatt asserted in the opening paragraph of their seminal work \cite{fatt1956}. Contemporary pore-network models portray the pore space as a network of geometrically simple voids, e.g., cylindrical ``pore throats'' and spherical ``pore bodies'', whose connectivities are either approximated with regular lattices \cite{celia1995,reeves1996} or extracted from 3-D images of real porous media \cite{bakke1997,blunt2001}. Predictions of continuum-scale properties are then made by numerically solving pore-scale equations written for elements in the pore-network \cite{dullien1991,celia1995,blunt2001,joekar-niasar2012}, as exemplified by the invasion percolation approach \cite{wilkinson1983}. In the same light, percolation theory has also been applied to model connectivity effects in porous media, and can yield analytical results based on calculations on simple lattices (e.g., the Bethe lattice or the Cayley tree) \cite{sahimi1994,stauffer2014,sahimi2011,selyakov2013,larson1981,mason1982,mason1983,parlar1988,seaton1991,kadet2013,pinson2014,masoero2015,pinson2015}.

Despite this significant body of research, we find it beneficial, both conceptually and practically, to propose a continuum-scale property of porous media that describes the microscopic connectivity of different sized pores. The proposed parameter would be analogous to established concepts like tortuosity, which is based on an intuitive yet imprecise pore-scale physical picture -- higher tortuosity means more ``tortuous'' pores (Eq. \eqref{eq:tortuosity_definition_geometric}), but interpreted more flexibly for real porous media based on macroscopic characteristics -- tortuosity is defined based on effective transport (Eq. \eqref{eq:tortuosity_definition_transport}). We name this new quantity the ``accessivity", and denote it by $\alpha$, with:
\begin{align}
    \alpha \in \left( 0,1 \right).
\end{align}

\begin{figure}[!h]
\centering
\includegraphics{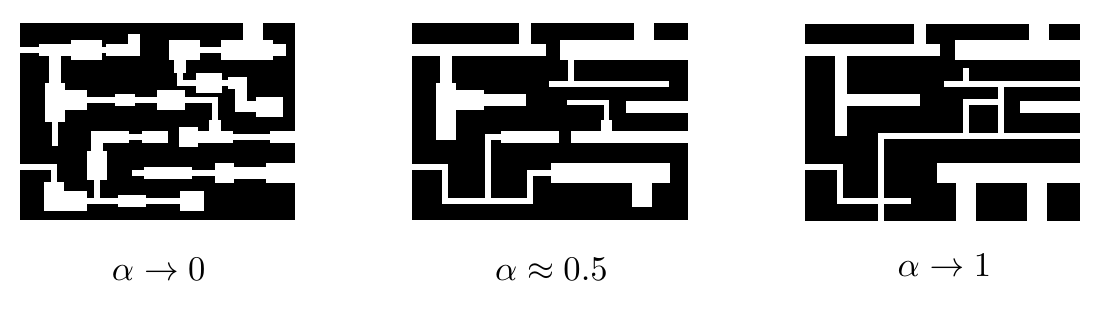}
\caption{Illustrations of a series of realizations of porous media with low, medium, and high accessivities. Black and white regions correspond to the solid matrix and the pore space, respectively.}\label{figure:accessivity_cartoon_mod}
\end{figure}

As with $\phi$ and $\tau$, the limiting values of $\alpha$ correspond to two extreme cases of pore-space connectivity. Figure \ref{figure:accessivity_cartoon_mod} illustrates the microscopic interpretation of accessivity for a hypothetical porous medium with a trimodal PSD. As $\alpha\to0$, different sized pore segments are overwhelmingly connected in series, which would lead to significant ink-bottle effects. As $\alpha\to1$, different sized pores are overwhelmingly arranged in parallel, coinciding with the capillary bundle model and thus eliminating connectivity effects. A larger $\alpha$ correlates with a slower rate of radius variation along a pore axis, which in turn reduces connectivity effects and any ensuing hysteresis.

Accessivity should be viewed as an approximate descriptor for the connectivity of different sized pores. Unlike pore-network models, a single parameter $\alpha$ cannot possibly capture the full details of pore-space connectivity, but that is not the objective of this work: we propose the concept as an intuitive extension to the capillary bundle model, where connectivity effects are entirely absent. While the capillary bundle model uses the PSD to discern the differences in behaviors of the different sized pores in isolation, accessivity broadens that analysis by considering how the different sized pores are arranged relative to one another.

The macroscopic implications of accessivity will become clear in the following sections through more quantitative descriptions. We will also discuss how the concept accessivity may be generalized.

\section{Characterization of fluid distribution}

In this section, we will consider a porous medium whose pore space now hosts multiple fluids. The fluid phases may distribute in any arbitrary fashion that is consistent with pore-scale physics. Our goal is to characterize the distribution of fluid phases at the macroscopic scale. The concepts we propose here are general and can be readily extended to systems involving any number of immiscible fluids, though for simplicity, we shall confine our discussion to the case of two fluid phases, which are denoted by $w$ and $n$, corresponding to the wetting and nonwetting phases, respectively.

\subsection{Conventional saturation}

The saturation of phase $w$ is the volume fraction of the wetting fluid in the pore space:
\begin{align}
    s_w = \frac{V_w}{V_p},
\end{align}
where $V_w$ and $V_p$ are the volume occupied by wetting phase and the total volume of the pore space in the same control volume in the porous medium; akin to the definition of porosity, we shall choose a macroscopic control volume that is representative of the porous medium \cite{bear1972}. We can similarly define saturation for the nonwetting phase, $s_n$. Both $s_w$ and $s_n$ vary between zero and unity, and it follows from their definitions that $s_w+s_n=1$.

In continuum descriptions of two-phase flow, sorption, mercury porosimetry, etc., saturation is a common a state variable that may vary in both space and time. When $s_w \to 1$ (or $0$) at a certain point in a macroscopic domain, some representative volume surrounding that point becomes exclusively filled with the wetting (or nonwetting) phase, and assuming we have perfect knowledge of the pore-space morphology, there is no ambiguity in the microscopic state of the medium near that point. By contrast, a fractional $s_w$ generally corresponds to a multitude of possible microscopic states: the two fluid phases can be distributed in the pore space in any arbitrary fashion, so long as the overall fraction of each phase corresponds to its saturation. Depending on the properties of the fluids and the solid matrix and the pore-scale physics in effect, some microscopic states may be more favorable than others; nevertheless, knowledge of saturation alone is generally far from sufficient for deducing the distribution of fluid phases at the pore scale.

\subsection{Radius-resolved saturation}

\begin{figure}[!h]
\centering
\includegraphics[scale=0.6]{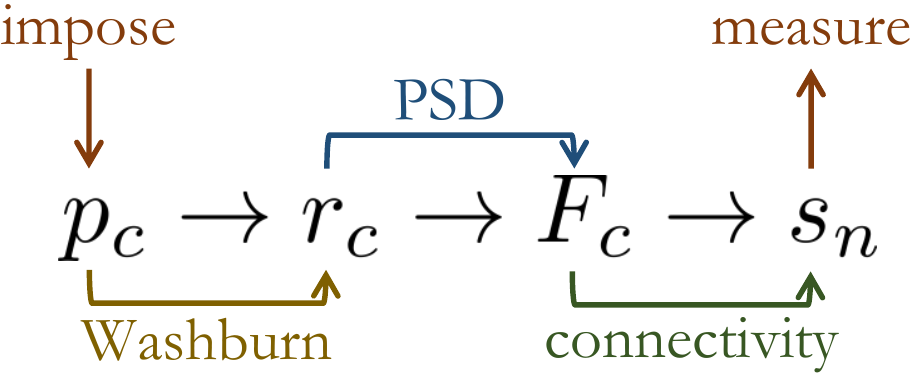}
\caption{Causal dependencies between macroscopic quantities during a conceptual invasion percolation process on a representative control volume in a porous medium. The measured $s_n(p_c)$ relationship results from a combination of capillary equilibrium conditions, the PSD, and pore-space connectivity, with two intermediate macroscopic variables, $r_c$ and $F_c$.}\label{figure:variables_causal}
\end{figure}

To motivate the definition of the ``radius-resolved saturation'', consider the conceptual process of invasion percolation \cite{wilkinson1983,sahimi1994,stauffer2014} in a representative control volume (small enough such that there is no internal variation of any continuum-scale properties) in a porous medium, which can be compared to carrying out mercury intrusion porosimetry or equilibrium sorption measurements on a small sample. We illustrate the causality between the quantities involved in such a process in Figure \ref{figure:variables_causal}. The pore space is initially filled with one fluid, say the wetting phase, such that $s_w=1$ in the control volume, while part of the outer surface is exposed to a reservoir containing the nonwetting fluid. At equilibrium, the pressure within either phase ($p_w,p_n$) is uniform, but there may exist a difference between the two pressures, which is referred to as the macroscopic capillary pressure, $p_c=p_n-p_w$. Suppose we raise $p_c$ quasistatically, whose effect is presumably felt at every $w$-$n$ interface in the porous medium. Each particular $p_c$ corresponds to an equilibrium capillary radius, $r_c$, given by the Washburn equation (Eq. \eqref{eq:Washburn}). Pores with effective radii smaller than $r_c$ will undergo imbibition if a meniscus is present, and those with radii larger than $r_c$ favor drainage. Given the PSD, each $r_c$ maps to a $F_c=F(r_c)$ (see earlier discussions on using the cumulative function of the PSD as a surrogate for pore size), such that $F_c$ gives the volume fraction of the pore space that would favor imbibition, while $1-F_c$ gives the volume fraction of pores that would favor drainage. As $p_c$ becomes higher, $r_c$ decreases, thus increasing $1-F_c$, predisposing a greater volume fraction of the pore space to invasion by the nonwetting fluid. The actual volume fraction of the nonwetting fluid in the pore space is measured by $s_n$, which may never exceed $1-F_c$ during the invasion process, but could be smaller than $1-F_c$ due to connectivity effects:
\begin{align}
    s_n \le 1-F_c.
\end{align}
That is, not all pores large enough to favor drainage ($1-F_c$) will actually undergo drainage ($s_n$) unless they are directly accessible by the nonwetting fluid -- consider the ink-bottle effect. If the pore space were well represented by a capillary bundle, then all pores would be directly accessible by the nonwetting phase at all times during invasion; the accessivity $\alpha\to1$ for the control volume, as different sized pores would be arranged entirely in parallel. In this case, we would have $s_n=1-F_c$. Conversely, for $\alpha<1$, we would generally expect $s_n<1-F_c$.

We propose the radius-resolved saturations, $\psi_w(F)$ and $\psi_n(F)$, to characterize the distribution of fluids across different pore sizes. Among all pores of a particular size $F_0 = F(r_0)$ (see earlier discussions), $\psi_w(F_0)$ is the volume fraction of pores filled with the wetting fluid, and $\psi_n(F_0)$ is the fraction of pores filled with the nonwetting fluid. Akin to conventional saturations, $\psi_w(F)+\psi_n(F)=1$ for all $0\le F \le 1$. Conventional saturations are easily recovered from the radius-resolved saturations:
\begin{align}
    s_w = \int_0^{1} \psi_w(F)\mathrm{d}F, \quad
    s_n = \int_0^{1} \psi_n(F)\mathrm{d}F.\label{eq:saturations_and_rrs}
\end{align}
Namely, conventional saturations are simply averaged radius-resolved saturations, weighted by the volume fraction of each pore size. Note that if we had expressed $\psi_w$ as a function of pore radius $r$, we would have written:
\begin{align}
    s_w = \int_0^{\infty} \psi_w(r)f(r)\mathrm{d}r
\end{align}
because $\mathrm{d}F=f(r)\mathrm{d}r$. Here, $\psi_w(r)f(r)$ would be the ``pore-size distribution'' density of just the wetting fluid-filled pores.

\begin{figure}[!h]
\centering
\includegraphics[width=\textwidth]{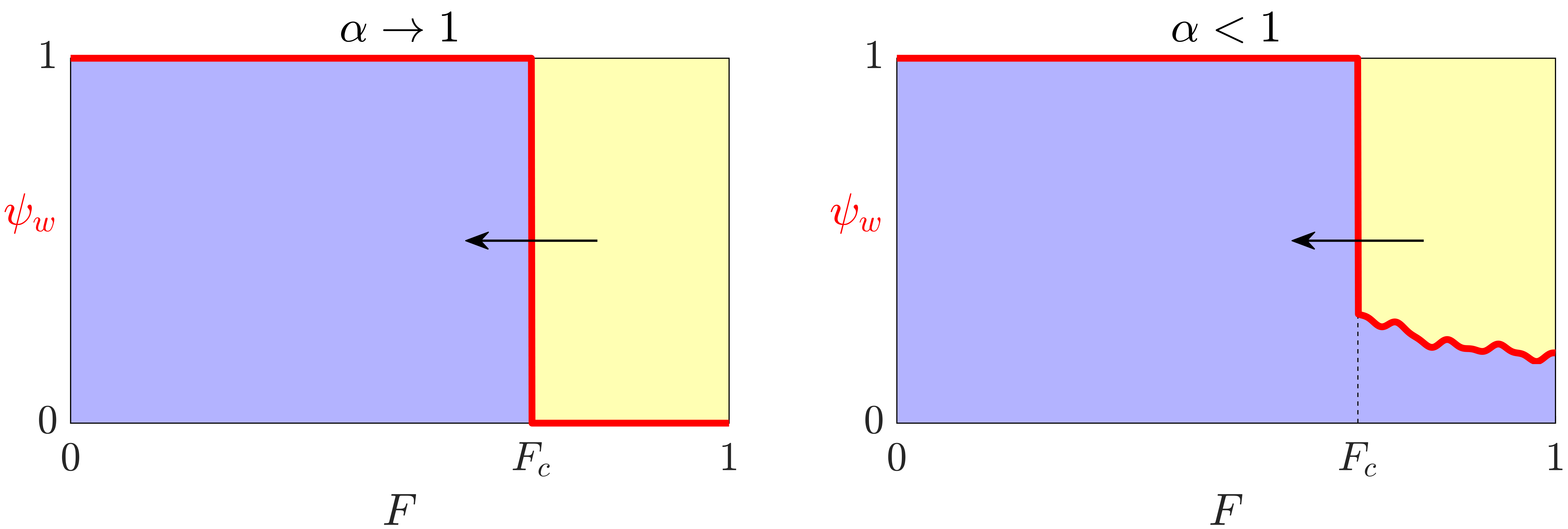}
\caption{Different qualitative features of the radius-resolved wetting-phase saturation, $\psi_w(F)$, during quasistatic primary drainage, where the nonwetting fluid replaces the wetting fluid as we raise $p_c$ (thus lowering $F_c$). When accessivity is nearly unity (left), all pores larger than $F_c$ are accessible by the nonwetting phase and will undergo drainage; when accessivity is less than unity (right), although all pores larger than $F_c$ will favor drainage, only some of them actually drains, the details of which are captured by the $\psi_w(F)$ function.}\label{figure:rrs_illustration}
\end{figure}

Now we apply the concept of radius-resolved saturation to invasion percolation. When $\alpha\to1$ (e.g., capillary bundle), as $p_c$ is quasistatically raised to a particular value, all pores larger than $r_c$ (and hence $F_c$) will undergo drainage, so we expect the following radius-resolved saturation:
\begin{align}
    \psi_w\left(F;F_c\right) =
    \begin{cases}
    1, & F < F_c \\
    0, & F > F_c,
    \end{cases}\label{eq:rrs_bundle}
\end{align}
which states that relative to the $F_c$ that corresponds to the imposed $p_c$, all pores that are smaller are $w$-filled ($\psi_w=1$), and all larger pores are $n$-filled ($\psi_w=0$). On left-hand side of Figure \ref{figure:rrs_illustration}, we plot the function $\psi_w\left(F\right)$ in red for some arbitrary $F_c$, for the case of $\alpha\to1$. The areas colored in blue and yellow correspond to the two integrals given in Eq. \eqref{eq:saturations_and_rrs}, hence $s_w$ and $s_n$, respectively. From the figure, it is clear that $s_w = F_c$ and $s_n = 1-F_c$ in this case, as expected.

On the other hand, if accessivity is lower, indicating some serial connectivity between different sized pores, then the radius-resolved saturation will generally evolve differently, as shown on the right-hand side of Figure \ref{figure:rrs_illustration}. We expect:
\begin{align}
    \psi_w\left(F;F_c\right) =
    \begin{cases}
    1, & F < F_c \\
    \text{varies between }0\text{ and }1, & F > F_c.\label{eq:primary_drainage_rrs_general}
    \end{cases}
\end{align}
As before, during primary drainage, all pores that are smaller than $F_c$ will remain $w$-filled, but with $\alpha<1$, some larger pores may remain $w$-filled, too. In fact, the integral $\int_{F_c}^1 \psi_w(F) \mathrm{d}F$ corresponds to wetting fluid in larger pores ($F>F_c$) favoring drainage yet cannot drain because they are blocked from the nonwetting fluid by smaller pores ($F<F_c$). In the context of primary drainage, this quantity is precisely Hilfer's $s_2$ (assuming the solid matrix is water-wet), or the saturation of the ``non-percolating water subphase'' \cite{hilfer1998,hilfer2006}. Because $s_2>0$, we have $s_w > F_c$ and $s_n < 1-F_c$, which is clear from Figure \ref{figure:rrs_illustration}.

\begin{figure}[!h]
\centering
\includegraphics[width=\textwidth]{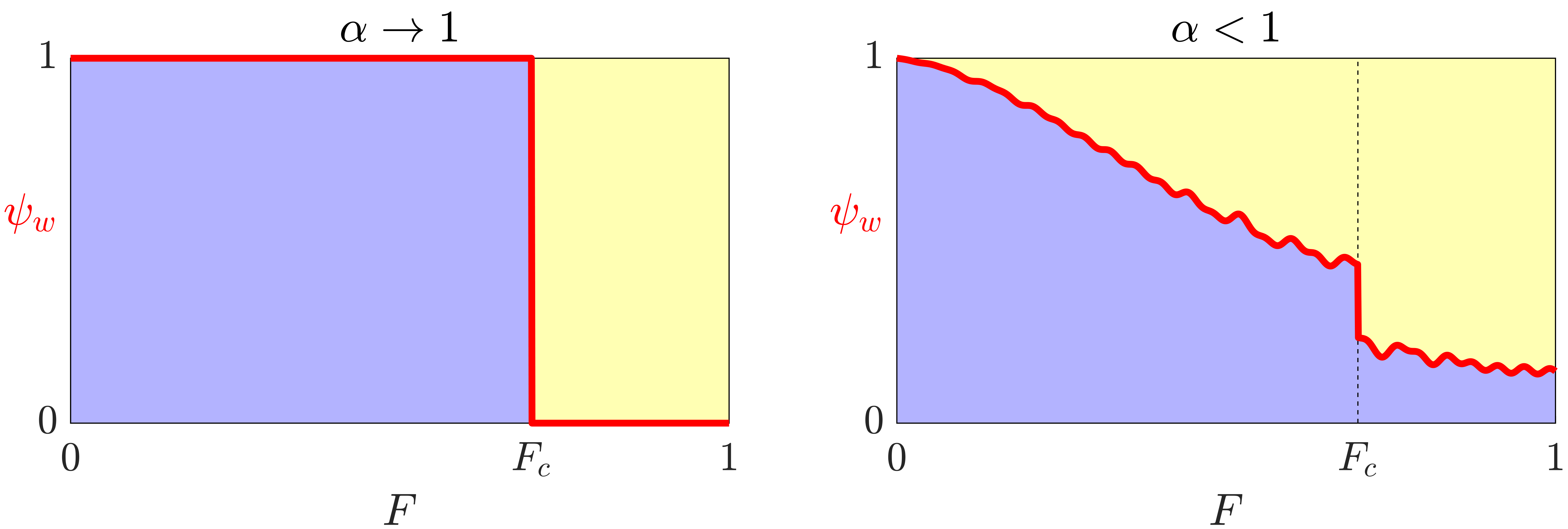}
\caption{Different qualitative features of the radius-resolved saturation, $\psi_w(F)$, after a number of arbitrary drainage-imbibition cycles. When accessivity is nearly unity (left), all pores larger than $F_c$ are accessible by the nonwetting phase and will undergo drainage; when accessivity is less than unity (right), although all pores larger than $F_c$ will favor drainage, only some of them actually drains, which is captured by the $\psi_w(F)$ function.}\label{figure:rrs_illustration_2}
\end{figure}

Now consider the general situation of a control volume that has undergone a number of arbitrary drainage-imbibition cycles before the capillary pressure is finally brought to $p_c$. Figure \ref{figure:rrs_illustration_2} depicts the expected $\psi_w(F)$ profiles for porous media with different accessivities. For $\alpha\to1$, the radius-resolved saturation $\psi_w(F)$ is the same as before (see Figure \ref{figure:rrs_illustration} and Eq. \ref{eq:rrs_bundle}) -- because the entire pore space is directly accessible by external fluids, all pores that are smaller (larger) than $F_c$ are filled with the wetting (non-wetting) phase. There is a one-to-one correspondence between $s_w$ and $\psi_w(F)$, i.e., one can reproduce the radius-resolved saturation function from $s_w$, which correctly implies that the capillary bundle model does not predict connectivity-based hysteresis in $p_c(s_w)$.

On the other hand, in a real porous medium with an accessivity that is lower than unity, $\psi_w(F)$ can acquire nontrivial shapes after arbitrary cycles of drainage and imbibition, an example of which is given on the right-hand side of Figure \ref{figure:rrs_illustration_2}. We see that $\psi_w(F)$ is a better representation of the microscopic state of porous medium than $s_w$ (given by the blue area), as the former registers the effects of the flow history on the current microscopic distribution of fluids.

\begin{figure}[!h]
\centering
\includegraphics[width=0.5\textwidth]{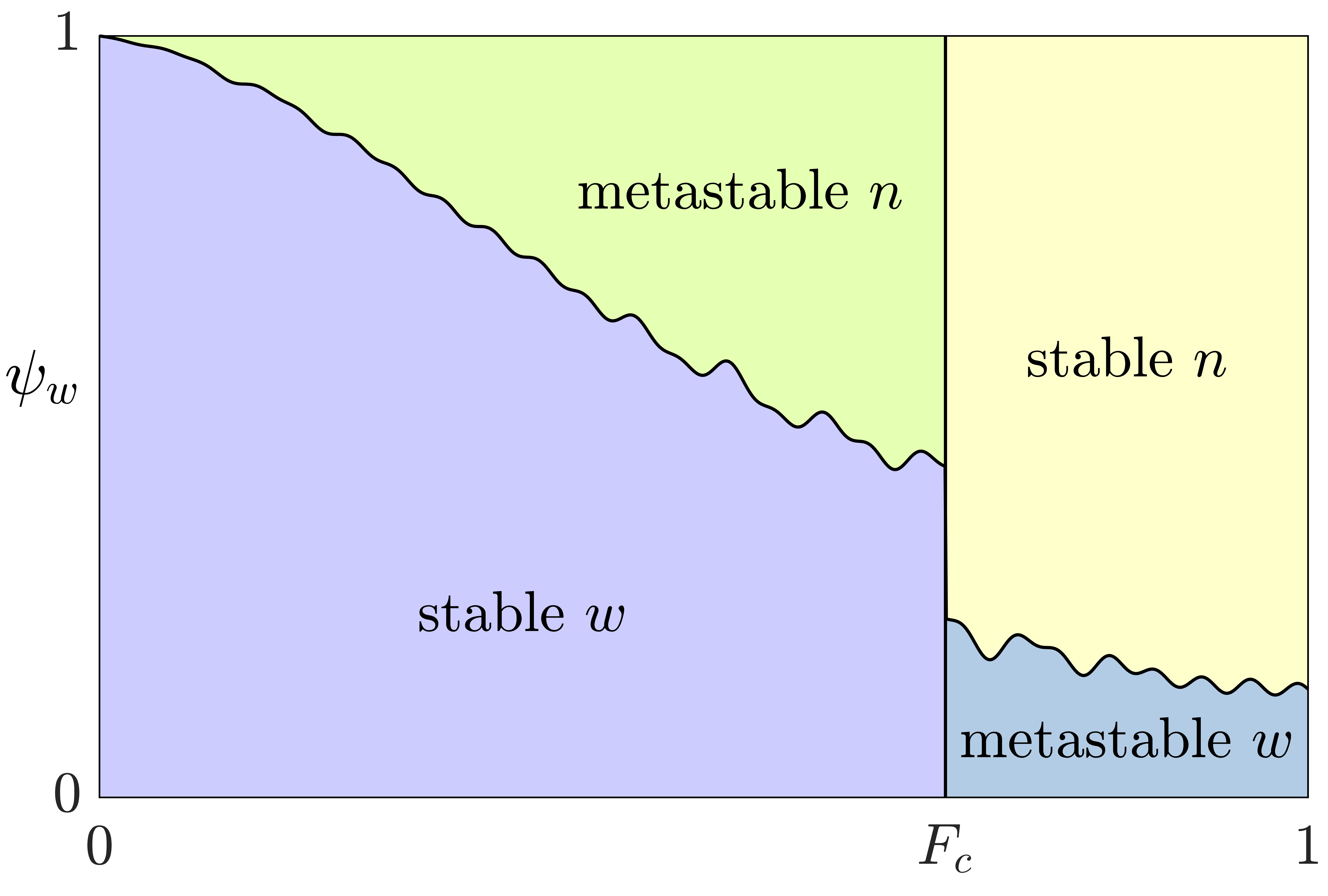}
\caption{The four ``quadrants'' in the right-hand side plot of Figure \ref{figure:rrs_illustration_2}, whose areas define four saturation variables with distinct physical meanings. These lumped saturation variables are reminiscent of (but differ from) Hilfer's four saturation variables based on percolation, yet lack the precise connection to the PSD found in the radius-resolved saturation, $\psi_w(F)$}\label{figure:rrs_quadrants}
\end{figure}

Interestingly, $\psi_w(F)$ may have some loose connection to Hilfer's four saturation variables \cite{hilfer1998,hilfer2006}, $s_1,\ldots,s_4$. We note that in the right-hand side plot of Figure \ref{figure:rrs_illustration_2}, the $[0,1]\times[0,1]$ domain is divided into four ``quadrants'' by the red $\psi_w(F)$ curve and the dashed vertical line at $F_c$. By recalling that a point in the blue (yellow) area corresponds to a pore filled with the wetting (nonwetting) fluid, and that a point to the left (right) of the $F=F_c$ line corresponds to a pore that favors imbibition (drainage), we find it appealing to interpret the areas of the four ``quadrants'' as follows:
\begin{align}
    \int_{0}^{F_c} \psi_w(F) \mathrm{d}F
    \quad
    \begin{matrix}\text{stable}\\w\text{ phase}\end{matrix} &\sim \text{Hilfer's } s_1, \label{eq:stable_w} \\
    \int_{F_c}^{1} \psi_w(F)\mathrm{d}F
    \quad
    \begin{matrix}\text{metastable}\\w\text{ phase}\end{matrix} &\sim \text{Hilfer's } s_2,
    \label{eq:metastable_w} \\
    \int_{F_c}^{1} 1-\psi_w(F) \mathrm{d}F
    \quad
    \begin{matrix}\text{stable}\\n\text{ phase}\end{matrix} &\sim \text{Hilfer's } s_3, \label{eq:stable_n} \\
        \int_{0}^{F_c} 1-\psi_w(F) \mathrm{d}F
    \quad
    \begin{matrix}\text{metastable}\\n\text{ phase}\end{matrix} &\sim \text{Hilfer's } s_4, \label{eq:metastable_n}
\end{align}
as demonstrated in Figure \ref{figure:rrs_quadrants}.

For instance, the integral in Eq. (\ref{eq:metastable_w}) corresponds to the area of the lower right ``quadrant'' in Figure \ref{figure:rrs_quadrants}, which gives the volume fraction of pores that are $w$-filled but large enough to favor drainage at the imposed $F_c$. We say these pores contain ``metastable $w$ phase'' because the wetting fluid would have drained out of those pores if they were accessed by mobile menisci. We note that the four integrals identified in Eq. \ref{eq:stable_w} -- \ref{eq:metastable_n} are not identical to Hilfer's definitions of $s_1,\ldots,s_4$, which are based on percolation rather than pore size. For instance, a pore with ``stable $n$ phase'' is not necessarily percolating to an external reservoir of the nonwetting fluid, and could be part of a trapped ganglion. Nevertheless, it is remarkable that the conventional saturations ($s_w$ and $s_n$) can be intuitively subdivided into four saturation-like variables in either case. Our framework is different in that it also affords an intuitive and precise connection to the PSD, and, in fact, goes beyond a four-variable description through characterizing the microscopic distribution of fluid phases across different pore sizes using the function $\psi_w(F)$.

\section{Statistical theory}

So far, we have proposed accessivity, $\alpha$, as a continuum-scale descriptor for the connectivity of different sized pores, and radius-resolved saturation, $\psi_w(F)$, for the distribution of immiscible fluid phases in the pore space. We have qualitatively demonstrated that, in a porous medium with $\alpha<1$, serial connections between different sized pores may contribute to the accumulation of ``metastable'' fluids (i.e., wetting phase in ``larger'' pores, or nonwetting phase in ``smaller'' pores, relative to a pore at capillary equilibrium) as a result of the ink-bottle effect, making $\psi_w(F)$ a more informative representation of the state of fluids in the pore space than the conventional saturation, $s_w$.

As much as we expect $\alpha$ and $\psi_w(F)$ to have much broader utility in the continuum modeling of multiphase processes in porous media, it is also desirable to develop a basic theoretical framework that captures their key ``mechanics'' based on pore-scale physical principles. For instance, we might expect a porous medium with a low $\alpha$ to experience much hysteresis in $p_c(s_w)$ due to significant ink-bottle effect, which should manifest as changes in $\psi_w(F)$ that are strongly history-dependent -- our framework should be able to reproduce these features by laying out a set of quantitative rules that are both physically intuitive and mathematically simple. The framework would also constitute an incremental improvement over the capillary bundle approach by remedying its most prominent shortcoming \cite{diamond2000,hunt2013}: its disregard for connectivity effects. Accuracy and predictability are not our objectives here -- think of ``Fick's law for connectivity effects''.

In regard to the scope of this paper, we will focus on computing quasistatic capillary pressure hysteresis in the context of two-phase flow, mercury porosimetry, and related processes. Dynamical effects, including simple models of relative permeability hysteresis, will be explored in future works. First, we model pore-space morphology as a statistical branching process, and regard the pore space within a representative control volume of the porous medium as an ensemble of ``pore instances''. Then, we analyze the movement of menisci in a pore instance based on either ordinary differential equations (ODEs) or algebraic arguments, which yield a set of simple formulae that govern how $\psi_w(F)$ evolves for arbitrary but quasistatic variations in $p_c$ in porous media with different $\alpha$. Finally, we demonstrate that hysteretic $p_c(s_w)$ loops are produced naturally by means of updating $\psi_w(F)$, while $\alpha$ acts as the sole parameter that controls the amount of hysteresis.

\subsection{Pore-space morphology}

\paragraph{Overview} The first step in developing our framework is to propose a procedure for conceptualizing the pore space in a porous medium with a given accessivity, $\alpha$. The procedure will be based on a statistical branching process, where $\alpha$ quantitatively controls the overall incidence of pore-radius variations. The idea here is similar to the rationale behind the microscopic interpretation of tortuosity through Eq. \eqref{eq:tortuosity_definition_geometric}, which is given for an idealized pore space and barely holds for generic porous media, although the physical interpretation of the quantity remains conceptually useful nevertheless.

\paragraph{Skeletal reduction of pore space} We assume that the entire pore space can be mapped to intersecting space curves, which may be accomplished by tracing out the ``skeleton'' of the pore space, defined as the set of points that are equidistant from nearby surfaces of the solid matrix \cite{bakke1997}. We refer to the individual segments of the space curves as pore branches, and their points of intersection as junctions.

\begin{figure}[!h]
\centering
\includegraphics{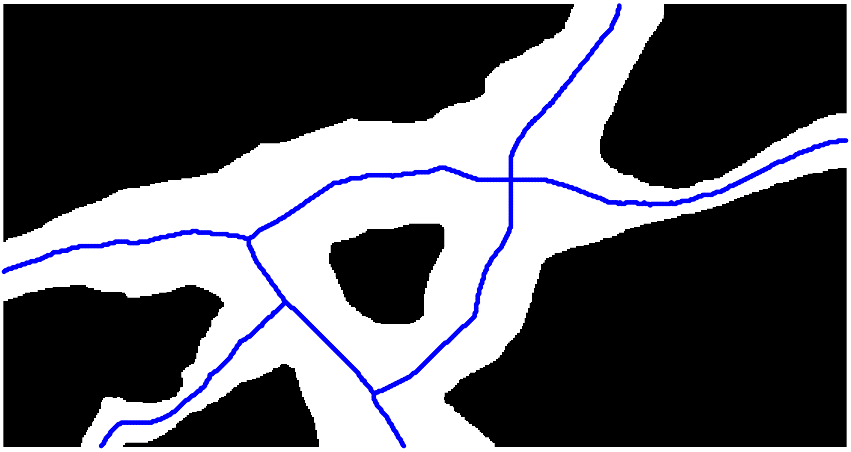}
\caption{A hypothetical 2-D porous sample (where black and white areas correspond to the solid matrix and the pore space, respectively) and its pore skeleton (blue curves). The various junctions display different coordination numbers.}\label{figure:pore_skeleton}
\end{figure}

The coordination number of a junction, denoted by $z$, is equal to the number of branches emanating from that junction. In general, we have:
\begin{align}
    z = 1,3,4,\dots,\label{eq:allowed_coordination_numbers}
\end{align}
where a 1-coordinate junction is either a dead end or a point on the boundary of the domain of the medium. Figure \ref{figure:pore_skeleton} shows the pore skeleton of a hypothetical 2-D porous sample.

Each pore branch is subsequently parameterized by some axial coordinate that corresponds to locations along the skeleton from one junction to another, which may be based on the arc length, the pore volume, or any other appropriate measure of size. We denote the size of a pore branch, i.e., its total arc length or total pore volume, by $b$.

\paragraph{Effective radius and PSD} We approximate the exact geometry of the pore space by assigning some effective radius as a function of the axial coordinate along the space curve segment for each pore branch. The meaning of ``effective'' depends on the application of interest: if we wish to study the displacement of immiscible fluids, for example, we may evaluate the equilibrium capillary pressure, $p_c$, across a stationary meniscus placed at a certain location along the pore branch, and convert it to an effective radius of $r_c$, the equilibrium capillary radius, using the Washburn equation (Eq. \eqref{eq:Washburn}).

By compiling radius measurements along all branches weighted by the pore volume associated with each radius, we obtain a PSD, whose cumulative distribution function we denote by $F\left(r\right)$.

\begin{figure}[!h]
\centering
\includegraphics[width=0.4\textwidth]{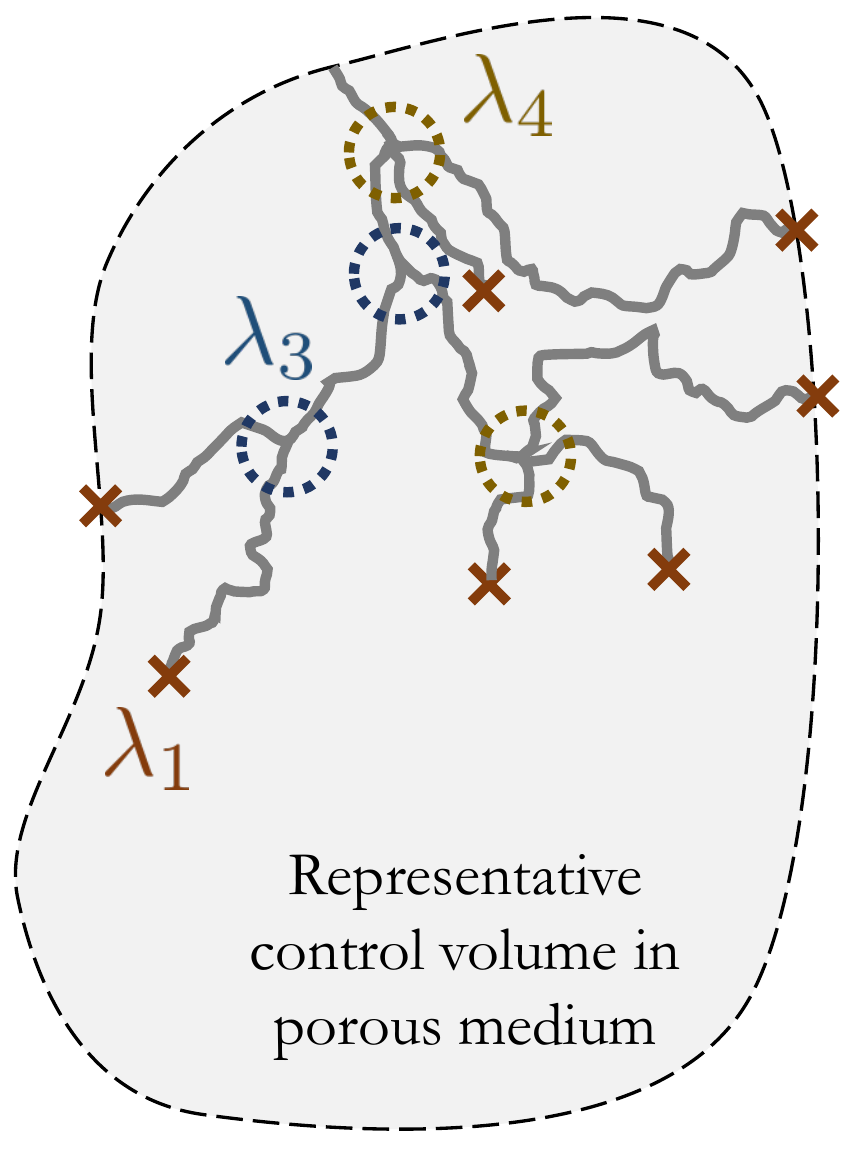}
\caption{An cartoon illustration of the branching process, which forms a random pore-space instance. Junctions of various coordination numbers occur at distinct frequencies, each following an independent Poisson process. Note the absence of loops in an instance.}\label{figure:branching_process}
\end{figure}

\paragraph{Probabilistic occurrence of junctions} We devise a probabilistic branching process by which we construct an conceptualized pore space. Imagining traversing the pore space along its skeleton beginning from an arbitrary location, we assume that the occurrence of junctions of each coordination number is described by a homogeneous Poisson point process. That is, we encounter $z$-coordinate junctions at a fixed rate of $\lambda_z$, which has the unit of the reciprocal of the axial coordinate, e.g., $\left[\mathrm{L}\right]$ or $\left[\mathrm{L}^{-3}\right]$. A random experiment following the above probabilistic rules yields a particular ``instance'' of the pore space, as shown in Figure \ref{figure:branching_process}. The overall pore space is described by an ensemble of all possible instances, probabilistically weighted. Because no loops are formed in this probabilistic branching process, the resulting pore instances resemble a Bethe lattice \cite{sahimi1994,stauffer2014}, although in our case the coordination number may vary from junction to junction.

The size of a branch, denoted by $b$ and measured in the unit of the axial coordinate, is equal to the distance travelled before encountering a junction of any coordination number, which occurs at a rate of:
\begin{align}
    \lambda_{\mathbb{Z}^+}
    = \sum\limits_{z\in\mathbb{Z}^+}\lambda_z,\label{eq:total_branching_rate}
\end{align}
where $\mathbb{Z}^+$ is the set of all positive integers; here, we let $\lambda_2=0$ (despite Eq. \eqref{eq:allowed_coordination_numbers}) for convenience of notation (or we could have replace $\mathbb{Z}^+$ with $\mathbb{Z}^+ \setminus \{2\}$, to the same effect). It is a property of homogeneous Poisson point processes that the size of a branch follows an exponential distribution, whose probability density is given by:
\begin{align}
    \mathrm{Pr}\left(b\right)
    = \lambda_{\mathbb{Z}^+}
    \exp{\left(-\lambda_{\mathbb{Z}^+}b\right)}.
\end{align}
The mean of the distribution, or the expected size of a branch, is given by:
\begin{align}
    \langle b \rangle = \frac{1}{\lambda_{\mathbb{Z}^+}}.\label{eq:expected_size_branch}
\end{align}
The expected size of an entire instance, denoted by $\langle c \rangle$, is the sum of the sizes of all the branches it contains. It follows the recursive relation:
\begin{align}
    \langle c \rangle = \langle b \rangle
    + \sum\limits_{z\in\mathbb{Z}^+}
    \left(\frac{\lambda_z}{\lambda_{\mathbb{Z}^+}}\right)
    \left( z-1 \right) \langle c \rangle,\label{eq:recursive_size_instance}
\end{align}
which states that encountering a $z$-coordinate junction, which occurs with conditional probability $\left(\lambda_z/\lambda_{\mathbb{Z}^+}\right)$, shall give rise to $\left(z-1\right)$ new branches, each of which behaving independently like a new instance, hence having an expected size of $\langle c \rangle$. Combining Eqs. \eqref{eq:total_branching_rate}, \eqref{eq:expected_size_branch}, and \eqref{eq:recursive_size_instance} gives a formula for the expected size of an instance:
\begin{align}
    \langle c \rangle
    &= \frac{1/\lambda_{\mathbb{Z}^+}}
    {1-\sum\limits_{z\in\mathbb{Z}^+}
    \left(\frac{\lambda_z}{\lambda_{\mathbb{Z}^+}}\right)
    \left( z-1 \right)}\nonumber\\
    &= \frac{1}
    {\lambda_{\mathbb{Z}^+}
    -\sum\limits_{z\in\mathbb{Z}^+}
    \lambda_z\left( z-1 \right)}\nonumber\\
    &= \frac{1}
    {\sum\limits_{z\in\mathbb{Z}^+}
    \left[ \lambda_z
    -\lambda_z\left( z-1 \right) \right]}\nonumber\\
    \langle c \rangle &= \frac{1}
    {\sum\limits_{z\in\mathbb{Z}^+}
    \left( 2-z \right)\lambda_z},\label{eq:expected_size_instance}
\end{align}
which is expected to be finite in a finite sized control volume, requiring that the denominator be greater than zero.

\paragraph{Probabilistic radius variation along branches}
Similarly, we assume that the radius along a pore branch varies according to a homogeneous Poisson point process at a constant rate of $\ell^{-1}$. Strictly speaking, this results in a series of constant-radius pore segments, whose sizes follow an exponential distribution with a mean of $\ell$, and whose radii are drawn at random from a prescribed PSD. Since pore radius generally varies smoothly in real porous media, we may interpret $\ell$ as the typical distance or volume over which the pore radius varies significantly along the axial coordinate. Conceptually speaking, if we examine $r$ as a function of the pore axial coordinate in the frequency domain, $\ell^{-1}$ should reflect the mean location of the peak signals. It is also possible for $\ell$ to depend on the instantaneous pore radius, e.g., the pore radius may vary more rapidly as it becomes smaller, but here we will assume that there exists some average $\ell$ that works across all pore sizes.

\paragraph{Geometric definition of accessivity} We now have all the ingredients to define accessivity based on geometric properties for the idealized case described above. Firstly, we define:
\begin{align}
    q = \frac{\ell}{\langle c \rangle}
    = \ell \sum\limits_{z\in\mathbb{Z}^+}
    \left( 2-z \right)\lambda_z,\label{eq:q_definition}
\end{align}
which requires Eq. \eqref{eq:expected_size_instance}. Next, accessivity is taken as:
\begin{align}
    \alpha = \frac{q}{1+q},\label{eq:q2alpha}
\end{align}
which we can also write as:
\begin{align}
    q = \frac{\alpha}{1-\alpha}.\label{eq:alpha2q}
\end{align}
Here, $q\in\left(0,\infty\right)$ is the ratio of the size of a constant-radius pore segment to that of an average instance of the pore space. Hence, $1/q\in\left(0,\infty\right)$ gives the expected number of pore-size changes per instance, and $1/\alpha=1+1/q\in\left(1,\infty\right)$ the expected number of different sized pores encountered per instance. Thus, $\alpha\in\left(0,1\right)$ is the fraction of an average instance that corresponds to constant-radius pore segments immediately accessible from the exterior of the control volume. As $\alpha\to1$, each instance contains only one pore size, making it entirely accessible from the exterior, while different pore sizes are only found in parallel instances. As $\alpha\to0$, an infinite number of distinct pore sizes are observed in an instance, meaning that a vanishing small fraction of each instance is accessible, and that different sized pores are organized in a highly serial manner. Increasing $\lambda_1$ in Eq. \eqref{eq:q_definition} increases the rate of encountering dead ends, which, at fixed $\ell$, decreases the expected size of the pore space, $\langle c \rangle$, hence increasing $\alpha$ and reducing connectivity effects. On the other hand, increasing $\lambda_z$ for $z\ge3$ increases the occurrences of high-coordinate junctions, which increases $\langle c \rangle$, hence decreasing $\alpha$ and promoting connectivity effects.

\subsection{Quasistatic fluid movements}

\paragraph{Overview} Our next step is to model quasistatic immiscible fluid displacement in a conceptualized pore space with to a given accessivity. To make our discussion more concrete, we will assume the context of mercury porosimetry, where the intrusion and extrusion of mercury correspond to quasistatic drainage and imbibition, respectively. In an intrusion experiment, we measure the intruded volume of mercury, $V$, as a function of its pressure, $p_c$. We normalize volume measurements to the maximum intruded volume, which defines the saturation of mercury, $s_n = {V}/{V_{\mathrm{max}}}$.
Our goal is to derive a formula for $s_n\left(p_c\right)$ given $F_c=F\left(r_c\right)$ and $\alpha$. Using Eq. \eqref{eq:Washburn}, we convert from $p_c$ to the effective radius of the smallest penetrable pores, $r_c$, assuming we know $\gamma$ and $\theta$. The remaining task in this analysis is to relate $s_n$ and $F_c$ for a certain $\alpha$, that is, given that a fraction $F_c$ of the pore space is too small in $r_c$ for mercury to invade at the imposed $p_c$, what volume fraction of the pore space, $s_n$, will become filled with mercury (see Figure \ref{figure:variables_causal})? 

As we have discussed, the answer is straightforward if the pore space is well represented by a bundle of straight capillaries, corresponding to $\alpha\to1$:
\begin{align}
    s_n = 1 - F_c,\label{eq:s_F_capillary_bundle}
\end{align}
which says that all pores that are large enough for mercury to penetrate will indeed become filled with mercury. Implicit in the standard approach to interpreting mercury intrusion data, Eq. \eqref{eq:s_F_capillary_bundle} is known to be ``wrong'' \cite{diamond2000}, but frequently used in practice nonetheless.

We hope to derive formulae for $s_n\left(F_c\right)$ that better captures connectivity effects than Eq. \eqref{eq:s_F_capillary_bundle}, but are still simple enough for practical use. This will be achieved through the inclusion of $\alpha$ as a model parameter, which would describe the effects of connectivity of different sized pores in a simple fashion.

To keep the derivation approachable, we will first based our discussion on the relatively simple case of mercury intrusion porosimetry, before generalizing the formulae to arbitrary drainage-imbibition cycles. During mercury intrusion, $F_c$ is lowered from 1 to 0, causing $s_n$ to increase from 0 to 1. We assume that any increase in $s_n$ can be attributed to the advancement of menisci into the ensemble of instances that make up the pore space. We use $\omega$ to denote the number of menisci per instance that may contribute to intrusion, and assume $\omega\left(F_c=1\right)=1$, i.e., there is one meniscus for every instance of the pore space at the beginning of intrusion. Now, we consider how $s_n$ and $\omega$ change in response to a differential change $\mathrm{d}F_c<0$ (since $F_c$ decreases during intrusion).

\paragraph{Number of advancing menisci} Since $s_n$ will not change if $F_c$ remains unchanged, we deduce that every meniscus must be immediately upstream to a pore whose radius is smaller than $r_c$, which prevents mercury from intruding further into that pore branch. We say such a meniscus is in the ``pinned'' state. When $F_c$ is reduced to $F_c+\mathrm{d}F_c$, only menisci that are adjacent to a pore whose radius falls in the interval $\left[r_c+\mathrm{d}r_c,r_c\right)$ will advance downstream. Thus, the number of menisci per instance that move in response to $\mathrm{d}F_c$ is equal to:
\begin{align}
    \delta\omega_{\text{adv}}
    = \omega\frac{-\mathrm{d}F_c}{F_c},\label{eq:omega_advancing}
\end{align}
where the fraction $-\mathrm{d}F_c/F_c$ is the conditional probability that the radius of the pore next to the meniscus is in $\left[r_c+\mathrm{d}r_c,r_c\right)$, given that it is in $\left(0,r_c\right)$.

\paragraph{Probabilistic events experienced by an advancing meniscus} A meniscus that begins moving in response to $\mathrm{d}F_c$ may experience either of the following two events as it travels along the pore branch:
\begin{itemize}
\item The meniscus encounters a pore with a radius smaller than $r_c$ and returns to the pinned state, which occurs at a rate of:
\begin{align}
    \lambda_{r_c^-} = \frac{F_c}{\ell}\label{eq:small_pore_rate},
\end{align}
where $F_c$ is the probability that a radius drawn randomly from the PSD is less than $r_c$, and $\ell^{-1}$ is the average rate for pore radius variation;
\item The meniscus encounters a junction (of any coordination number), which occurs at a rate of $\lambda_{\mathbb{Z}^+}$ (see Eq. \eqref{eq:total_branching_rate}).
\end{itemize}

\paragraph{Mean displacement of an advancing meniscus} We denote the mean total displacement of an advancing meniscus and all its descendants by $\langle d \rangle$. It must follow the recursive relation (cf. Eq. \eqref{eq:recursive_size_instance}):
\begin{align}
    \langle d \rangle
    = \frac{1}{\lambda_{r_c^-}+\lambda_{\mathbb{Z}^+}}
    + \sum\limits_{z\in\mathbb{Z}^+}
    \left(\frac{\lambda_z}
    {\lambda_{r_c^-}+\lambda_{\mathbb{Z}^+}}\right)
    \left( z-1 \right)\langle d \rangle. \label{eq:recursive_size_displacement}
\end{align}
Here, $1/\left(\lambda_{r_c^-}+\lambda_{\mathbb{Z}^+}\right)$ is the rate at which the advancing meniscus either becomes pinned or encounters a junction. Given that, it may be the case that the meniscus encounters a $z$-coordinate junction before it gets pinned, which occurs with conditional probability $\lambda_z/\left(\lambda_{r_c^-}+\lambda_{\mathbb{Z}^+}\right)$; this would transform the meniscus into $\left(z-1\right)$ independent menisci, each of which independently traverses one of the additional pore branches, and is subject to the same two events described above. (Note that in the case of $z=1$, the meniscus either encounters a dead end or exits the domain of the porous medium, thereby reducing the number menisci responsible for further mercury intrusion to zero.) On the other hand, if the meniscus gets pinned before it encounters a junction, which occurs with conditional probability, $\lambda_{r_c^-}/\left(\lambda_{r_c^-}+\lambda_{\mathbb{Z}^+}\right)$, intrusion will stop, resulting in no further increase in $\langle d \rangle$.

We can solve Eq. \eqref{eq:recursive_size_displacement} for $\langle d \rangle$, similar to how we arrived at Eq. \eqref{eq:expected_size_instance}. We obtain:
\begin{align}
    \langle d \rangle = \frac{1}
    {\lambda_{r_c^-} + \sum\limits_{z\in\mathbb{Z}^+}
    \left( 2-z \right)\lambda_z},
\end{align}
which we simplify by inserting Eqs. \eqref{eq:q_definition} and \eqref{eq:small_pore_rate}:
\begin{align}
    \langle d \rangle
    = \frac{1}{F/\ell + q/\ell}.\label{eq:expected_size_displacement}
\end{align}

\paragraph{ODE for saturation} As $F_c$ is reduced to $F_c+\mathrm{d}F_c$, the differential change in the saturation of mercury, $\mathrm{d}s_n$, must satisfy:
\begin{align}
    \langle c \rangle\mathrm{d}s_n = \delta\omega_{\text{adv}}
    \langle d \rangle.\label{eq:ODE_saturation_raw}
\end{align}
In words, $\langle c \rangle\mathrm{d}s_n$ is the differential amount of mercury intrusion observed per instance in response to $\mathrm{d}F_c$, measured in the unit of the pore axial coordinate. It is equal to the number of advancing menisci per instance, $\delta\omega_{\text{adv}}$, multiplied by the mean total displacement of each meniscus, $\langle d \rangle$.\\
Substituting Eqs. \eqref{eq:q_definition}, \eqref{eq:omega_advancing}, and \eqref{eq:expected_size_displacement} into Eq. \eqref{eq:ODE_saturation_raw}, we obtain an ODE for $s_n\left(F_c\right)$:
\begin{align}
    \left(\frac{\ell}{q}\right)\mathrm{d}s_n &= \left(\omega\frac{-\mathrm{d}F_c}{F_c}\right)
    \left(\frac{1}{F_c/\ell + q/\ell}\right)\nonumber\\
    \frac{\mathrm{d}s_n}{\mathrm{d}F_c}
    &= -\frac{\omega}{F_c}
    \frac{q}{F_c + q}.
    \label{eq:ODE_saturation_final}
\end{align}

\paragraph{Mean number of descendants of an advancing meniscus} An advancing meniscus can potentially transform into many descendent menisci. The mean number of descendants per meniscus (including their progenitor), which we denote by $\langle n \rangle$, is given by the recursive relation:
\begin{align}
    \langle n \rangle
    = \left(\frac{\lambda_{r_c^-}}
    {\lambda_{r_c^-}+\lambda_{\mathbb{Z}^+}}\right)
    \left(1\right)
    + \sum\limits_{z\in\mathbb{Z}^+}
    \left(\frac{\lambda_z}
    {\lambda_{r_c^-}+\lambda_{\mathbb{Z}^+}}\right)
    \left( z-1 \right)\langle n \rangle,\label{eq:recursive_number_descendants}
\end{align}
which says that an advancing meniscus remains one meniscus in the case that it becomes pinned, but turns into $\left( z-1 \right)\langle n \rangle$ menisci if it encounters a $z$-coordinate junction. Solving for $\langle n \rangle$:
\begin{align}
    \langle n \rangle = \frac{\lambda_{r_c^-}}
    {\lambda_{r_c^-} + \sum\limits_{z\in\mathbb{Z}^+}
    \left( 2-z \right)\lambda_z},
\end{align}
which we simplify by inserting Eqs. \eqref{eq:q_definition} and \eqref{eq:small_pore_rate} to arrive at:
\begin{align}
    \langle n \rangle
    = \frac{F_c/\ell}{F_c/\ell + q/\ell}
    = \frac{F_c}{F_c + q}.\label{eq:expected_number_descendants}
\end{align}

\paragraph{ODE for number of available menisci} As $F_c$ is reduced to $F_c+\mathrm{d}F_c$, the differential change in the number of menisci per instance available for further mercury intrusion, $\mathrm{d}\omega$, is equal to:
\begin{align}
    \mathrm{d}\omega = \delta\omega_{\text{adv}}
    \left( \langle n \rangle - 1 \right),\label{eq:ODE_number_menisci_raw}
\end{align}
where $\left( \langle n \rangle - 1 \right)$ represents the net growth in the number of menisci for each advancing meniscus. Substituting Eqs. \eqref{eq:omega_advancing} and \eqref{eq:expected_number_descendants} into Eq. \eqref{eq:ODE_saturation_raw}, we obtain an ODE for $\omega\left(F_c\right)$:
\begin{align}
    \mathrm{d}\omega &= \left(\omega\frac{-\mathrm{d}F_c}{F_c}\right)
    \left(\frac{F_c}{F_c + q}-1\right)\nonumber\\
    \frac{\mathrm{d}\omega}{\mathrm{d}F_c}
    &= \frac{\omega}{F_c}
    \frac{q}{F_c + q}.
    \label{eq:ODE_number_menisci_final}
\end{align}

\paragraph{Analytical solutions} To recapitulate, we have derived a system of ODEs for $s_n$ and $\omega$ consisting of Eqs. \eqref{eq:ODE_saturation_final} and
\eqref{eq:ODE_number_menisci_final}. Firstly, we solve Eq. \eqref{eq:ODE_number_menisci_final}, subject to $\omega\left(F_c=1\right)=1$:
\begin{align}
    \frac{\mathrm{d}\omega}{\omega}
    &= \frac{q}{{F_c\left(F_c + q\right)}} \mathrm{d}F_c\nonumber\\
    \int_1^{\omega}\frac{\mathrm{d}\widehat{\omega}}{\widehat{\omega}}
    &= \int_1^{F_c}\left( \frac{1}{\widehat{F}} - \frac{1}{\widehat{F}+q} \right) \mathrm{d}\widehat{F}\nonumber\\
    \ln{\omega} &= \ln{F_c} - \ln{\left(\frac{F_c+q}{q+1}\right)}\nonumber\\
    \omega &= \frac{\left(q+1\right)F_c}
    {F_c+q}.\label{eq:number_menisci_solution}
\end{align}
Secondly, by comparing Eqs. \eqref{eq:ODE_saturation_final} and \eqref{eq:ODE_number_menisci_final}, we find:
\begin{align}
    \frac{\mathrm{d}s_n}{\mathrm{d}F_c}
    = -\frac{\mathrm{d}\omega}{\mathrm{d}F_c} \implies
    \frac{\mathrm{d}s_n}{\mathrm{d}\omega}
    = -1\nonumber.
\end{align}
Since $s_n\left(F_c=1\right)=0$, we have $s_n\left(\omega=1\right)=0$. Hence:
\begin{align}
    s_n &= 1 - \omega\nonumber\\
    s_n &= \frac{q\left(1-F_c\right)}
    {F_c+q}\nonumber\\
    s_n(F_c) &= \frac{\alpha\left(1-F_c\right)}
    {(1-\alpha)F_c+\alpha}.\label{eq:saturation_intrusion_solution}
\end{align}
Eq. \eqref{eq:saturation_intrusion_solution} is our formula for the $s_n\left(F_c\right)$ relationship during mercury intrusion (or primary drainage), where the accessivity, $\alpha$, serves as a model parameter for controlling connectivity effects. Its limiting behavior when $\alpha\to1$:
\begin{align}
    \lim_{\alpha\to1} s_n(F_c)
    = \lim_{\alpha\to1} \frac{\alpha\left(1-F_c\right)}
    {(1-\alpha)F_c+\alpha}
    = 1-F_c\label{eq:saturation_intrusion_solution_limiting}
\end{align}
coincides with Eq. \eqref{eq:s_F_capillary_bundle}, which is the prediction of the capillary bundle model.

We may write $s_n\left(F_c\right)$ for quasistatic mercury extrusion (or primary imbibition) by simply replacing $F_c$ with $\left(1-F_c\right)$ and $s_n$ with $\left(1-s_n\right)$ in Eq. \eqref{eq:saturation_intrusion_solution}, which gives:
\begin{align}
    s_n(F_c) &= 1 - \frac{\alpha F_c}
    {(1-\alpha)\left(1-F_c\right)+\alpha}.\label{eq:saturation_extrusion_solution}
\end{align}
where exactly the same reasoning used to derive Eq. \eqref{eq:saturation_intrusion_solution} applies, except that $F_c$ is now quasistatically raised instead of lowered in mercury extrusion, causing increasingly larger pores to empty and hence reducing $s_n$. Note that by this simply analogy with mercury intrusion, Eq. \eqref{eq:saturation_extrusion_solution} will not predict mercury entrapment during mercury extrusion, which is due to other pore-scale mechanisms such as snap-off \cite{lenormand1983} unaccounted for in this simple analysis.

\paragraph{An algebraic derivation} Under the assumption of quasistatic fluid movements, it is possible to derive Eq. \eqref{eq:saturation_intrusion_solution} algebraically without considering ODEs. Instead of analyzing how $s_n$ responds to differential changes in $F_c$, we lower $F_c$ from an initial value of $1$ to its final value directly, and consider the independent displacements of all advancing menisci at once. Menisci pinned at pores whose radii falls within $\left[r_c,\infty\right)$ will begin advancing when $F_c$ is lowered. This occurs in a fraction of all instances, which is given by:
\begin{align}
    \Delta\omega_{\text{adv}}
    = 1-F_c.\label{eq:cumulative_omega_advancing}
\end{align}
We have already derived the mean displacement of an advancing meniscus that becomes pinned at a rate of $F_c/\ell$; the result is given by Eq. \eqref{eq:expected_size_displacement}. The amount of mercury intrusion in response to the abrupt lowering of $F_c$ is equal to (cf. Eq. \eqref{eq:ODE_saturation_raw} for the differential lowering of $F_c$):
\begin{align}
    \langle c \rangle s = \Delta\omega_{\text{adv}}
    \langle d \rangle.\label{eq:algebraic_saturation_raw}
\end{align}
Substituting Eqs. \eqref{eq:q_definition}, \eqref{eq:cumulative_omega_advancing}, and \eqref{eq:expected_size_displacement} into Eq. \eqref{eq:algebraic_saturation_raw} gives Eq. \eqref{eq:saturation_intrusion_solution} again.

\begin{figure}[!h]
\centering
\includegraphics{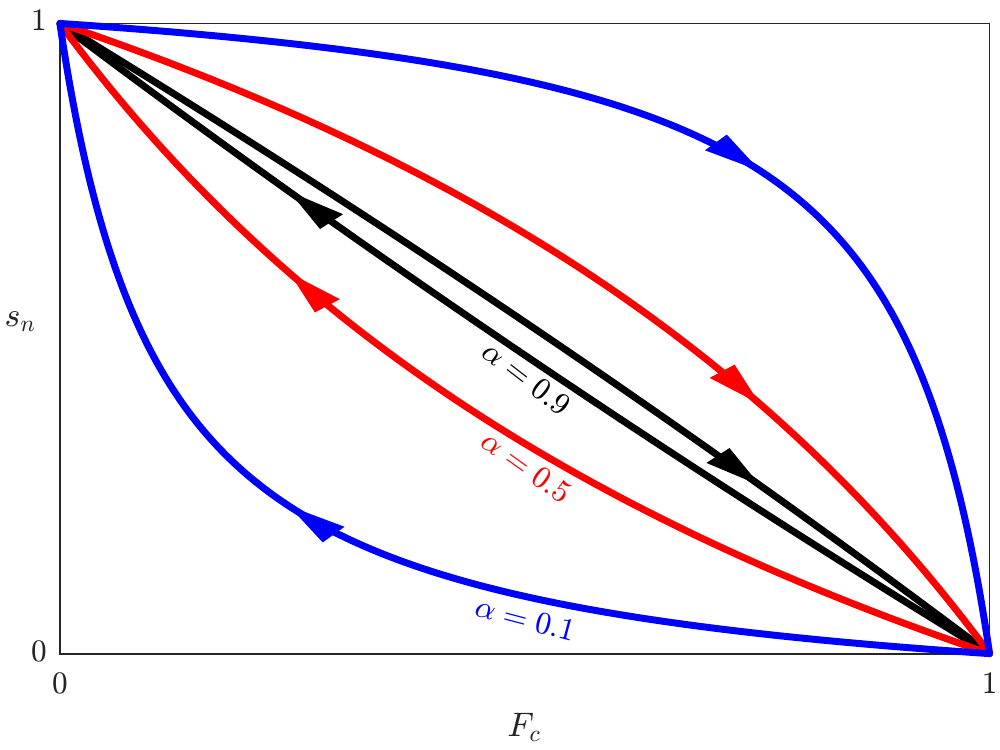}
\caption{Plots of $s_n\left(F_c\right)$ for quasistatic primary drainage (decreasing $F_c$, see Eq. \eqref{eq:saturation_intrusion_solution}) and primary imbibition (increasing $F_c$, see Eq. \eqref{eq:saturation_extrusion_solution}) for $\alpha=0.1$ (blue curves), $\alpha=0.5$ (red curves), and $\alpha=0.9$ (black curves). Increasing $\alpha$ results in less hysteresis, which is manifested in the reduced degrees of serial connections between different sized pores. We recover $s_n\left(F_c\right)=1-F_c$ as $\alpha\to1$, which is given by the capillary bundle model (see Eq. \eqref{eq:s_F_capillary_bundle}).}\label{figure:s_vs_F}
\end{figure}

\paragraph{Plots of formulae} Figure \ref{figure:s_vs_F} shows the $s_n\left(F_c\right)$ relationship for both quasistatic drainage (given by Eq. \eqref{eq:saturation_intrusion_solution}) and imbibition (given by Eq. \eqref{eq:saturation_extrusion_solution}), for porous media of low, medium, and high accessivities.

During quasistatic primary drainage, e.g., mercury intrusion, as we raise $p_c$ in the invading fluid, the effective radius of the largest penetrable pores, $r_c$, becomes lower. This gives a smaller volume fraction of pores that are too small to be filled, which is given by the cumulative function of the PSD, $F_c$. As $F_c$ decreases, the volume fraction of pores that are penetrable, which is given by $\left(1-F_c\right)$, increases, leading to higher saturations of the invading fluid, $s_n$.

Except when $\alpha\to1$, we always have $s_n<1-F_c$ (disregarding the end points). As discussed previously, this results from connectivity effects: the actual volume fraction of mercury-filled pores, $s_n$, is always less than what it would have been should pores of all sizes be directly accessible from the exterior, $\left(1-F_c\right)$.

Connectivity effects are the strongest when $\alpha\to0$, which corresponds to highly serial connections between different sized pores. Intrusion does not occur to an appreciable extent when $F_c$ is lowered at first because many larger pores are only accessible through smaller ones. As $F_c$ approaches zero, a larger fraction of pores become penetrable, resulting in a rapid rise in $s_n$, which is reminiscent of a 1-D critical percolation transition \cite{sahimi1994,stauffer2014}. The extrusion curves mirror these behaviors for increasing $F_c$.

As $\alpha$ increases, different sized pores become arranged in a more parallel fashion, which weakens connectivity effects, as evinced in the narrower gap between $s_n\left(F_c\right)$ and the $s_n=1-F_c$ line. As $\alpha\to1$, pores of all sizes become equally and indefinitely accessible from the outer surface of the porous medium, thereby recovering the capillary bundle model and eliminating any hysteresis.

\paragraph{Rules for updating radius-resolved saturation} To truthfully account for hysteresis, we must return to radius-resolved saturation for the calculation of $s_n\left(F_c\right)$. Based on the our statistical model of fluid movements, during intrusion, we expect:
\begin{align}
    \psi_n\left(F;F_c\right) =
    \begin{cases}
    0, & F < F_c \\
    \psi_0\left(F_c\right), & F > F_c
    \end{cases},\label{eq:radius_resolve_saturation_intrusion_basic}
\end{align}
Note that $F$ is the independent variable for $\psi_n(F)$, while $F_c$ is a parameter that corresponds to the imposed $p_c$. Eq. \eqref{eq:radius_resolve_saturation_intrusion_basic} says that at a given $F_c$, none of the smaller pores will be filled, while the larger pores, which are effectively indistinguishable to an advancing meniscus, must all be filled to the same extent, given by $\psi_0$ (cf. the more general statement given by Eq. \eqref{eq:primary_drainage_rrs_general}). We can evaluate $\psi_0$ by applying the condition in Eq. \eqref{eq:saturations_and_rrs} and recalling the formula for $s_n\left(F_c\right)$ (Eq. \eqref{eq:saturation_intrusion_solution}):
\begin{align}
    \frac{\alpha\left(1-F_c\right)}{(1-\alpha)F_c+\alpha}
    &= \int_{F_c}^1
    \psi_0\left(F_c\right)\mathrm{d}F\nonumber\\
    \frac{\alpha\left(1-F_c\right)}{(1-\alpha)F_c+\alpha}
    &= \left(1-F_c\right)
    \psi_0\left(F_c\right)\nonumber\\
    \psi_0\left(F_c\right)
    &= \frac{\alpha}{(1-\alpha)F_c+\alpha}.\label{eq:psi0_vs_F_intrusion}
\end{align}
Substituting this result into Eq. \eqref{eq:radius_resolve_saturation_intrusion_basic}, we have, for primary intrusion:
\begin{align}
    \psi_n\left(F;F_c\right) =
    \begin{cases}
    0, & F < F_c \\
    \alpha/\left[(1-\alpha)F_c+\alpha\right], & F > F_c
    \end{cases}.\label{eq:radius_resolve_saturation_intrusion_final}
\end{align}
Similarly, for primary extrusion we find:
\begin{align}
    \psi_n\left(F;F_c\right) =
    \begin{cases}
    1-\alpha/\left[(1-\alpha)(1-F_c)+\alpha\right], & F < F_c \\
    1, & F > F_c
    \end{cases},\label{eq:radius_resolve_saturation_extrusion_final}
\end{align}
which is, again, obtained by replacing $F_c$ with $\left(1-F_c\right)$, $F$ with $\left(1-F\right)$, and $\psi_n$ with $\left(1-\psi_n\right)$ in Eq. \eqref{eq:radius_resolve_saturation_intrusion_final}.

\begin{figure}[!p]
\centering
\includegraphics[width=\textwidth]{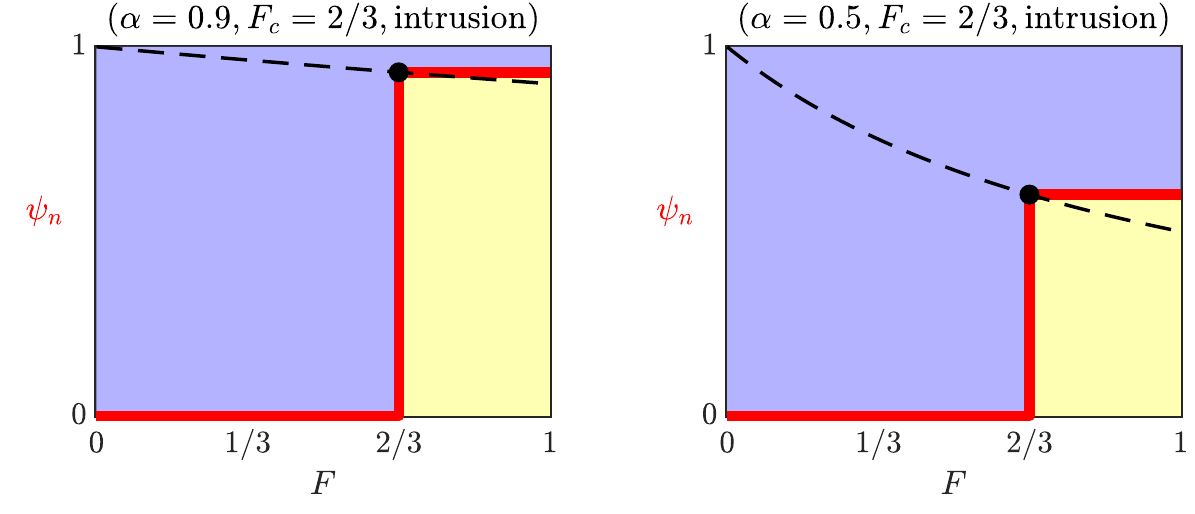} \\~\\
\includegraphics[width=\textwidth]{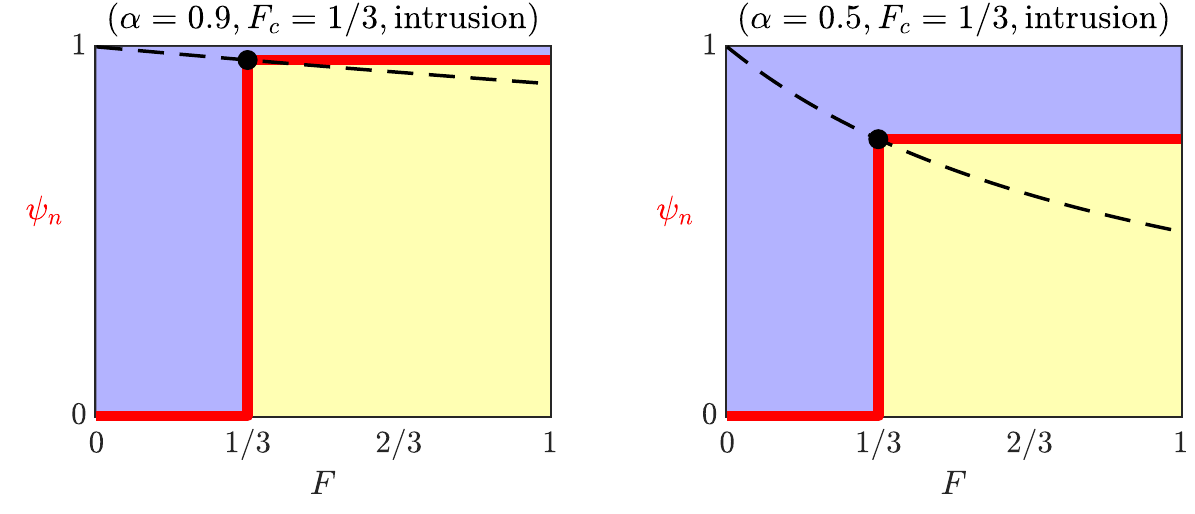} \\~\\
\includegraphics[width=\textwidth]{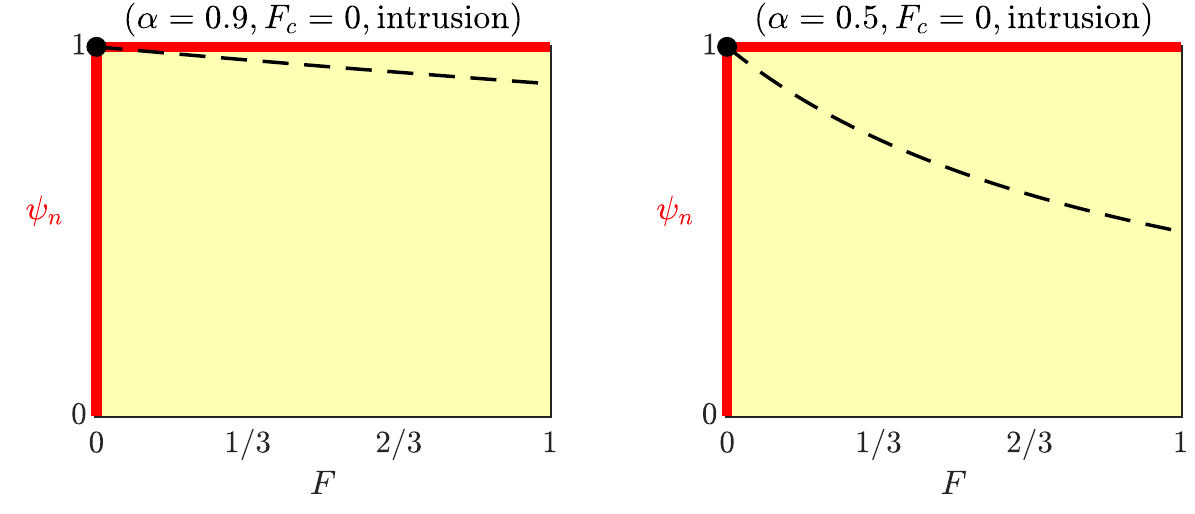}
\caption{In each column (left: $\alpha=0.9$, right: $\alpha=0.5$), the radius-resolved saturation profile, $\psi_n\left(F\right)$, is plotted for incrementally decreasing $F_c$ during mercury intrusion or primary drainage (see Eq. \eqref{eq:radius_resolve_saturation_intrusion_final}). The shaded areas represent saturations of the two phases (see Eq. \eqref{eq:saturations_and_rrs}). The black dashed curve shows the trajectory of the value of $\psi_n$ for $F_c<F \le 1$ as a function of $F_c$ (see Eq. \eqref{eq:radius_resolve_saturation_intrusion_final}).}\label{figure:radius_resolved_intrusion}
\end{figure}

\begin{figure}[!p]
\centering
\includegraphics[width=\textwidth]{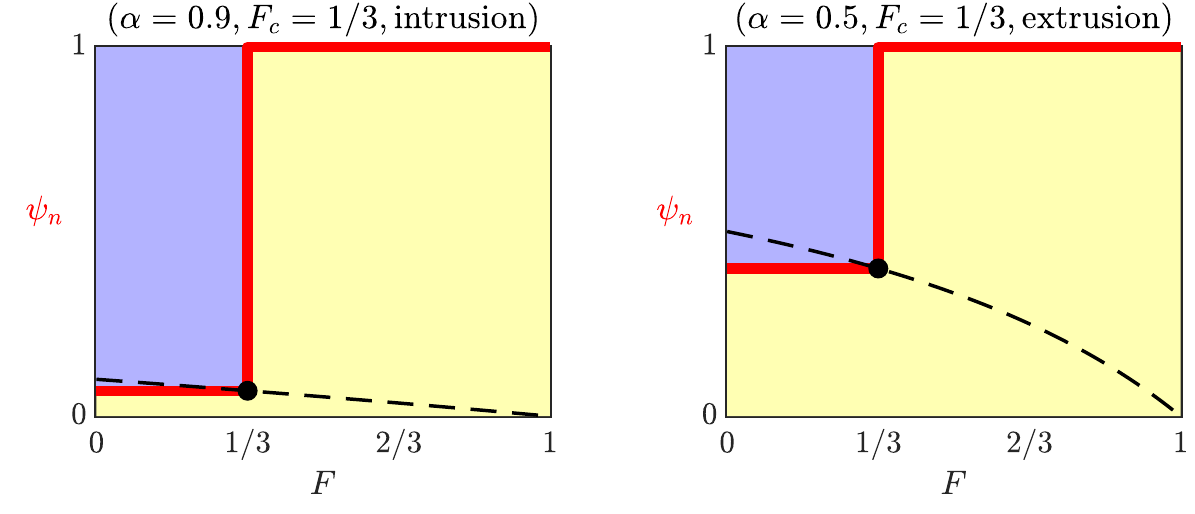} \\~\\
\includegraphics[width=\textwidth]{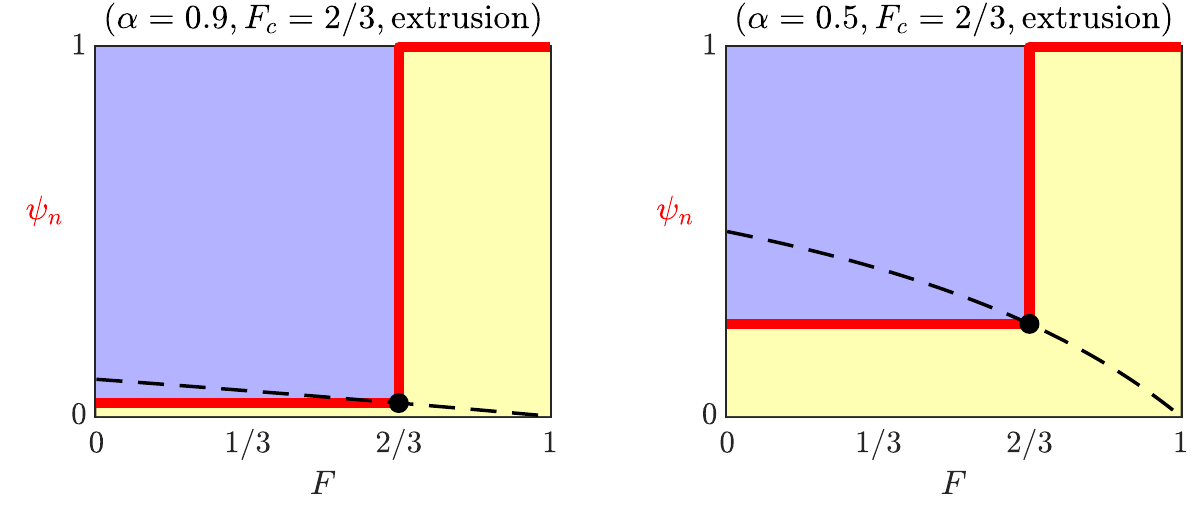} \\~\\
\includegraphics[width=\textwidth]{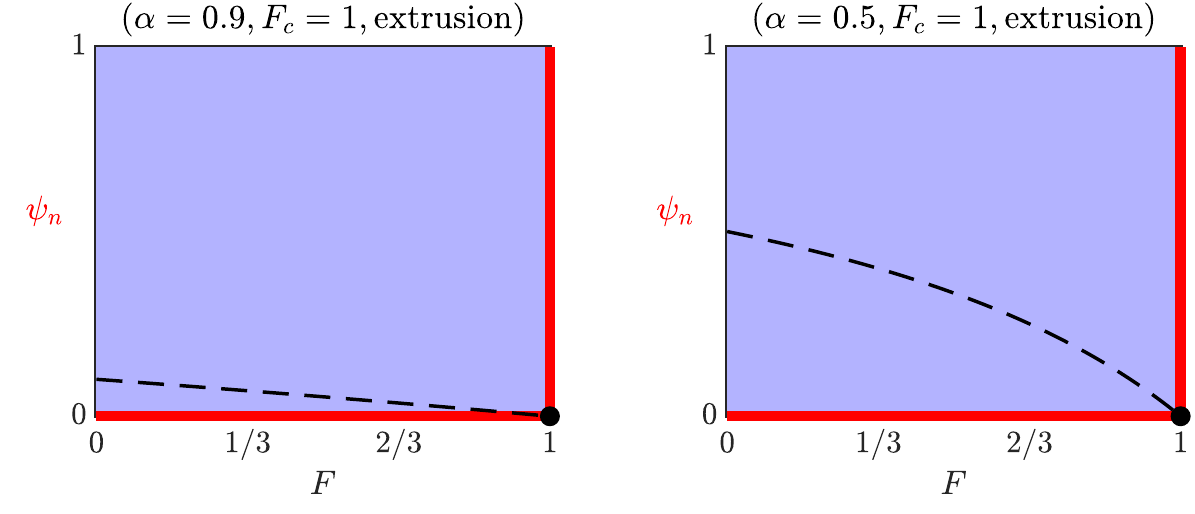}
\caption{In each column (left: $\alpha=0.9$, right: $\alpha=0.5$), the radius-resolved saturation profile, $\psi_n\left(F\right)$, is plotted for incrementally increasing $F_c$ during mercury extrusion or primary imbibition (see Eq. \eqref{eq:radius_resolve_saturation_extrusion_final}). The shaded area areas represent saturations of the two phases (see Eq. \eqref{eq:saturations_and_rrs}). The black dashed curve shows the trajectory of the value of $\psi_n$ for $0 \le F<F_c$ as a function of $F_c$ (see Eq. \eqref{eq:radius_resolve_saturation_extrusion_final}).}\label{figure:radius_resolved_extrusion}
\end{figure}

Figures \ref{figure:radius_resolved_intrusion} and \ref{figure:radius_resolved_extrusion} display the radius-resolved saturation profiles, $\psi_n\left(F\right)$, at several values of $F_c$ during intrusion and extrusion, respectively, for either $\alpha=0.9$ (left column) or $\alpha=0.5$ (right column).

During intrusion, in a porous sample with high accessivity, at any $F_c$, nearly all pore segments larger than the smallest penetrable pore become filled, and the area of the shaded region is nearly $\left(1-F_c\right)$ at all times. On the other hand, for a lower $\alpha$, intrusion occurs in a smaller fraction of the penetrable pores at the same $F_c$, leading to a lower $s_n$ when compared to the previous case.

\paragraph{Arbitrary scanning cycles} Having analyzed primary drainage and primary imbibition, we can generalize the above results to describe arbitrary quasistatic drainage-imbibition cycles using radius-resolved saturation. Suppose a porous sample acquires a certain $\psi_n(F;F_c)$ after begin subject to arbitrary drainage and imbibition steps. We are interested in predicting changes in $\psi_n(F;F_c)$ from its current state for quasistatic changes in $F_c$. Specifically, we will modify Eqs. \eqref{eq:radius_resolve_saturation_intrusion_final} and \eqref{eq:radius_resolve_saturation_extrusion_final} by considering how general drainage and imbibition compare with their primary counterparts.

The first point to consider is that, under quasistatic conditions, pores smaller (or larger) than $F_c$ may only undergo imbibition (or drainage), respectively, due to local capillary equilibria. Thus, when and only when the imposed $F_c$ changes, $\psi_n(F)$ may only decrease (or remain unchanged) for any $F<F_c$, and may only increase (or remain unchanged) for $F>F_c$. These observations are consistent with Eqs. \eqref{eq:saturation_intrusion_solution} and \eqref{eq:saturation_extrusion_solution}, respectively.

The second point concerns the rate of meniscus pinning. In our primary intrusion (or drainage) formula, Eq. \eqref{eq:radius_resolve_saturation_intrusion_final}, $F_c$ is thought to be associated with the rate at which a moving meniscus becomes ``pinned'' -- see Eq. \eqref{eq:small_pore_rate}. Since all pores with $F<F_c$ are occupied by the wetting phase during primary drainage (as we demonstrated in Eqs. \eqref{eq:primary_drainage_rrs_general} and \eqref{eq:radius_resolve_saturation_intrusion_final}), a moving meniscus that encounters a pore segment smaller than $F_c$ during primary drainage will indeed be blocked by the wetting fluid that it contains. However, in a material with an arbitrary history of fluid displacements, only a fraction of the pore segments smaller than $F_c$ are filled with the wetting phase and are hence able to block an advancing meniscus during drainage. This fraction can be calculated given the current $\psi_n(F)$ from the definite integral in Eq. \eqref{eq:stable_w}, which would replace $F_c$ to describe the true rate of pinning of advancing menisci. Accordingly, for pores with $F>F_c$, at a newly imposed $F_c$, we expect $\psi_n(F)$ to increase to a value of:
\begin{align}
    \psi_{n,\text{dr}}=\frac{\alpha}{(1-\alpha) \int_0^{F_c} \left(1-\psi_n\right) \mathrm{d}F +\alpha}.\label{eq:dr_bound}
\end{align}
Analogously, for quasistatic imbibition starting from an arbitrary $\psi_n(F)$, we expect pores with $F<F_c$ to acquire lower values of $\psi_n(F)$ given by:
\begin{align}
    \psi_{n,\text{im}}=1 - \frac{\alpha}{(1-\alpha) \int_{F_c}^{1} \psi_n \mathrm{d}F +\alpha},\label{eq:im_bound}
\end{align}
where, the definite integral, as given by Eq. \eqref{eq:stable_n}, represents the fraction of pore segments that are larger than $F_c$ and filled with the nonwetting phase, and, likewise, replaces $(1-F_c)$ in Eq. \eqref{eq:radius_resolve_saturation_extrusion_final}, which would be the value of the definite integral for primary imbibition.

Lastly, we combine the above results to arrive at an algebraic formula for updating the radius-resolved saturation in response to quasistatic capillary pressure variations during an arbitrary drainage-imbibition (or intrusion-extrusion) scanning cycle:
\begin{align}
    \psi_n\left(F;F_c\right) =
    \begin{cases}
    \text{min}\left\{ \psi_n,\psi_{n,\text{im}} \right\}, & F < F_c \\
    \text{max}\left\{ \psi_n,\psi_{n,\text{dr}} \right\}, & F > F_c
    \end{cases},\label{eq:rrs_n_general}
\end{align}
where $\psi_{n,\text{dr}}$ and $\psi_{n,\text{im}}$ are given by Eqs. \eqref{eq:dr_bound} and \eqref{eq:im_bound}, respectively. As we vary $p_c$ and hence $F_c$ quasistatically, $\psi_{n,\text{dr}}$ and $\psi_{n,\text{im}}$ change correspondingly, which indicate the degree to which pores larger and smaller than $F_c$ can undergo drainage and imbibition, respectively, based on the connectivity of the pore space that is indicated by the accessivity, $\alpha$. We update $\psi_n(F)$ at the new $F_c$ by comparing its previous value at each $F$ with either $\psi_{n,\text{dr}}$ or $\psi_{n,\text{im}}$, depending on whether $F>F_c$ or $F<F_c$. That is, since pores larger than $F_c$ may only undergo drainage under quasistatic conditions, $\psi_n$ for any $F>F_c$ may only increase from its previous value, without exceeding $\psi_{n,\text{dr}}$, which is the upper bound for the extent of drainage based on pore-space accessivity; on the other hand, pores smaller smaller than $F_c$ may only undergo imbibition, so $\psi_n$ for any $F<F_c$ may only decrease, but never dropping below a lower bound given by $\psi_{n,\text{im}}$.

Eq. \eqref{eq:rrs_n_general} simplifies to Eqs. \eqref{eq:radius_resolve_saturation_intrusion_final} and \eqref{eq:radius_resolve_saturation_extrusion_final} for primary drainage and primary imbibition, respectively. For example, we have $\psi_{n}(F)=0$ at the start of primary drainage, when the pore space is filled exclusively with the wetting phase at all pore sizes, and the capillary pressure is at a minimum, or $F_c=1$. As we reduce $F_c$ from $1$ to $0$, $\psi_{n}$ for any $F<F_c$ remains unchanged at $0$, while $\psi_{n}$ for $F>F_c$ readily increases to $\psi_{n,\text{dr}}$, which itself increases as $F_c$ decreases according to Eq. \eqref{eq:dr_bound}, where the definite integral simply evaluates to $F_c$. We recover the simpler result given by Eq. \eqref{eq:radius_resolve_saturation_intrusion_final} as a result.

\begin{align}
    \psi_w\left(F;F_c\right) =
    \begin{cases}
    \text{max}\left\{ \psi_w , \frac{\alpha}{ (1-\alpha)\int_{F_c}^1 (1-\psi_w)\mathrm{d}F+\alpha} \right\}, & F < F_c \\
    \text{min}\left\{ \psi_w , 1-\frac{\alpha}{ (1-\alpha)\int_0^{F_c} \psi_w \mathrm{d}F+\alpha} \right\}, & F > F_c
    \end{cases}.\label{eq:rrs_w_general}
\end{align}

\section{Illustrative examples}

In this section, we present several exploratory case studies to illustrate the implications of our theory. We will examine the behaviors of the key formulae derived, including: rules for updating radius-resolved saturations during arbitrary drainage-imbibition cycles, Eq. \eqref{eq:rrs_n_general} and \eqref{eq:rrs_w_general}; their simplified forms for primary drainage and imbibition, Eq. \eqref{eq:radius_resolve_saturation_intrusion_final} and
\eqref{eq:radius_resolve_saturation_extrusion_final}; and the resulting formulae for conventional saturations, \eqref{eq:saturation_intrusion_solution} and \eqref{eq:saturation_extrusion_solution}, respectively. Of particular interest is the role of accessivity, $\alpha$, in this framework, including how it controls hysteresis, as well as its connection to pore-space morphology.

\subsection{New constitutive law for capillary pressure hysteresis}
Since conventional saturations, $s_w$ and $s_n$, are easily determined given the radius-resolved saturations, $\psi_w(F)$ and $\psi_n(F)$, by Eq. \eqref{eq:saturations_and_rrs}, any results based on radius-resolved saturations can be easily expressed in terms of conventional saturations. Meanwhile, $\psi_w(F)$ and $\psi_n(F)$ better represent the distribution of fluid phases in the pore space, thus naturally capable of producing hysteresis.

For instance, Eqs. \eqref{eq:rrs_n_general} and \eqref{eq:rrs_w_general}, which express the rules for updating radius-resolved saturations in response to arbitrary quasistatic changes in the capillary pressure, already constitute a new constitutive law for capillary pressure hysteresis. In fact, since the formulae derived are algebraic in nature, they can be readily incorporated into calculations where it is desirable to include hysteresis in the $p_c(s_w)$ relationship, at the expense of the inclusion of extra state variables.

To apply our model as a constitutive law for capillary pressure hysteresis, we would need the following information: the cumulative function of the PSD, $F(r)$; the condition for pore-scale capillary equilibrium based on effective pore radius, which would map $p_c$ to $r_c$, and hence to $F_c$ if the PSD is known (see Figure \ref{figure:variables_causal}); the accessivity of the porous medium, $\alpha$, which would control the amount of hysteresis in the resulting $p_c(s_w)$ curves. Compared to conventional $p_c(s_w)$ models, the only additional parameter introduced here is $\alpha$. Therefore, this new constitutive law may be regarded as an augmentation of any conventional $p_c(s_w)$ formula that does not explicitly model hysteresis, such as parametric laws \cite{brooks1964,van_genuchten1980} and capillary bundle models based on the PSD \cite{thomson1872,washburn1921}.

\begin{figure}[!h]
\centering
\includegraphics[width=0.48\textwidth]{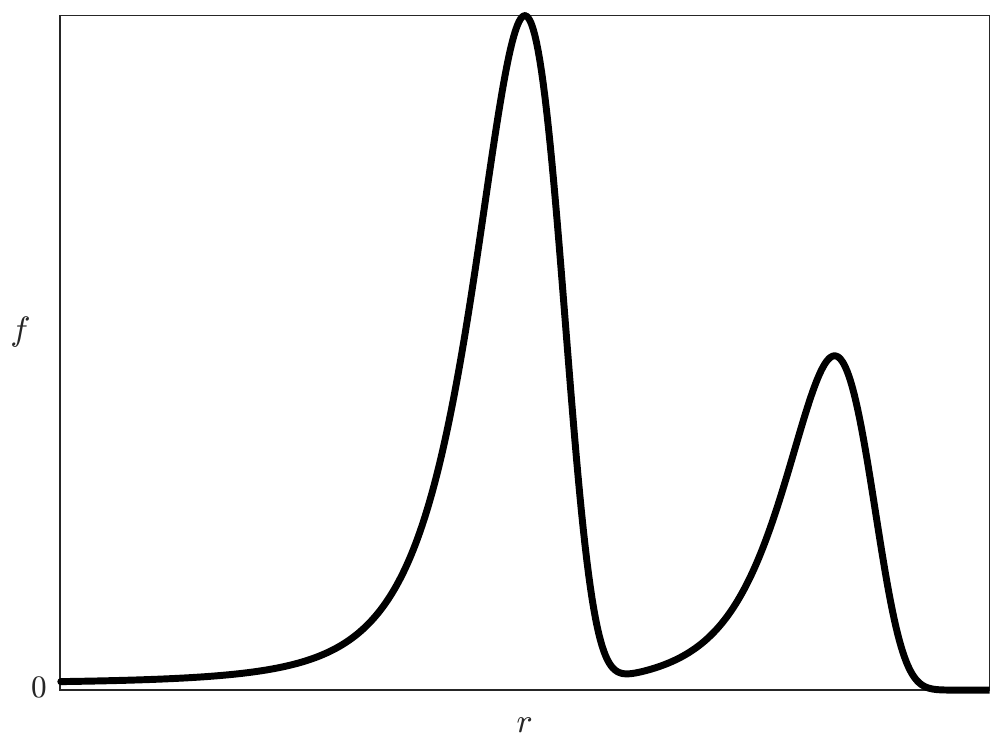}~\includegraphics[width=0.48\textwidth]{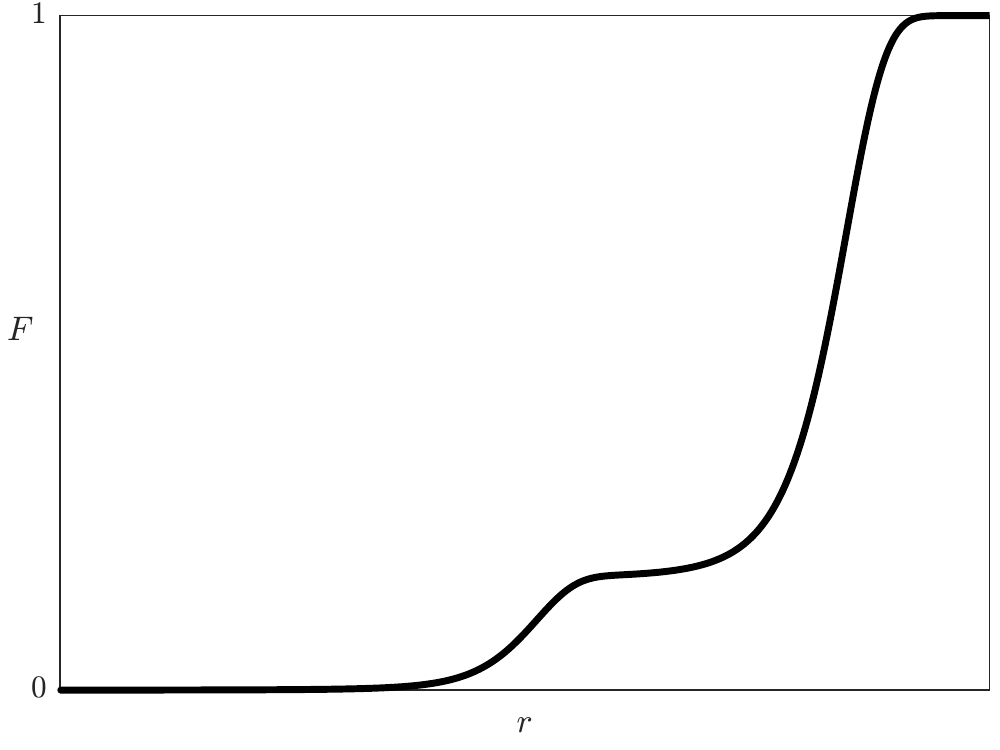}
\caption{The density function (left) and cumulative function (right) of a hypothetical bimodal pore-size distribution. The horizontal axis is plotted on a logarithmic scale.}\label{figure:hypothetical_PSD}
\end{figure}

To illustrate, consider a hypothetical pore-size distribution, whose density function $f(r)$ and cumulative function $F(r)$ are displayed in Figure \ref{figure:hypothetical_PSD}. For simplicity, assume for now that pore-scale capillary equilibrium follows $r_c \propto 1/p_c$, e.g., Eq. \eqref{eq:Washburn}, and that there is no contact-angle hysteresis.

\begin{figure}[!p]
\centering
\includegraphics[width=0.55\textwidth]{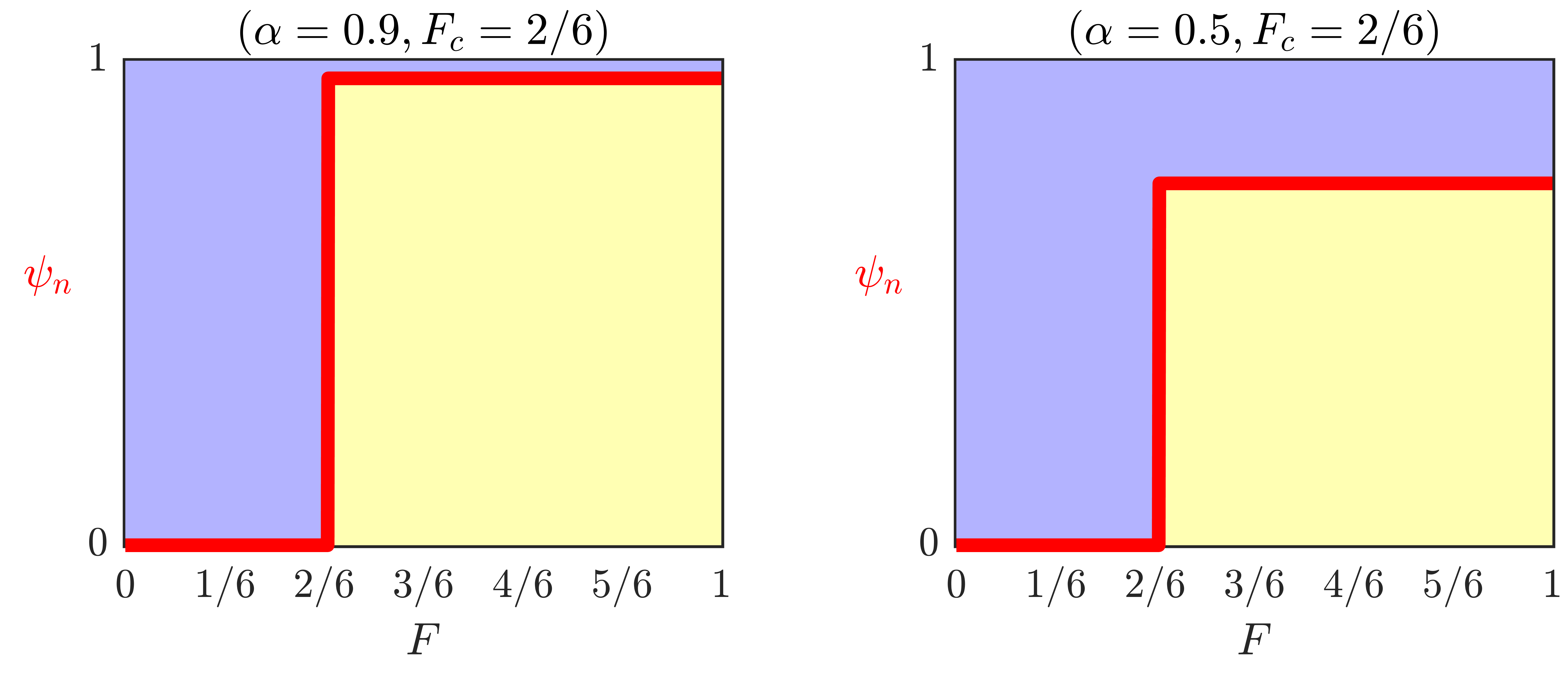} \\~\\
\includegraphics[width=0.55\textwidth]{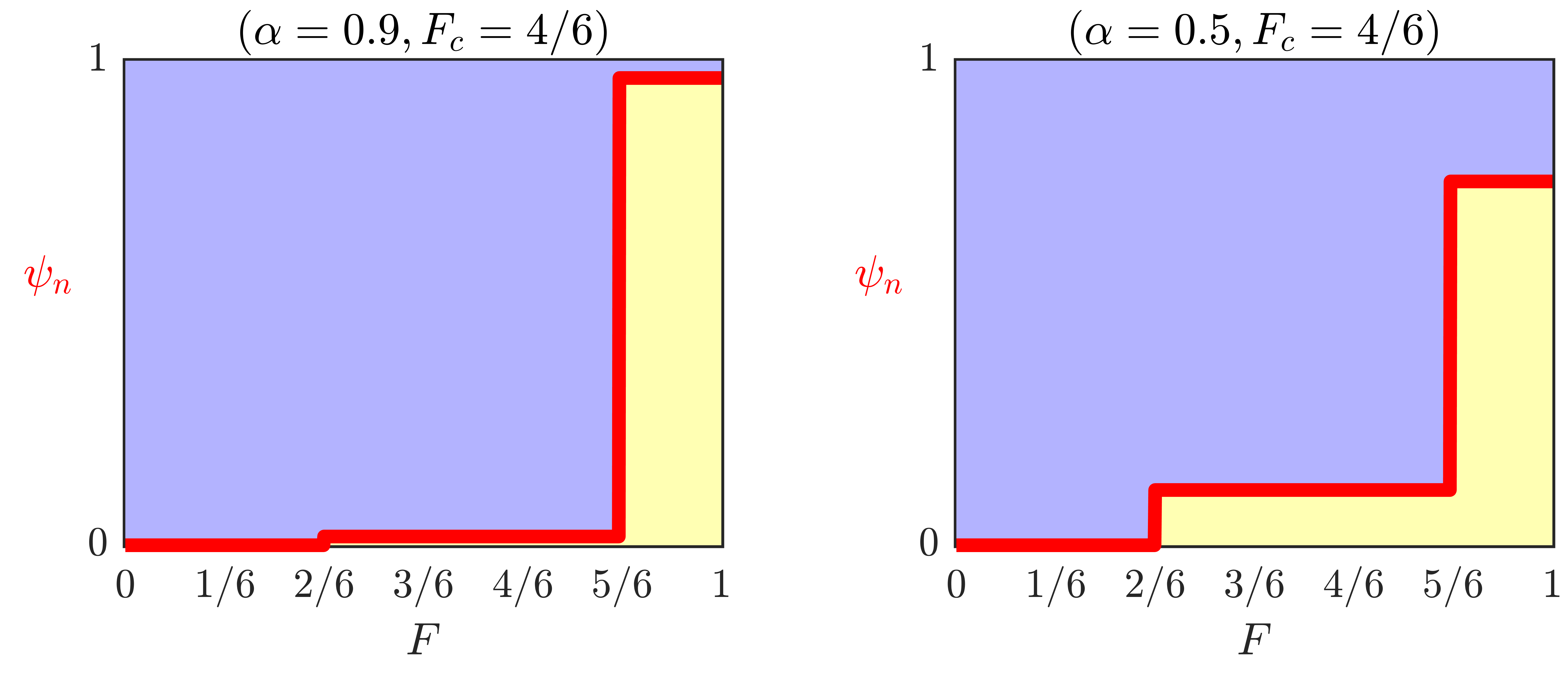} \\~\\
\includegraphics[width=0.55\textwidth]{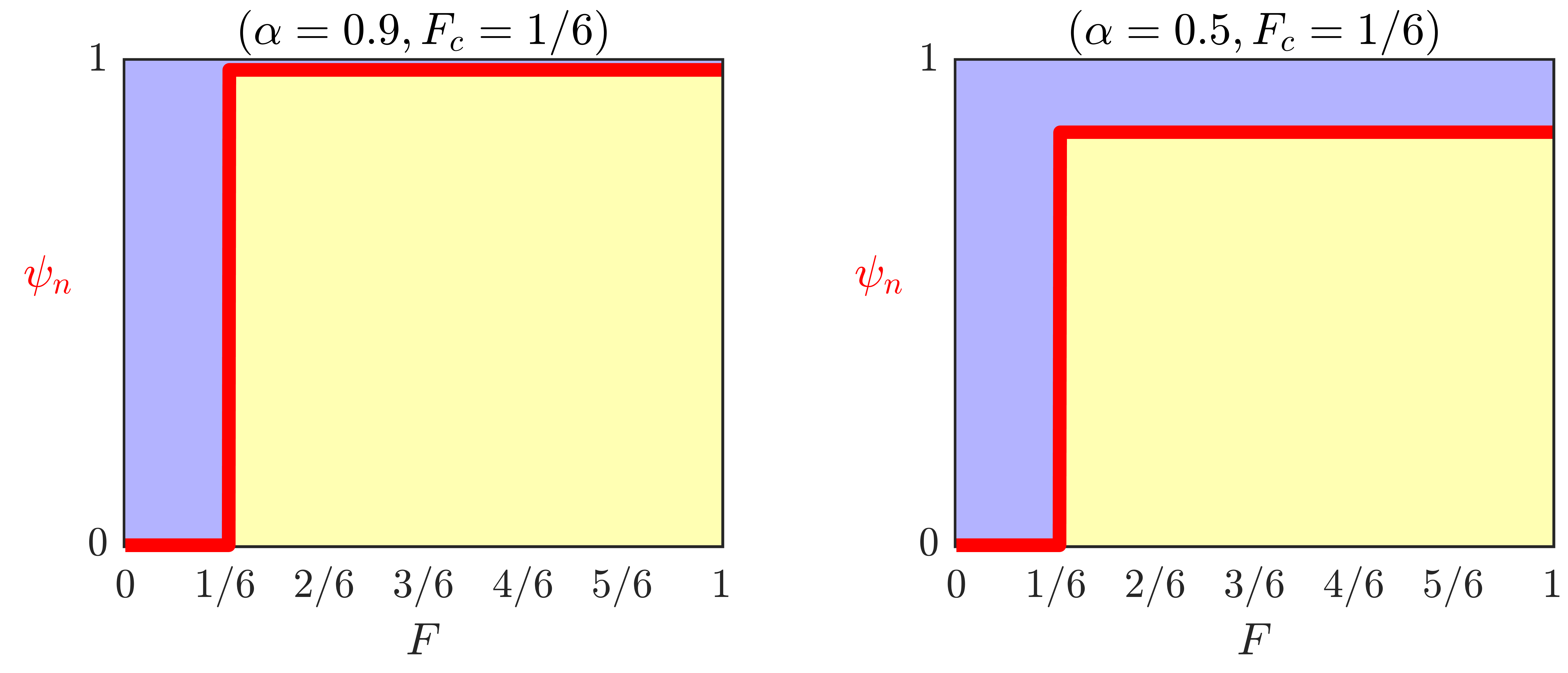}
\\~\\
\includegraphics[width=0.55\textwidth]{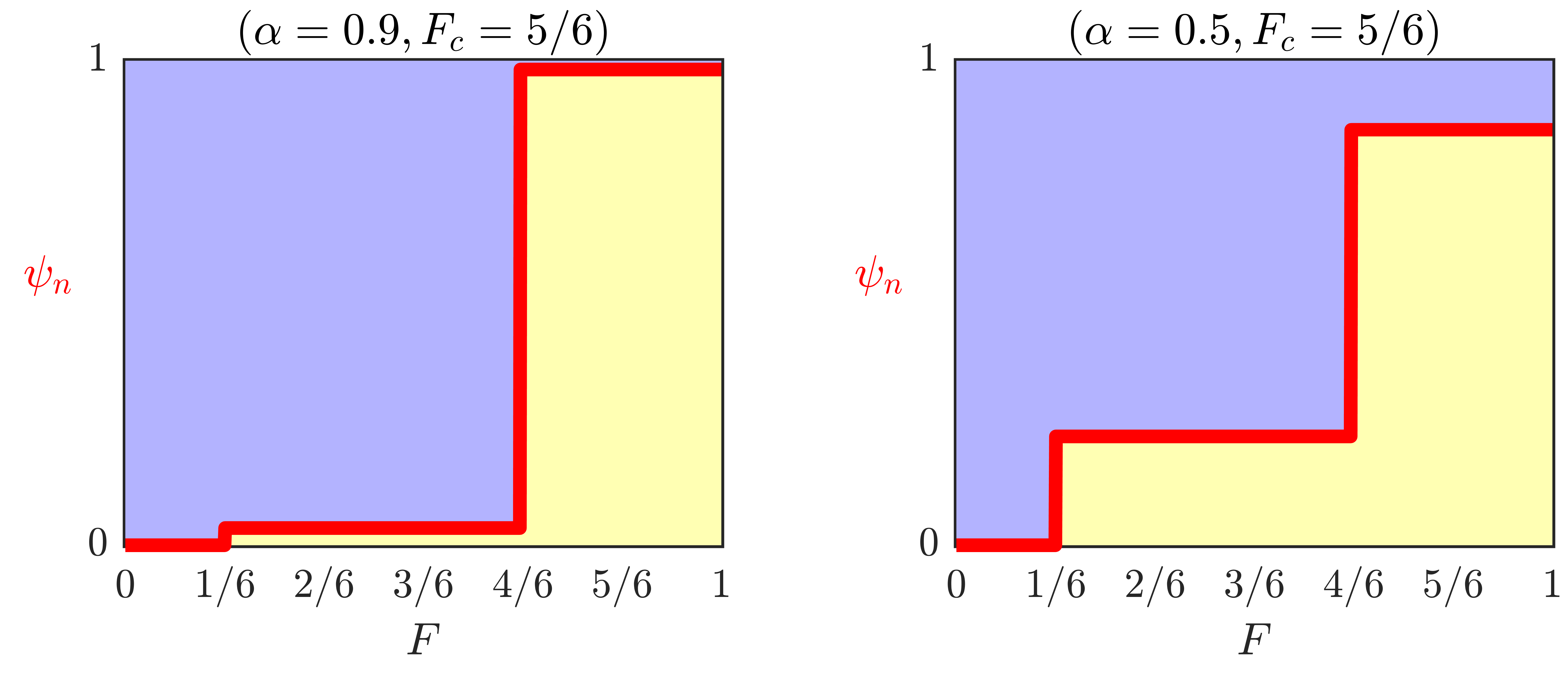}
\\~\\
\includegraphics[width=0.55\textwidth]{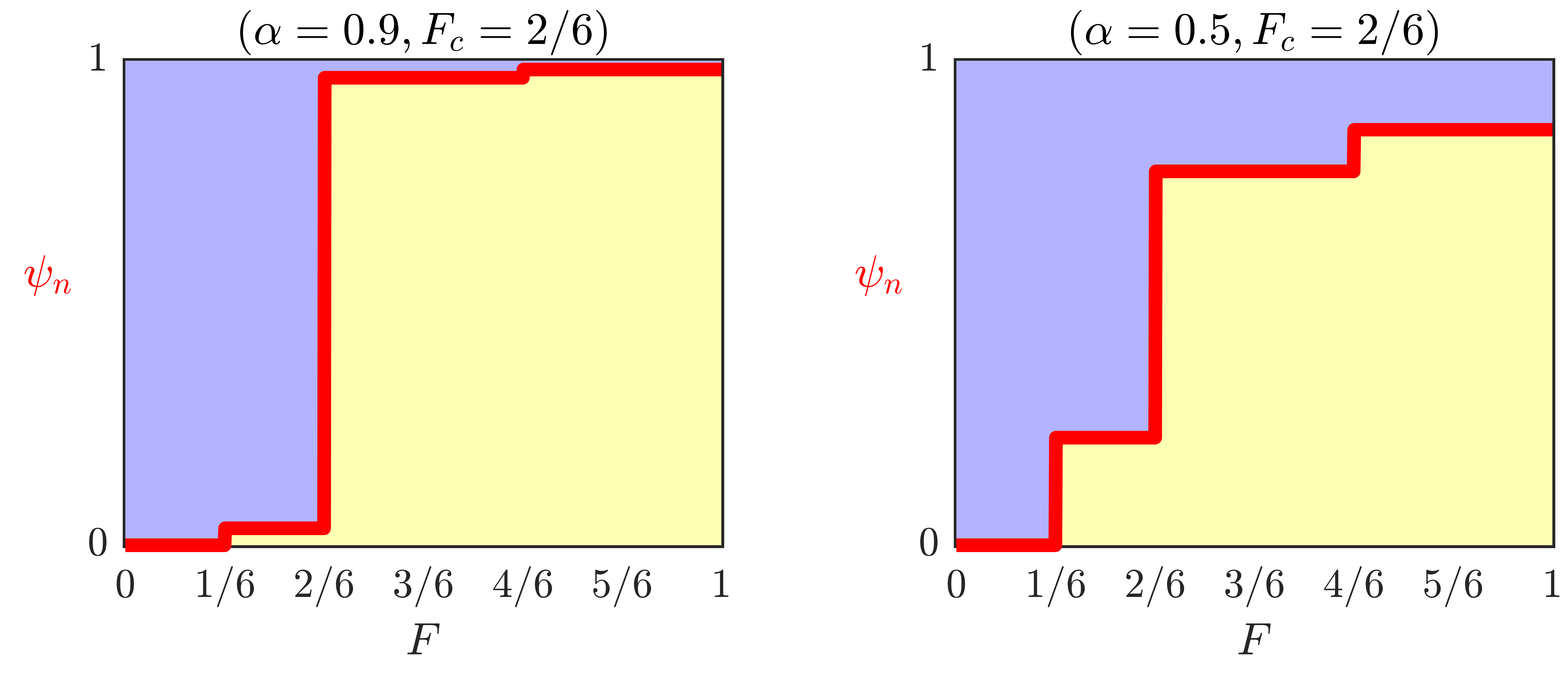}
\caption{In each column (left: $\alpha=0.9$, right: $\alpha=0.5$), the radius-resolved saturation profile, $\psi_n\left(F_c\right)$, is plotted at various $F_c$ during a quasistatic drainage-imbibition cycle, following Eq. \eqref{eq:rrs_n_general}. The shaded areas represent saturations of the two phases (see Eq. \eqref{eq:saturations_and_rrs}).}\label{figure:rrs_scanning}
\end{figure}

Now, suppose we carry out a quasistatic drainage-imbibition cycle by varying $p_c$, such that the corresponding $F_c$ surveys the following turning points or endpoints: $1$, $2/6$, $4/6$, $1/6$, $5/6$, and $2/6$ (that is, $F_c$ is initially at $1$, then reduced to $2/6$, then raised to $4/6$, and so on). At the beginning of the cycle, because $F_c=1$, we assume $s_n=0$, which implies $\psi_n(F)=0$ for all $F$. Applying Eq. \eqref{eq:rrs_n_general}, for a given $\alpha$, we update $\psi_n(F)$ for incrementally varying $F_c$, and obtain a series of $\psi_n(F)$ profiles. Figure \ref{figure:rrs_scanning} shows $\psi_n(F)$ at the $F_c$ endpoints of the scanning cycle, whereas the two columns correspond to two different accessivities.

\begin{figure}[!h]
\centering
\includegraphics[width=0.48\textwidth]{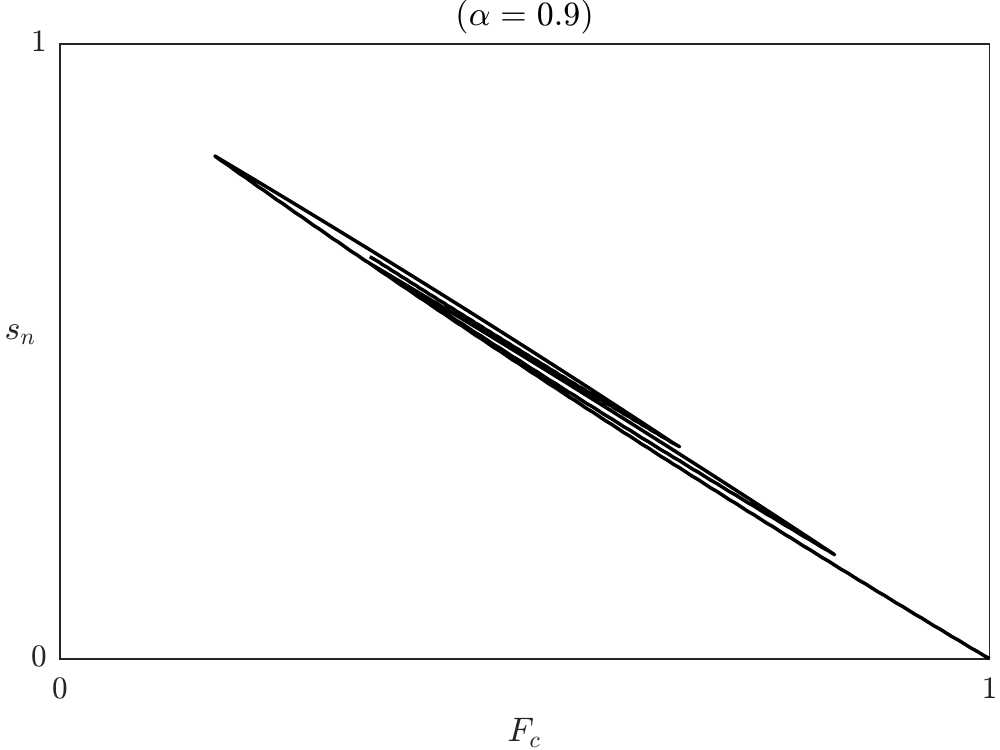}~\includegraphics[width=0.48\textwidth]{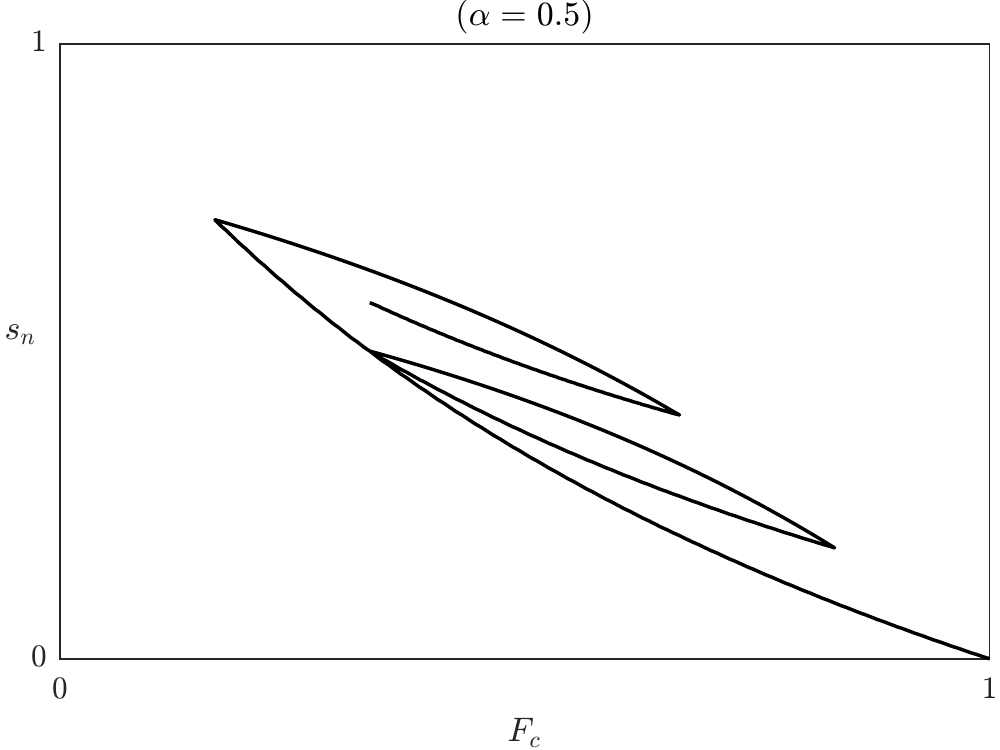}\\~\\
\includegraphics[width=0.48\textwidth]{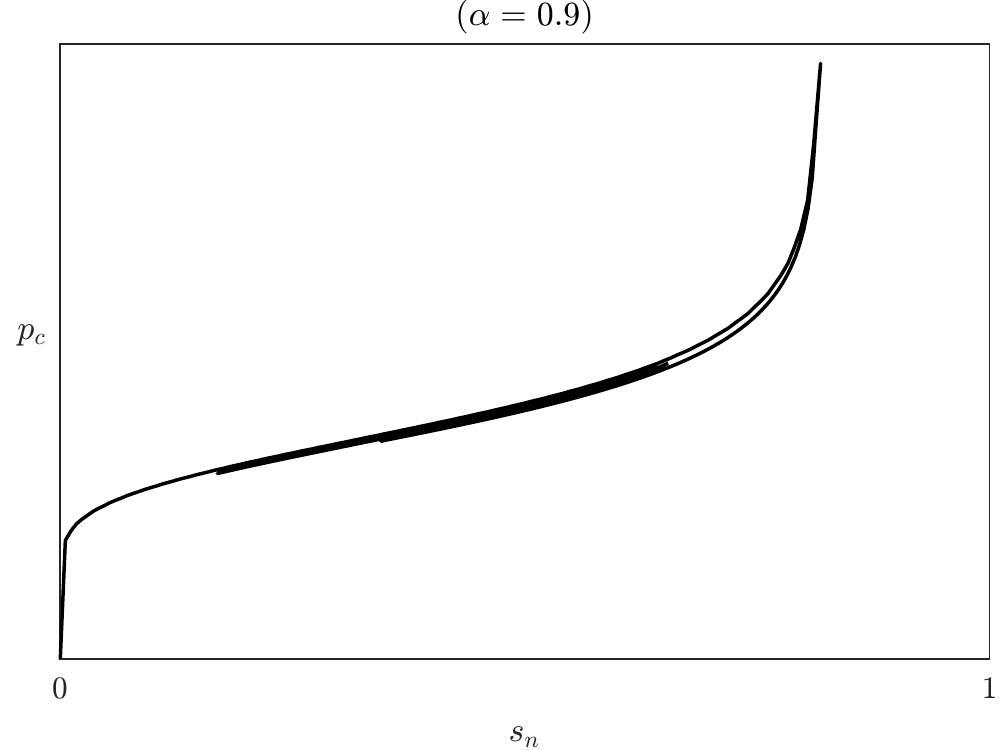}~\includegraphics[width=0.48\textwidth]{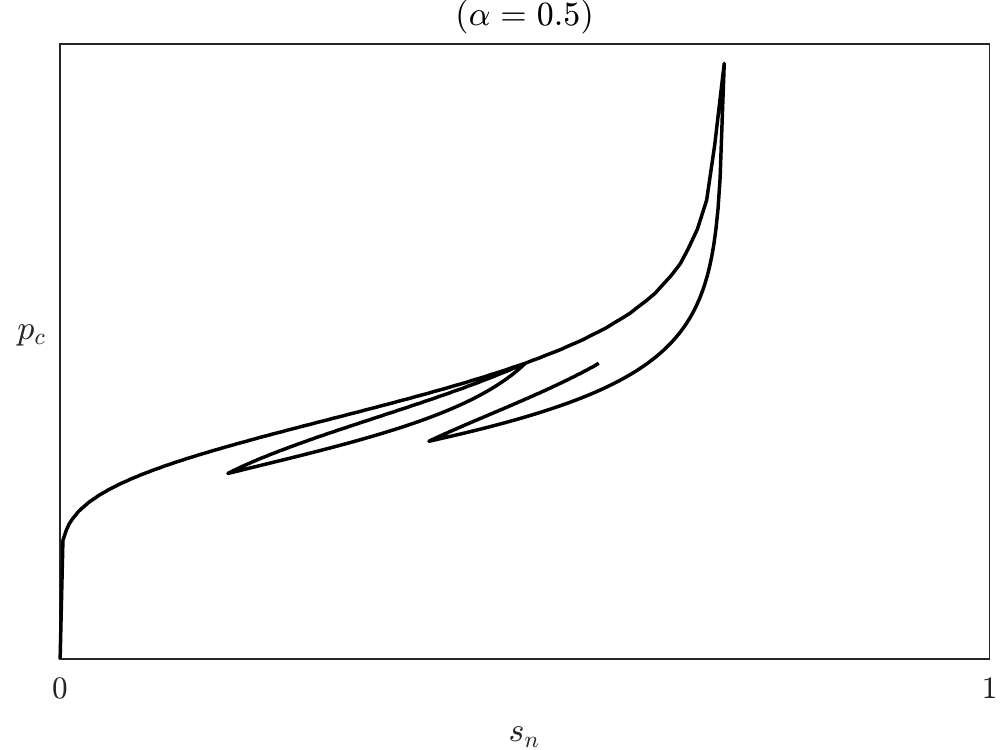}
\caption{Capillary pressure curves with hysteresis during a drainage-imbibition cycle, produced from our constitutive law based on the evolution of $\psi_n(F)$. The left and right columns correspond to high ($\alpha=0.9$) and moderate ($\alpha=0.5$) accessivities, yielding negligible and moderate degrees of hysteresis, respectively. The vertical axes in the bottom two figures are on logarithmic scales.}\label{figure:scanning_curves}
\end{figure}

Subsequently, we use Eq. \eqref{eq:saturations_and_rrs} to find $s_n$ by calculating the yellow shaded area under the $\psi_n(F)$ curves in Figure \ref{figure:rrs_scanning}. The $s_n$ trajectory is plotted against the imposed $F_c$ in the upper two figures of Figure \ref{figure:scanning_curves}. Next, we map $F_c$ to $p_c$ as we discussed earlier, and plot $p_c$ versus $s_n$ in the lower two figures of Figure \ref{figure:scanning_curves}, as capillary pressure data are ordinarily reported.

From Figure \ref{figure:scanning_curves}, it is apparent that $\alpha$ indeed controls the amount of hysteresis in the $p_c(s_n)$ curves produced from our constitutive law. This is explained by the different  patterns of evolution of $\psi_n(F)$ in Figure \ref{figure:rrs_scanning}. At $\alpha=0.9$, the radius-resolved saturation evolves in such a way that, at any particular $F_c$, $\psi_n(F)$ resembles a step change from $0$ to $1$ at $F=F_c$, analogous to the $\alpha\to1$ case in Figure \ref{figure:rrs_illustration_2}, regardless of the history of $F_c$. As a result, we have $s_n=1-F_c$ at all $F_c$ during the cycle, as shown in the upper left plot in Figure \ref{figure:scanning_curves}. On the other hand, at $\alpha=0.5$, as the cycling of $F_c$ continues, we observe more complex profiles of $\psi_n(F)$, whose evolution is now highly history dependent -- for instance, the $\psi_n(F)$ profiles in the upper right and lower right corners of Figure \ref{figure:rrs_scanning} look notably different, despite the fact that $F_c=2/6$ in both cases. The resulting $s_n(F_c)$ and $p_c(s_n)$ curves, shown in the right column of Figure \ref{figure:scanning_curves}, are more hysteretic.

We have demonstrated that our new constitutive law, Eq. \eqref{eq:rrs_n_general}, captures capillary pressure hysteresis by recording changes in the distribution of immiscible fluid phases in different sized pores through the radius-resolved saturation, $\psi_n(F)$. The only additional parameter, $\alpha$, controls the amount of hysteresis in the resulting capillary pressure curves by affecting the patterns of evolution of $\psi_n(F)$ as $F_c$ changes. At high accessivities, as most pores become directly accessible by both fluid phases, $\psi_n(F)$ takes similar patterns throughout a drainage-imbibition cycle, irrespective of the history of $F_c$, leading to little capillary pressure hysteresis. At lower accessivities, a greater degree of serial connectivity between different sized pores renders the evolution of $\psi_n(F)$ history dependent, resulting in more pronounced capillary pressure hysteresis.

\subsection{Application to mercury intrusion-extrusion porosimetry}
The simple algebraic formulae for primary drainage and imbibition, Eqs. \eqref{eq:saturation_intrusion_solution} and \eqref{eq:saturation_extrusion_solution}, can aid the interpretation of mercury intrusion-extrusion porosimetry data, even without the explicit use of radius-resolved saturations. Here, we present a simple example to highlight the contrast between the effects of pore-space connectivity and contact-angle hysteresis on intrusion-extrusion measurements. A thorough analysis involving experimental validation will be presented in a future work.

Suppose that a porous material has the same PSD that we have considered in the preceding analysis, given by Figure \ref{figure:hypothetical_PSD}. Even with the PSD kept unchanged, mercury intrusion-extrusion measurements on the sample may yield different results, associated with varying amounts of hysteresis, the causes of which include contact-angle hysteresis and connectivity effects.

In regard to contact-angle hysteresis, it is believed that the contact angle, measured in the mercury phase, could be smaller during extrusion than during intrusion \cite{lowell2012}. We may use the ratio:
\begin{align}
    \kappa = \frac{\cos{\theta_{\mathrm{extr}}}}
    {\cos{\theta_{\mathrm{intr}}}}
    \in \left(0,1\right]
\end{align}
as a descriptor for the significance of this well-known phenomenon, where $\theta_{\mathrm{intr}}$ and $\theta_{\mathrm{extr}}$ are the contact angles during intrusion and extrusion, respectively. For the typical case of $\pi/2<\theta_{\mathrm{extr}}<\theta_{\mathrm{intr}}<\pi$, we have $\kappa<1$; if $\theta_{\mathrm{intr}}=\theta_{\mathrm{extr}}$, then we have $\kappa=1$. According to Eq. \eqref{eq:Washburn}, variations in the contact angle would affect the pore-scale capillary equilibrium condition: pores of a certain size would correspond to a lower equilibrium capillary pressure during extrusion than during intrusion. In regard to connectivity effects, our framework uses a single parameter, the accessivity, $\alpha$, to describe the arrangement of different sized pores. Recall that the smaller the accessivity of a porous sample, the more serial the connection between different sized pores, and the more prominent the ink-bottle effect.

Contact-angle hysteresis may be the primary cause of hysteresis in mercury intrusion-extrusion porosimetry in at least some cases \cite{lowell2012}, but certainly not all cases \cite{salmas2001,giesche2006}. Here, we shall demonstrate using our simple formulae to take into account both connectivity effects (described with $\alpha$) and contact-angle variations (described with $\kappa$) to explain hysteresis in porosimetry measurements.

\begin{figure}[!h]
\centering
\includegraphics[width=0.48\textwidth]{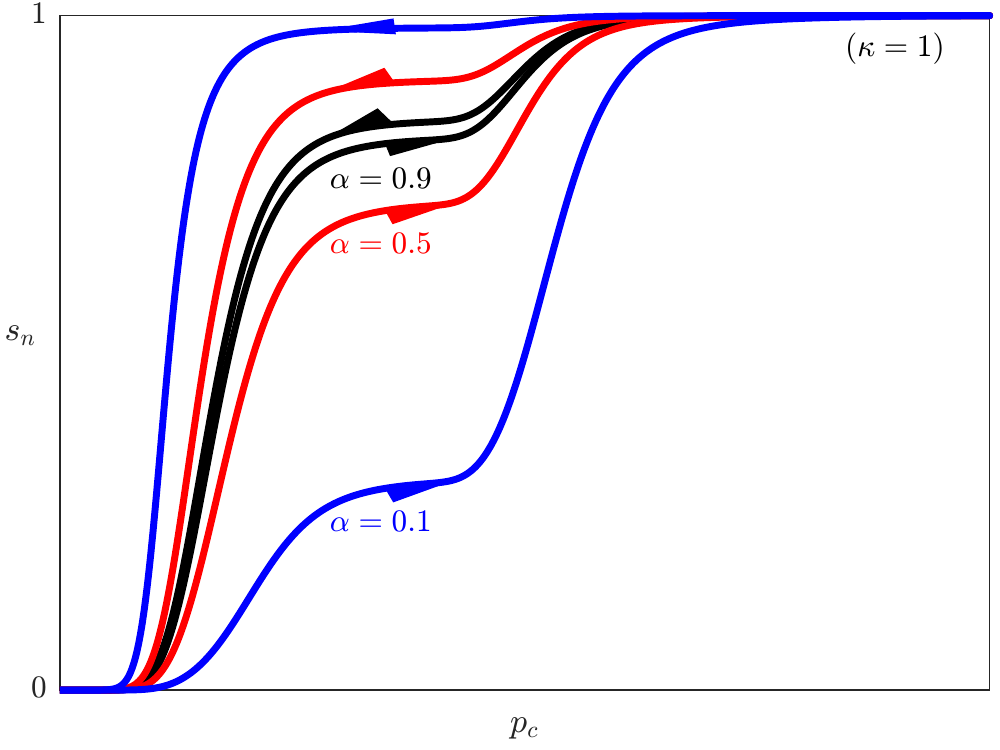}~\includegraphics[width=0.48\textwidth]{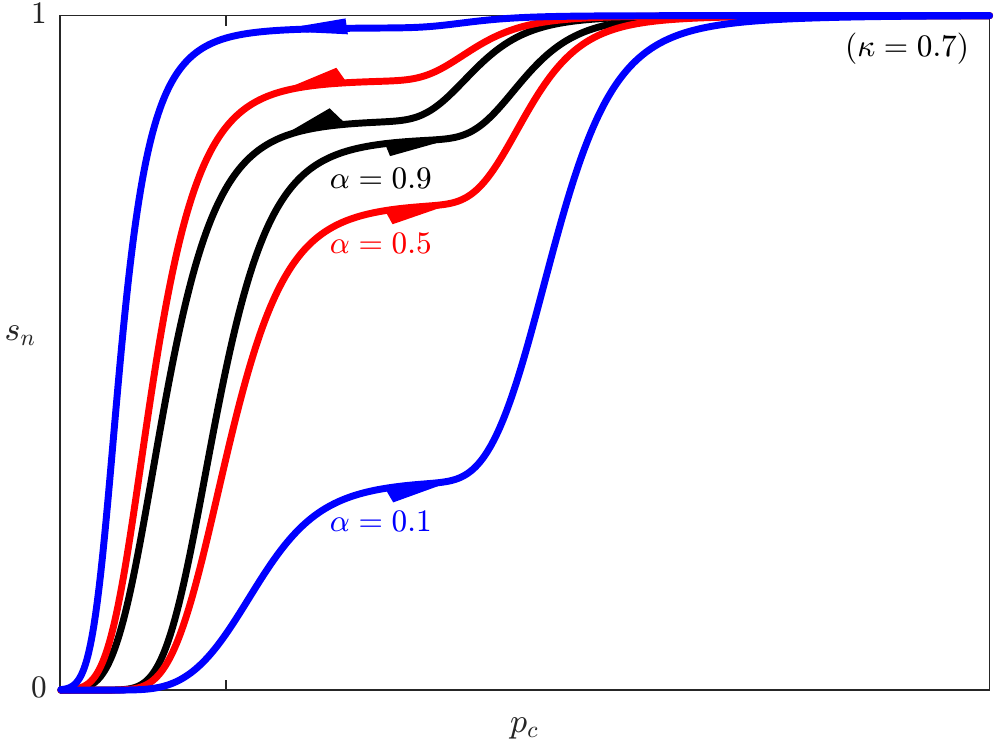}
\caption{Mercury intrusion-extrusion porosimetry curves for porous samples with $\alpha=0.1$ (blue curves), $\alpha=0.5$ (red curves), or $\alpha=0.9$ (black curves), assuming $\kappa=1$ (no contact-angle hysteresis) and $\kappa=0.7$ (moderate contact-angle hysteresis, e.g., $140^\circ$ during intrusion to $122^\circ$ during extrusion) in the left and right plots, respectively.}\label{figure:hypothetical_porosimetry}
\end{figure}

Figure \ref{figure:hypothetical_porosimetry} shows the intrusion-extrusion curves produced from Eqs. \eqref{eq:saturation_intrusion_solution} and \eqref{eq:saturation_extrusion_solution} at various $\alpha$ and $\kappa$ for the assumed PSD. On the left-hand side of Figure \ref{figure:hypothetical_porosimetry}, since contact-angle hysteresis is absent ($\kappa=1$), any hysteresis observed is due to connectivity effects alone. The hysteresis loop widens as $\alpha$ becomes lower, which can be attributed to the increase in serial connectivity between different sized pores. As $\alpha\to1$, the intrusion and extrusion curves collapse into a single curve, which resembles the shape of $F\left(r\right)$ in Figure \ref{figure:hypothetical_PSD}, as the capillary bundle model would entail. On the right-hand side of Figure \ref{figure:hypothetical_porosimetry}, we lower $\kappa$ to 0.7. The inclusion of contact-angle hysteresis shifts each extrusion curve to the left, towards lower $p_c$, relative to the corresponding intrusion curve, which is in contrast with the effect of increasing $\alpha$, the latter stretching and ``opening up'' the intrusion and extrusion curves in the vertical direction instead to form a hysteresis loop. This suggests that the effects of $\alpha$ and $\kappa$ on porosimetry curves are dissimilar, and by considering both of them using the simple formulae proposed in this work, we may generate porosimetry intrusion-extrusion cycles of a greater variety of shapes than just considering contact-angle hysteresis alone. Similarly, the same principles could also apply to other characterization techniques for porous materials, e.g., vapor sorption-desorption \cite{mason1982,mason1983,parlar1988,seaton1991,pinson2014,masoero2015,pinson2015}, water intrusion-withdrawal in gas diffusion layers \cite{gostick2008,forner-cuenca2016,lamibrac2016,sabharwal2018,tranter2018}, etc..

We should note that Eqs. \eqref{eq:saturation_intrusion_solution} and \eqref{eq:saturation_extrusion_solution} may not be sufficient for interpreting all kinds of mercury intrusion-extrusion porosimetry data observed in practice. For instance, mercury entrapment is not considered here, which would involve additional pore-scale physics \cite{hill1960}. Nevertheless, by incorporating the parameter $\alpha$ as a continuum descriptor for pore-space connectivity, these simple formulae represent an incremental improvement upon the capillary bundle approach, which implicitly assumes Eq. \eqref{eq:s_F_capillary_bundle}, or either of our Eqs. \eqref{eq:saturation_intrusion_solution} and \eqref{eq:saturation_extrusion_solution} with $\alpha=1$.

\subsection{Connection to invasion percolation}
Despite the various simplifying assumptions underlying our statistical theory, it is nevertheless of interest to investigate the proposed concepts in more realistic contexts. In this paper, we will examine invasion percolation on two-dimensional square lattices as a specific example of pore-network simulations.

\begin{figure}[!p]
\centering
\includegraphics[width=0.25\textwidth]{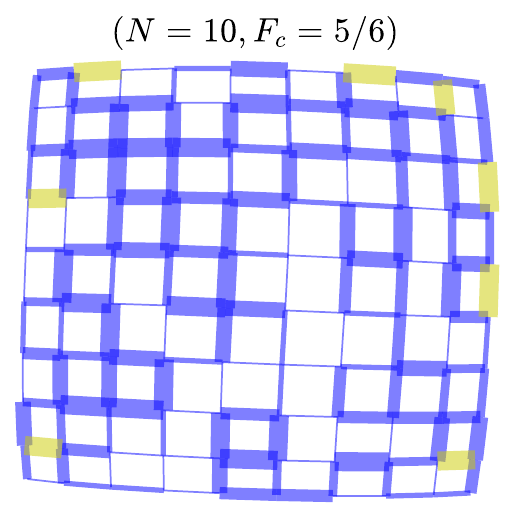} ~ \includegraphics[width=0.25\textwidth]{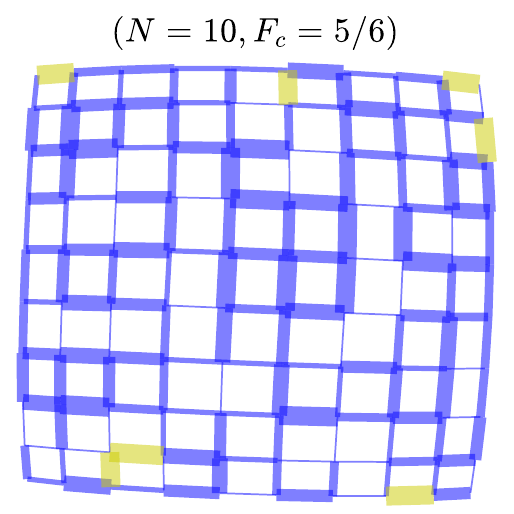} ~ \includegraphics[width=0.30\textwidth]{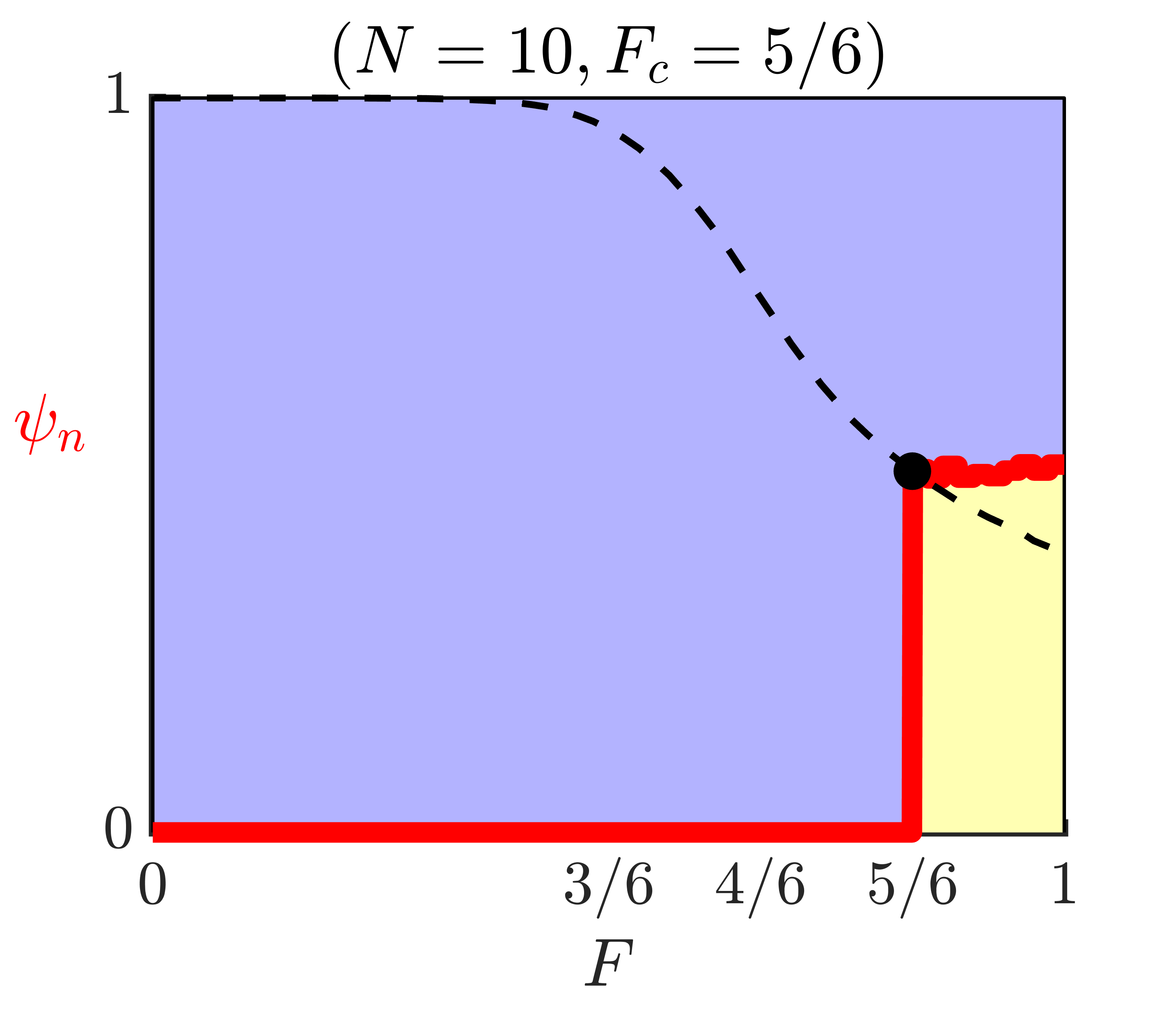} \\~\\
\includegraphics[width=0.25\textwidth]{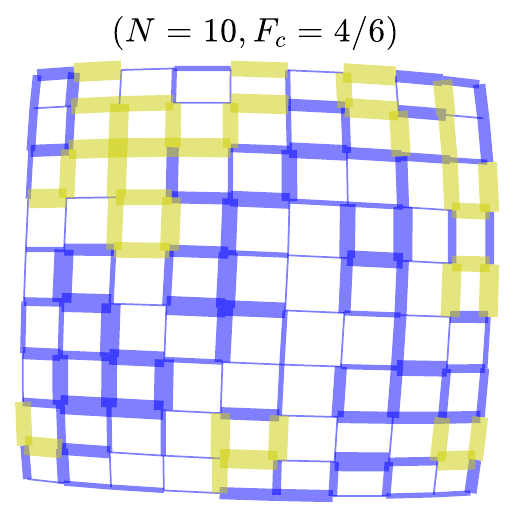} ~ \includegraphics[width=0.25\textwidth]{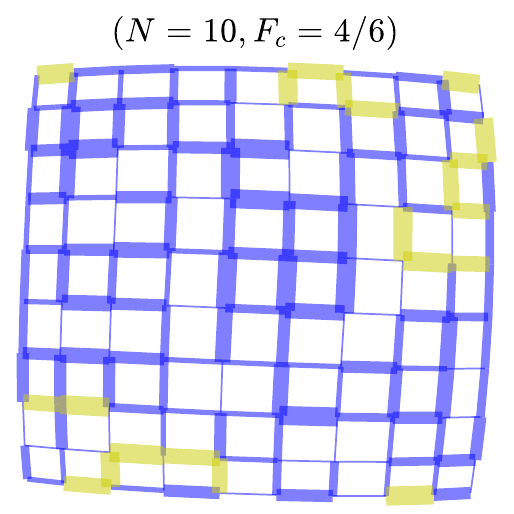} ~ \includegraphics[width=0.30\textwidth]{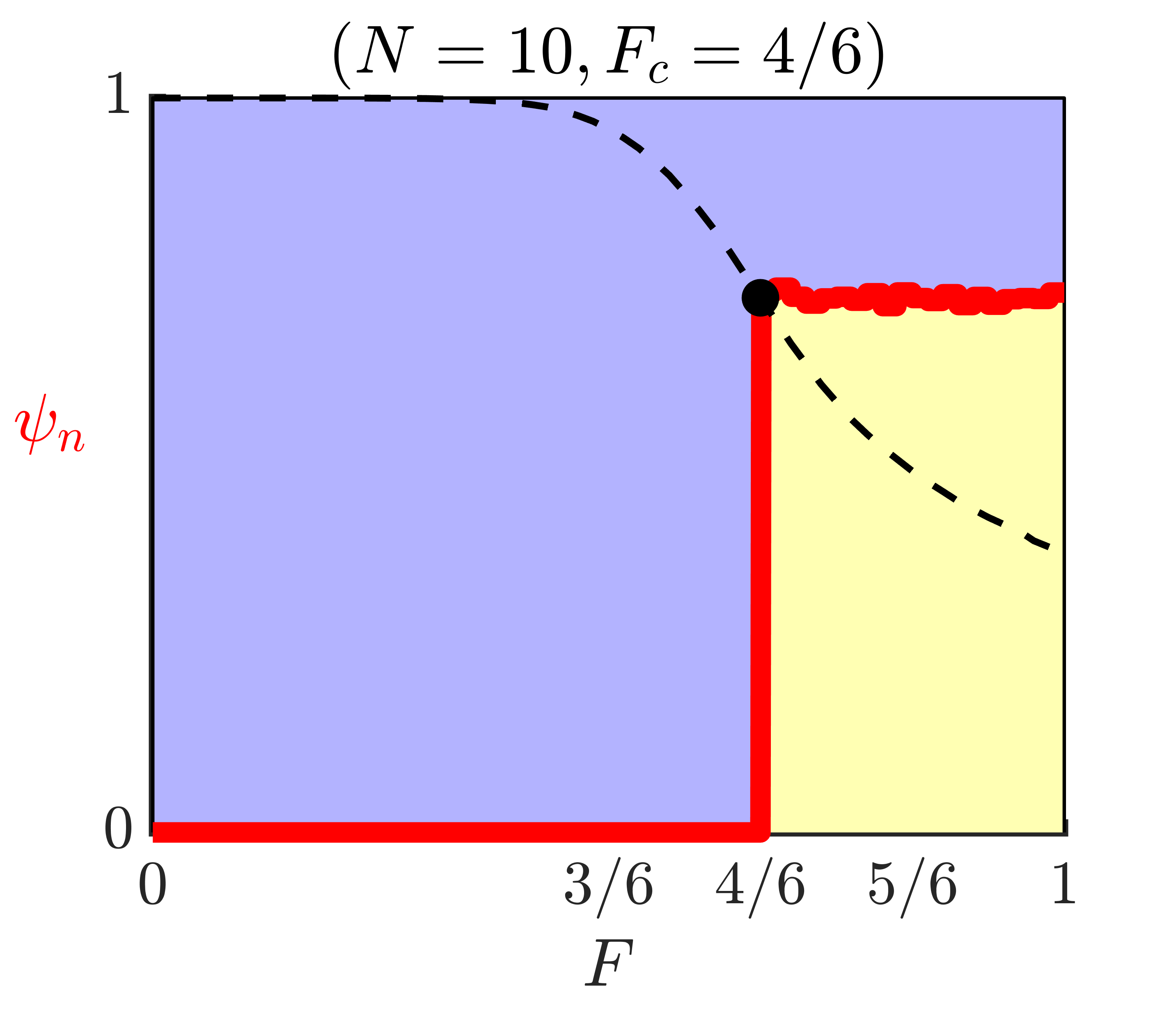} \\~\\
\includegraphics[width=0.25\textwidth]{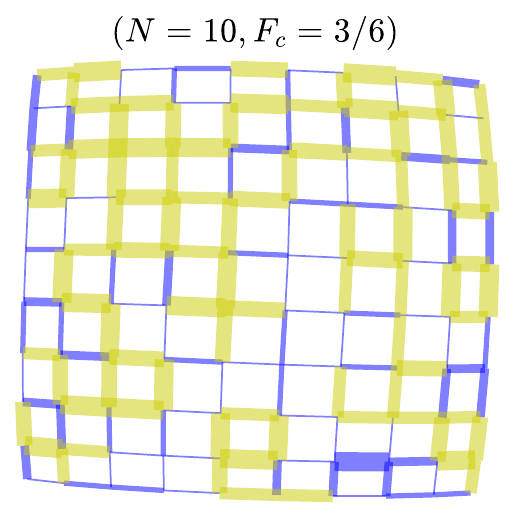} ~ \includegraphics[width=0.25\textwidth]{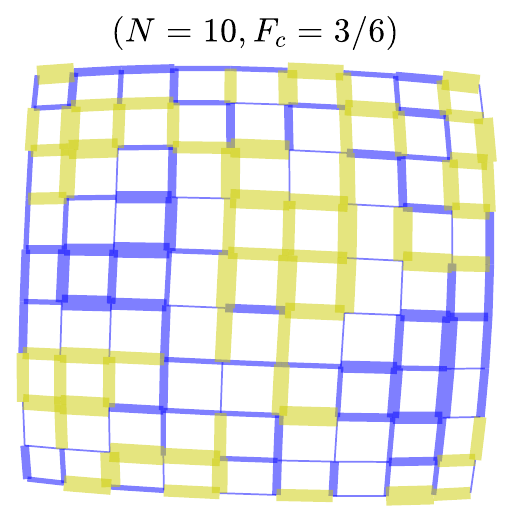} ~ \includegraphics[width=0.30\textwidth]{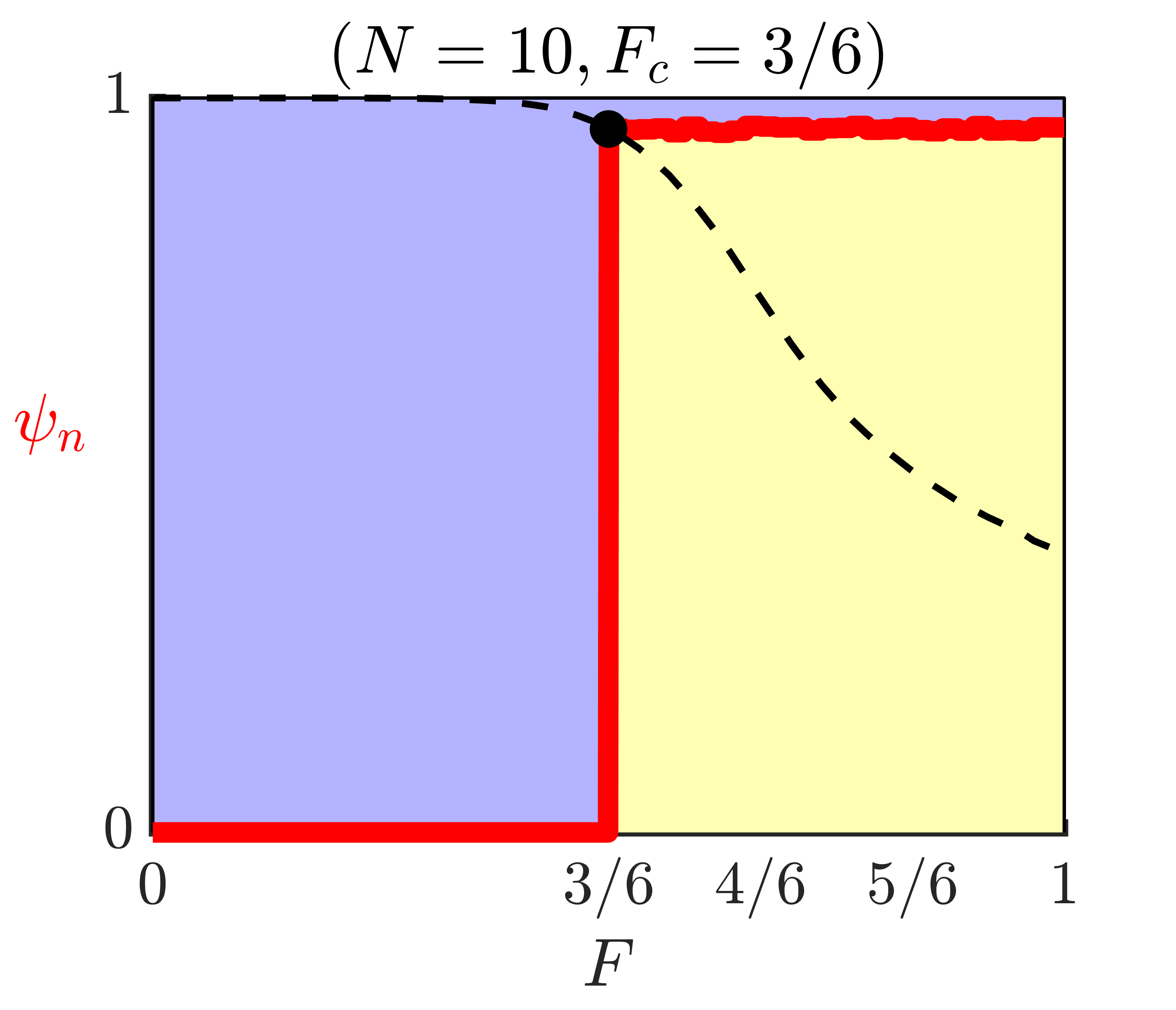}
\\~\\
\includegraphics[width=0.25\textwidth]{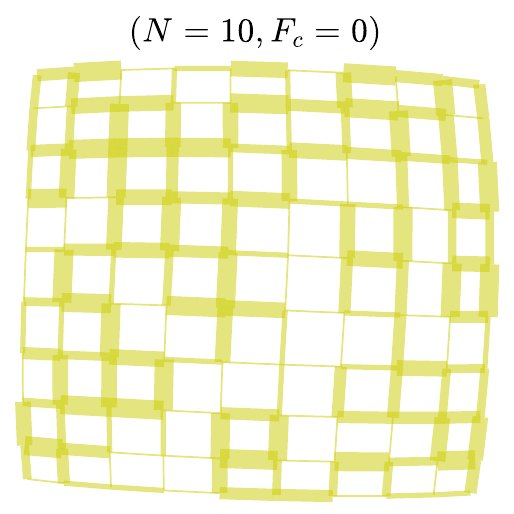} ~ \includegraphics[width=0.25\textwidth]{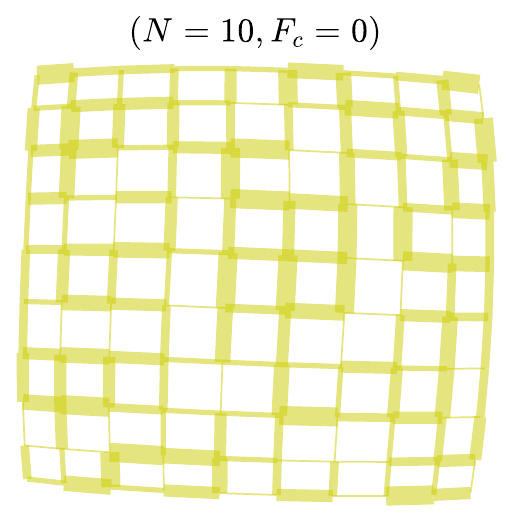} ~ \includegraphics[width=0.30\textwidth]{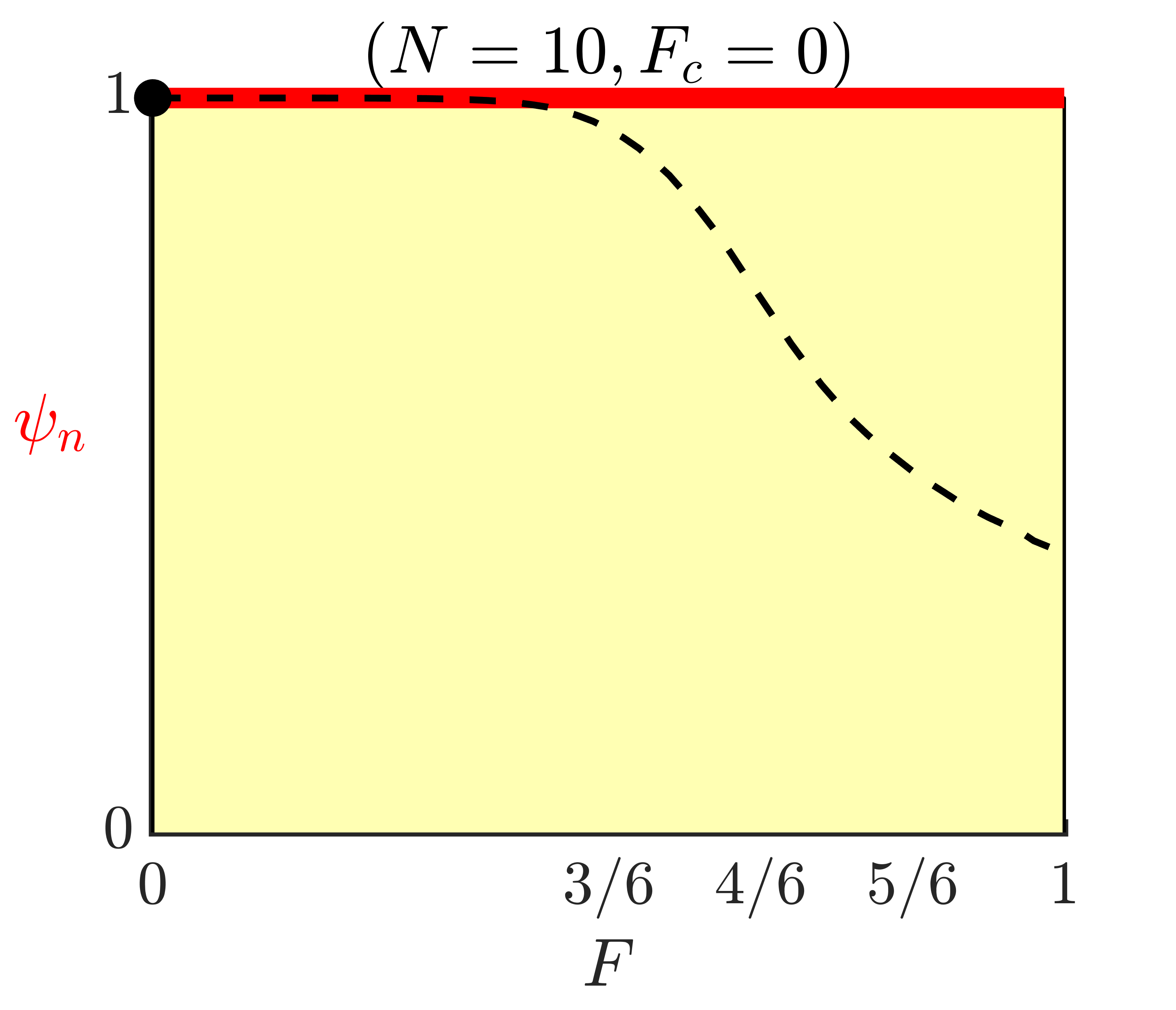}
\caption{Each of the leftmost two columns depicts quasistatic drainage on a two-dimensional square lattice with side length $N=10$, showing the fluid distribution at various $F_c$ as it decreases from $1$ to $0$. The thickness of each edge corresponds to its $F(r)$, where $r$ is its randomly assigned pore radius. The rightmost column displays the corresponding radius-resolved saturation profile, $\psi_n(F)$, as red curves, based on a total of $2\,000$ independent trials. The shaded areas represent saturations of the two phases (see Eq. \eqref{eq:saturations_and_rrs}). The black dashed curve shows the trajectory of the mean value of $\psi_n$ for $F_c<F \le 1$ as a function of $F_c$ (see Eq. \eqref{eq:radius_resolve_saturation_intrusion_final}).}\label{figure:invasion_rrs}
\end{figure}

Firstly, we construct an $N$-by-$N$ square lattice whose edges are assigned random ``pore radii''. Recall from earlier discussions that we find it advantageous to refer to a particular pore size by $F$, or the volume fraction of all pores in the sample that are smaller than that size. This way, we may simply assign to each edge on the lattice an $F$ drawn randomly and uniformly from the interval $[0,1]$, obviating the need for prescribing a PSD. Repeating this process a great number of times would result in an ensemble of realizations of lattices of the same dimensions, where the arrangement of different sized edges is statistically similar, but distinct in each realization. The leftmost two columns of Figure \ref{figure:invasion_rrs} represent two such realizations with $N=10$.

Secondly, we suppose that the pore space is initially filled with the wetting phase only (represented by the blue fluid in Figure \ref{figure:invasion_rrs}), which would undergo drainage as it is replaced by the nonwetting phase (represented by the yellow fluid in Figure \ref{figure:invasion_rrs}) in response to $F_c$ decreasing in small increments from $1$ to $0$. We assume that all vertices on the perimeter of the lattice have direct access to the invading nonwetting fluid, and that all interior vertices are connected to sinks, into which the defending wetting fluid may drain freely (alternatively, we may assume that the defending wetting fluid is indefinitely compressible, like the vacuum phase in mercury intrusion porosimetry). Under quasistatic conditions, at any prescribed $F_c$, each edge may be filled with either the wetting or the nonwetting phase, but not both. We neglect the capacity of vertices on the lattice, and consider bond percolation only (similar to \cite{fatt1956}). Like in typical invasion percolation calculations, an edge filled with the wetting phase will drain if and only if both of the following two conditions are satisfied: (1) its size $F$ is larger than the imposed $F_c$; (2) at least one of its vertices belongs to an edge that is filled with the nonwetting phase. As we see in either of the first two columns of Figure \ref{figure:invasion_rrs}, the $n$ phase replaces the $w$ phase in an increasing number of edges as $F_c$ decreases, until all edges are filled with the $n$ phase when $F_c$ goes to $0$, although the exact invasion percolation pattern is not the same in each realization due to the randomness in the arrangement of different sized pores.

Thirdly, we compute the radius-resolved saturation of the nonwetting phase at each $F_c$, $\psi_n(F;F_c)$, by tallying edges of each size (given by $F$) filled with either fluid ($w$ or $n$), across all realizations of lattices with the same prescribed side length. For example, at $F_c=4/6$, we find that among all edges of sizes near $F=0.81$ (i.e., in some small interval centered at this value depending on the discretization level) across a total of $2\,000$ realizations of $N=10$ lattices, $71.7\%$ are filled with the nonwetting phase and the rest are filled with the wetting phase; therefore, $\psi_n(F=0.81;F_c=4/6)=0.717$, as we may identify in the second plot in the rightmost column of Figure \ref{figure:invasion_rrs}. Similarly, the figure also contains $\psi_n(F)$ at other selected $F_c$. We observe that in each profile, $\psi_n=0$ for all $F<F_c$, and that $\psi_n$ is nearly constant for $F>F_c$. This is consistent with our proposition in Eq. \eqref{eq:radius_resolve_saturation_intrusion_basic}. Denoting the mean value of $\psi_n(F;F_c)$ for $F>F_c$ by $\psi_0\left(F_c\right)$ (similar to Eq. \eqref{eq:radius_resolve_saturation_intrusion_basic}), we plot its trajectory as black dashed curves overlaying the radius-resolved saturation profiles in Figure \ref{figure:invasion_rrs} (similar to Figure \ref{figure:radius_resolved_intrusion}). Evidently, the $\psi_0\left(F_c\right)$ trajectory is shaped differently in either case; notably, $\psi_0\left(F_c\right)$ approaches unity around $F_c=1/2$ on two-dimensional square lattices (see Figure \ref{figure:invasion_rrs}), rather than at $F_c=0$ in our statistical theory based on pore branching (see Eq. \eqref{eq:radius_resolve_saturation_intrusion_final} and Figure \ref{figure:radius_resolved_intrusion}). These correspond to the critical occupation probabilities (percolation thresholds) for bond percolation on a 2-D square lattice and in 1-D, which are $1/2$ and $1$, respectively \cite{sahimi1994,stauffer2014}.

\begin{figure}[!h]
\centering
\includegraphics[width=0.48\textwidth]{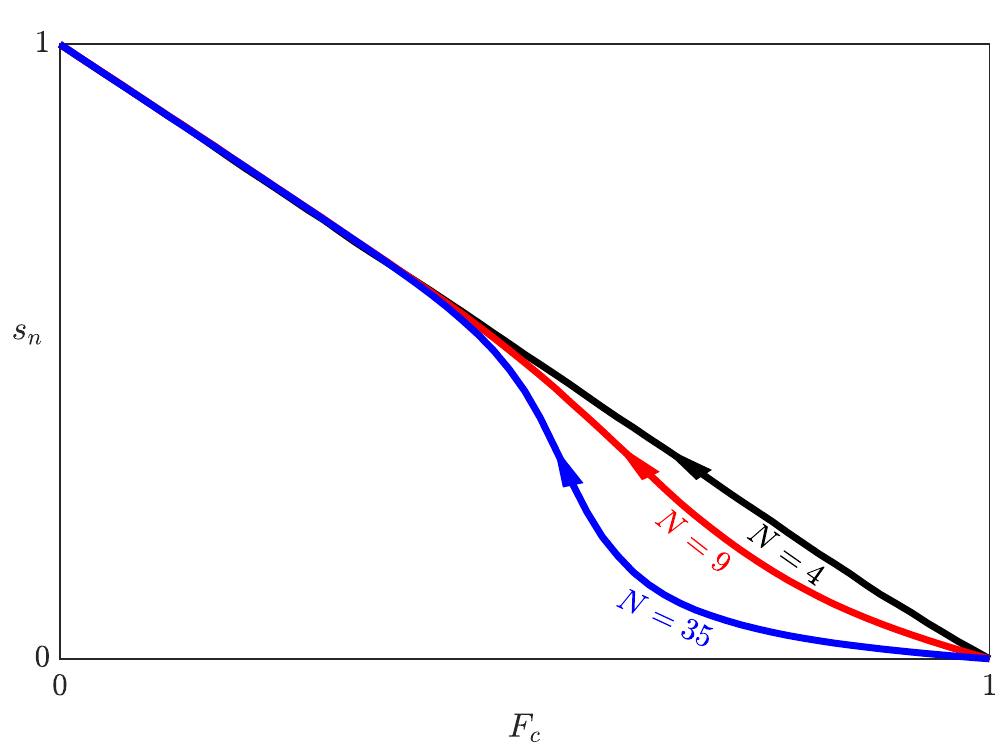}~\includegraphics[width=0.48\textwidth]{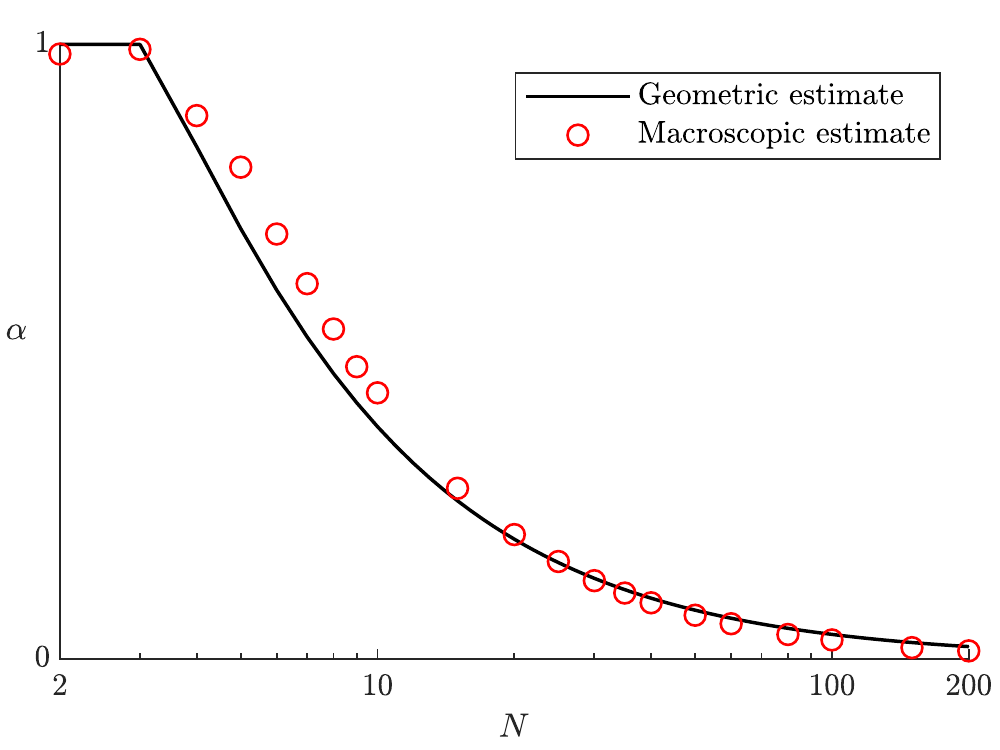}
\caption{The left-hand side plot shows the $s_n\left(F_c\right)$ curves during primary drainage on 2-D square lattices of side lengths $N=4$, $N=9$, and $N=35$, which deviates further from the $s_n=1-F_c$ (capillary bundle limit) as $N$ grows larger. The right-hand side plot compares the geometric and macroscopic estimates of accessivities of lattices with various $N$ (the horizontal axis uses a logarithmic scale), which are calculated from pore-scale data and continuum-scale measurements, respectively, and seem to agree well.}\label{figure:invasion_analysis}
\end{figure}

Fourthly, we use Eq. \eqref{eq:saturations_and_rrs} to compute $s_n$ from $\psi_n(F)$ at each $F_c$, so as to obtain the $s_n\left(F_c\right)$ relationship for quasistatic primary drainage. This calculation is then repeated for various $N$ ranging from $2$ to $200$, with selected results shown on the left-hand side of Figure \ref{figure:invasion_analysis}. For small $N$, nearly all pore segments on the lattice are directly accessible by the invading fluid, and we have $s_n \approx 1-F_c$, as the capillary bundle model would predict. As $N$ becomes larger, $s_n\left(F_c\right)$ deviates further from the  $s_n=1-F_c$ line, indicating a more prominent role of the ink-bottle effect. However, in all cases, $s_n$ rapidly approaches $1-F_c$ past the critical probability of $1/2$, which is reminiscent of the observations in \cite{larson1981}, and differs from those shown in Figure \ref{figure:s_vs_F}. Nevertheless, it is clear that when we increase $N$ here, like when we decrease $\alpha$ in our statistical theory, the porous sample behave in a way that is more and more distinct from that of a capillary bundle, as a result of a greater degree of serial connections between different sized pores.

Lastly, we capitalize on these observations to arrive at estimates for the accessivities of porous samples represented by the various square lattices considered. On the one hand, according to our microscopic statistical theory, $1/\alpha$ can be interpreted geometrically as the mean number of different sized pores encountered per pore instance (see Eqs. \eqref{eq:q_definition} -- \eqref{eq:alpha2q}). Although this interpretation of accessivity, as we have discussed, is only strictly valid when the pore space contains no loops, we may still consider it in the context of square lattices to obtain a ``geometric estimate'' for $\alpha$. One can verify that a 2-D square lattice with side length $N=2,3,\ldots$, as depicted in Figure \ref{figure:invasion_rrs}, has $2N(N-1)$ edges, and that of those edges, $4(2N-3)$ are connected to a vertex on the perimeter. If we claim that each edge that is accessible from the perimeter constitutes a pore instance for fluid displacement, and all the different sized edges are shared among these instances, we may write:
\begin{align}
    \frac{1}{\alpha_{\text{geom}}} &= \frac{\text{total \# edges on lattice}}{\text{\# edges on perimeter}}\nonumber\\
    &= \frac{2N(N-1)}{4(2N-3)}\nonumber\\
    \alpha_{\text{geom}} &= \frac{2(2N-3)}{N(N-1)}.
\end{align}
These geometric estimates, shown as the black curve in the right-hand side plot of Figure \ref{figure:invasion_analysis}, are valid for all integers greater than or equal to $2$. Note that $\alpha_{\text{geom}}=1$ for both $N=2$ and $N=3$ because in either case all edges on the lattice are also directly accessible from its perimeter. On the other hand, from a macroscopic perspective, we expect $\alpha$ to be correlated with the area of a hysteresis loop in a drainage-imbibition cycle, which we shall denote by $H$. Based on our simple formulae for primary drainage and imbibition, Eqs. \eqref{eq:saturation_intrusion_solution} and \eqref{eq:saturation_extrusion_solution}, the area between each pair of curves for a given $\alpha$ in Figure \ref{figure:s_vs_F} is:
\begin{align}
    H &= 1-2\int_0^1 \frac{\alpha\left(1-F_c\right)}
    {(1-\alpha)F_c+\alpha} \mathrm{d}F_c \nonumber\\
    \implies H &= 1+\frac{2\alpha}{1-\alpha}\left(\frac{\ln{\alpha}}{1-\alpha}+1\right), \label{eq:alpha2H}
\end{align}
which implies that the area of the hysteresis loop on a $s_n\left(F_c\right)$ graph (e.g., Figure \ref{figure:s_vs_F}) would vary from $0$ to $1$ as $\alpha$ changes from $1$ to $0$. In contrast, in the left-hand side plot in Figure \ref{figure:invasion_analysis}, it appears that the area between the $s_n\left(F_c\right)$ curve and the $s_n=1-F_c$ line only increases up to $1/8$ as $N\to\infty$. The area of a full hysteresis loop, which we shall denote by $H'$, would hence only increase up to $1/4$. If we consider $H'=1/4$ and $H=1$ analogous in the sense that they are both the maximum possible areas of a drainage-imbibition hysteresis loop in either scenario, it is plausible to estimate accessivity from invasion percolation data by substituting $H=4H'$ into Eq. \eqref{eq:alpha2H} and solving for $\alpha$, which we shall we refer to as $\alpha_{\text{marco}}$ because it is based on measurements in terms of macroscopic quantities only. The results for various $N$ are shown as red circles on the right-hand side of Figure \ref{figure:invasion_analysis}. Remarkably, $\alpha_{\text{geom}}$ and $\alpha_{\text{marco}}$ agree quite well with each other, even though they are both crude estimates based on our statistical theory involving pore morphologies that are notably simpler than those considered in these invasion percolation simulations. These results substantiate the view that accessivity as a continuum property of porous media does indeed have a physically intuitive pore-scale interpretation: because $\alpha$ measures the degree to which different sized pores are arranged in parallel or series, it must correlate with the area of hysteresis loops that arise due to connectivity effects, which holds true beyond the premise of our simple statistical theory based on pore branching, at least in a qualitative sense, even in pore networks that are plagued with loops.

\section{Further discussions}

In this section, we will further expound the conceptual usefulness as well as limitations of our theory. We will discuss the broader utility of accessivity and radius-resolved saturation, including how they may be generalized.

\subsection{Analyzing experimental and numerical data}

It is of interest to analyze simulation and experimental results based on accessivity and radius-resolved saturations, which better captures the microscopic distribution of fluid phases than the conventional saturations. Using these concepts affords an intuitive connection from macroscopic observations to the PSD, without resorting to the capillary bundle model.

For instance, suppose we would like to explain to someone for the first time why $p_c(s_w)$, $k_{rw}(s_w)$, and $k_{rn}(s_w)$ curves take their usual shapes. It would be impossible to do so without referring to how the wetting and nonwetting fluids are distributed in different sized pores. For example, in \cite{pinder2008}, the authors attributes the asymmetry between $k_{rw}(s_w)$ and $k_{rn}(s_w)$ curves (i.e., $k_{rw}(s_w=s_0) < k_{rn}(s_w=1-s_0)$ for some given $s_0$) to the fact that ``the wetting phase preferentially occupies the small pores'', which are associated with lower conductances (see Eq. \eqref{eq:poiseuille}). We can add to this intuitive explanation by considering $\psi_w(F)$ and $\psi_n(F)$: in a porous material with high accessivity, nearly all pores smaller than $F_c$ are indeed filled with the wetting phase; however, as $\alpha$ becomes lower, although the wetting phase still ``prefers'' smaller pores, only a fraction of all pores smaller than $F_c$ actually contains the wetting fluid. Furthermore, this line of reasoning would also naturally attribute hysteresis in $k_{rw}(s_w)$ and $k_{rn}(s_w)$ to the history dependent evolution patterns in $\psi_w(F)$ and $\psi_n(F)$, at the same time predicting less hysteresis in a sample with high accessivity.

Thus, it is conceptually advantageous to apply our theory to analyze experimental and simulation data where both pore-scale information and continuum-scale measurements are available. Such efforts may improve our understanding of how $\psi_w(F)$ and $\psi_n(F)$ evolve under different circumstances, and could facilitate the upscaling of pore-scale data to continuum-scale results. Additionally, empirical observations may suggest useful improvements upon the current formulae. For example, we may add relaxation dynamics and fluid entrapment to the laws governing the evolution of radius-resolved saturation, or consider the spatial correlation of different sized pores by making accessivity dependent on pore size, rather constant for the whole medium. Similar to efforts examining the role of the ``specific interfacial area'' \cite{kalaydjian1987,hassanizadeh1990,hassanizadeh1993,beliaev2001,hassanizadeh2002} as a state variable in continuum models of multiphase flow \cite{cheng2004,chen2007,pyrak-nolte2008,karadimitriou2014,porter2009,reeves1996,held2001,joekar-niasar2008,joekar-niasar2012-1,helland2007}, these investigations could lead to new constitutive relationships in the continuum description of multiphase flow.

\subsection{Improving continuum simulations of multiphase flow}

We have discussed how our statistical theory constitutes a constitutive law for capillary pressure hysteresis for use in continuum simulations of multiphase flow. Similarly, new constitutive relationships may be derived for relative permeability hysteresis too, which will be explored in future works. These new constitutive laws would complement traditional formulations of multiphase flow with the addition of $\psi_w(F)$ and $\psi_n(F)$ as state variables that better describe the pore-scale distribution of fluid phases, and $\alpha$ as a material property that tunes the amount of hysteresis in drainage-imbibition cycles.

More broadly speaking, it is conceivable that the simple probabilistic arguments that we used in deriving the ODEs for quasistatic drainage-imbibition could motivate new PDE models of greater generality for the dynamics of multiphase flow in porous media, which would be comparable to Hilfer's approach \cite{hilfer1998,hilfer2006,doster2010} but with an intuitive connection to microscopic physics.

\subsection{Establishing correlations}

\begin{figure}[!h]
\centering
\includegraphics{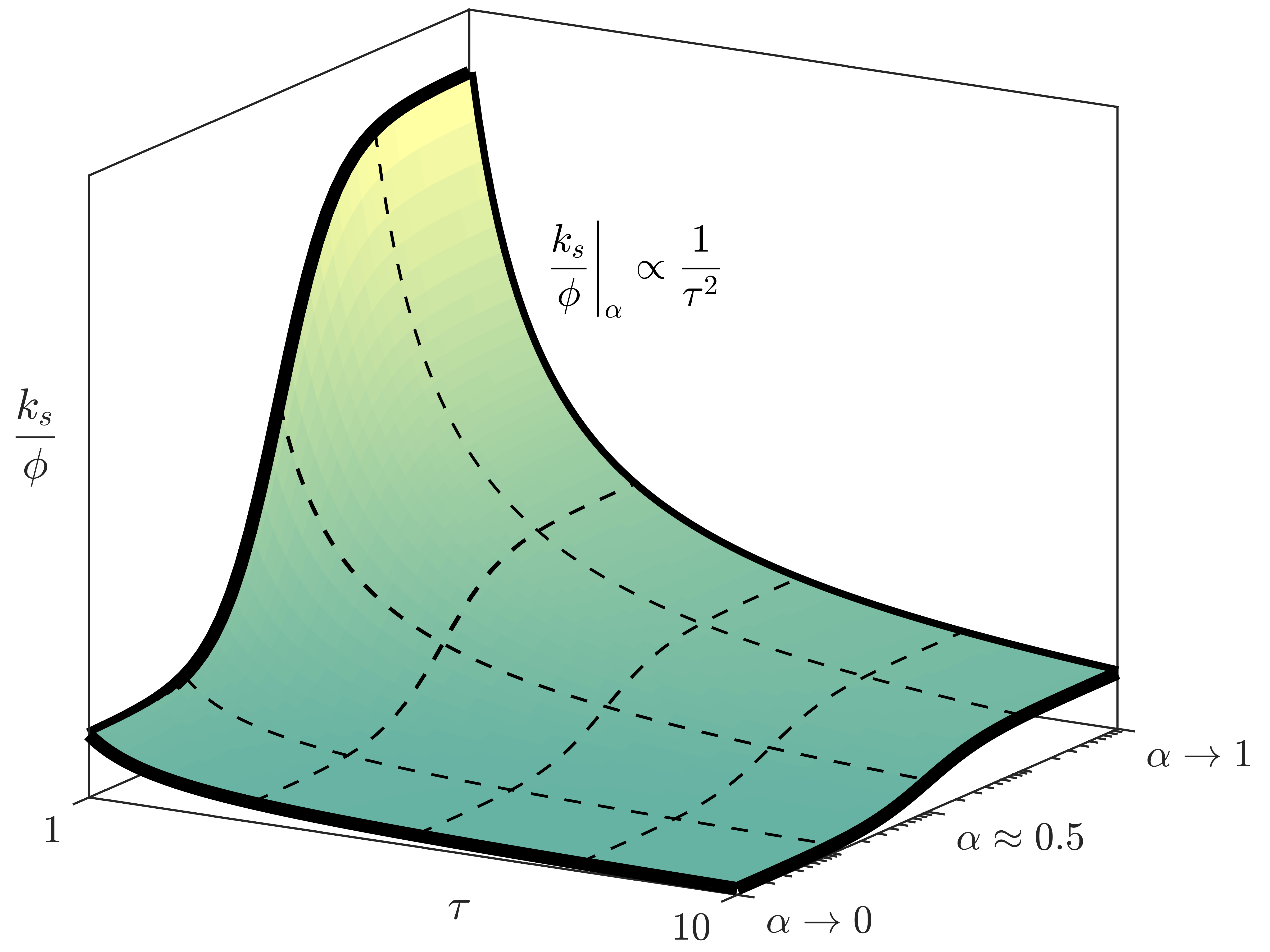}
\caption{Continuum-scale properties such as porosity ($\phi$), tortuosity ($\tau$), and accessivity ($\alpha$) distinguish between porous media with dissimilar pore-scale features. Introducing accessivity, which describes the connectivity of different sized pores in a simple fashion, expands our ability to conceptually and quantitatively characterize porous media (such as depicted in qualitative terms here for the intrinsic permeability, $k_s$), which would be relevant in continuum modeling.}\label{figure:permeability_surface}
\end{figure}

Finally, accessivity as a new property of porous materials can be incorporated into conceptual discussions of macroscopic processes where the connection of different sized pores plays a role. As a specific example, in the classical literature \cite{kozeny1927,carman1937,bear1972}, the intrinsic permeability, $k_s$, of a porous medium for single-phase flow is often linked to $\phi$ and $\tau$. Under the assumption that the pore space behaves like a bundle of nearly identical channels, all with the same cross-sectional shape and effective radius, the intrinsic permeability is modeled as:
\begin{align}
    k_s = C \frac{\phi}{\tau},\label{eq:permeability_classical}
\end{align}
where the proportionality constant $C$ has units of area and is dependent on the assumed cross-sectional shape and pore radius. Scheidegger \cite{scheidegger1953} extended this picture by allowing the pores to vary in size, deriving formulae for $C$ in Eq. \eqref{eq:permeability_classical} given the PSD, for two limiting cases -- where different sized pores are arranged completely in parallel, and where they are arranged completely in series. Unsurprisingly, the model predicts a higher $C$, and hence $k_s$, for the parallel case, where the narrower pores contribute less to the overall resistance to flow. This hints at incorporating $\alpha$ into our description for $k_s$, which we qualitatively portray in Figure \ref{figure:permeability_surface}, with higher $\alpha$ leading to a higher $k_s$ at constant $\phi$ and $\tau$, assuming the PSD remains the same. Note that, for a certain type of porous media, we may not be able to vary all of $\phi$, $\tau$, and $\alpha$ completely independently, and it is an interesting research question as to what the various constraints might be.

\section{Conclusions}
We have proposed the pore-space accessivity, $\alpha\in(0,1)$, as a continuum-scale parameter for describing the arrangement of different sized in porous media. Defined quantitatively for an idealized pore space but interpreted conceptually similarly for real porous media, $\alpha$ compares the length (or volume) scale for pore radius variation to the typical size of a pore instance (the pore space explored by an average meniscus in a control volume); in other words, we may interpret $1/\alpha$ as the average number of different sized pore segments encountered in a pore instance. In the limit of $\alpha\to0$, different sized pores are overwhelmingly connected in series, leading to significant ink-bottle effect, thereby causing hysteresis loops in drainage-imbibition cycles to widen. As $\alpha\to1$, different sized pores become overwhelmingly arranged in parallel, hence eliminating connectivity effects and recovering the classical model of a bundle of straight capillaries.

We have proposed the radius-resolved saturations, $\psi_w(F)$ and $\psi_n(F)$, to replace convectional saturations, $s_w$ and $s_n$. Because $\psi_w(F)$ and $\psi_n(F)$ explicitly assign fluid saturations to different sized pores, they serve as a better state variable for describing multiphase processes in porous media than conventional saturation variables. Based on our statistical framework, we derived a simple set of algebraic formulae for updating $\psi_w(F)$ and $\psi_n(F)$ during arbitrary cycles of quasistatic drainage and imbibition. The formalism naturally predicts capillary pressure hysteresis caused by connectivity effects, while the amount of hysteresis is controlled by accessivity. As $\alpha\to1$, at a certain imposed capillary pressure, we have $\psi_n(F)\approx1$ for all pores large enough to favor occupation by the nonwetting fluid, and $\psi_n(F)\approx0$ for all smaller pores. At a lower $\alpha$, connectivity effects create more complex patterns in $\psi_n(F)$, whose history dependence leads to hysteresis in conventional macroscopic state variables.

The statistical framework we considered uses a probabilistic process to construct an idealized pore space, where pore branching and effective pore radius variation are modeled as independent homogeneous Poisson point processes with respect to some pore axial coordinate. Considering quasistatic immiscible drainage-imbibition in a porous medium, we have examined events experienced by a meniscus in an ensemble of instances of probabilistically constructed pore spaces, and have subsequently derived simple algebraic formulae (Eqs. \eqref{eq:saturation_intrusion_solution}, \eqref{eq:saturation_extrusion_solution}, \eqref{eq:rrs_n_general}, \eqref{eq:rrs_w_general}, \eqref{eq:radius_resolve_saturation_intrusion_final}, and \eqref{eq:radius_resolve_saturation_extrusion_final}) for capillary pressure hysteresis for porous media with different $\alpha$, expressed in terms of either conventional saturations or radius-resolved saturations. All formulae converge to the capillary bundle model in the $\alpha\to1$ limit.

We have demonstrated that our simple algebraic formulae could serve as a new constitutive law for capillary pressure hysteresis for use in continuum simulations, at the expense of the inclusion additional state variables. The formulae may also be applied to interpret mercury intrusion-extrusion porosimetry measurements, where $\alpha$ accounts for connectivity effects in a simple fashion to complement the usual consideration of contact angle hysteresis. Using results from a simple invasion percolation study, we have interpreted accessivity both in terms of pore-scale, geometric information and continuum-scale measurements, which suggest the conceptual generality of our theoretical framework.

Like established concepts including porosity ($\phi$), tortuosity ($\tau$), and pore-size distributions, we expect accessivity ($\alpha$) and radius-resolved saturations ($\psi_w(F),\psi_n(F)$) to have much broader utility in the continuum modeling of multiphase processes in porous media at larger. Our framework conceptualizes and quantifies the arrangement of different sized pores and the microscopic distribution of fluid phases within, and makes predictions of continuum-scale behaviors of the porous medium accordingly. We expect the concepts proposed in this work to motivate new constitutive laws, correlations, and PDE models for continuum-scale processes in porous media that captures connectivity effects in a mathematically simple and physically intuitive fashion.

\section*{Acknowledgement}
The authors wish to acknowledge funding from Saudi Aramco, a Founding Member of the MIT Energy Initiative. We are also grateful to Samuel J. Cooper for useful discussions.


\section*{References}
\begin{thebibliography}{100}
\expandafter\ifx\csname url\endcsname\relax
  \def\url#1{\texttt{#1}}\fi
\expandafter\ifx\csname urlprefix\endcsname\relax\def\urlprefix{URL }\fi
\expandafter\ifx\csname href\endcsname\relax
  \def\href#1#2{#2} \def\path#1{#1}\fi

\bibitem{bear1972}
J.~Bear, Dynamics of fluids in porous media, New York : American Elsevier,
  1972.

\bibitem{torquato2013}
S.~Torquato, Random heterogeneous materials: microstructure and macroscopic
  properties, Vol.~16, Springer Science \& Business Media, 2013.

\bibitem{sahimi2011}
M.~Sahimi, Flow and {Transport} in {Porous} {Media} and {Fractured} {Rock}:
  {From} {Classical} {Methods} to {Modern} {Approaches}, John Wiley \& Sons,
  2011.

\bibitem{allaire1989}
G.~Allaire, Homogenization of the {Stokes} flow in a connected porous medium,
  Asymptotic Analysis 2~(3) (1989) 203--222.

\bibitem{hornung2012}
U.~Hornung, Homogenization and porous media, Vol.~6, Springer Science \&
  Business Media, 2012.

\bibitem{loeb1954}
A.~L. Loeb, Thermal conductivity: {VIII}, a theory of thermal conductivity of
  porous materials, Journal of the American Ceramic Society 37~(2) (1954)
  96--99.

\bibitem{carson2005}
J.~K. Carson, S.~J. Lovatt, D.~J. Tanner, A.~C. Cleland, Thermal conductivity
  bounds for isotropic, porous materials, International Journal of Heat and
  Mass Transfer 48~(11) (2005) 2150--2158.

\bibitem{ranut2016}
P.~Ranut, On the effective thermal conductivity of aluminum metal foams:
  {Review} and improvement of the available empirical and analytical models,
  Applied Thermal Engineering 101 (2016) 496--524.

\bibitem{pabst2017}
W.~Pabst, E.~Gregorova, A generalized cross-property relation between the
  elastic moduli and conductivity of isotropic porous materials with spheroidal
  pores, Ceram. Silik 61~(1) (2017) 74--80.

\bibitem{neithalath2010}
N.~Neithalath, M.~S. Sumanasooriya, O.~Deo, Characterizing pore volume, sizes,
  and connectivity in pervious concretes for permeability prediction, Materials
  characterization 61~(8) (2010) 802--813.

\bibitem{chareyre2012}
B.~Chareyre, A.~Cortis, E.~Catalano, E.~Barthélemy, Pore-scale modeling of
  viscous flow and induced forces in dense sphere packings, Transport in porous
  media 94~(2) (2012) 595--615.

\bibitem{mostaghimi2013}
P.~Mostaghimi, M.~J. Blunt, B.~Bijeljic, Computations of absolute permeability
  on micro-{CT} images, Mathematical Geosciences 45~(1) (2013) 103--125.

\bibitem{scheidegger1974}
A.~E. Scheidegger, The physics of flow through porous media, 3rd Edition,
  University of Toronto Press: Toronto, 1974.

\bibitem{richards1931}
L.~A. Richards, Capillary conduction of liquids through porous mediums, physics
  1~(5) (1931) 318--333.

\bibitem{buckley1942}
S.~E. Buckley, M.~C. Leverett, Mechanism of fluid displacement in sands, T.
  AIME 146~(01) (1942) 107--116.

\bibitem{childs1950}
E.~C. Childs, N.~Collis-George, The permeability of porous materials, in:
  Proceedings of the {Royal} {Society} of {London} {A}: {Mathematical},
  {Physical} and {Engineering} {Sciences}, Vol. 201, The Royal Society, 1950,
  pp. 392--405.

\bibitem{klute1952}
A.~Klute, Some {Theoretical} {Aspects} of the {Flow} of {Water} in
  {Unsaturated} {Soils} 1, Soil Science Society of America Journal 16~(2)
  (1952) 144--148.

\bibitem{tamon1981}
H.~Tamon, M.~Okazaki, R.~Toei, Flow mechanism of adsorbate through porous media
  in presence of capillary condensation, AIChE Journal 27~(2) (1981) 271--277.

\bibitem{lee1986}
K.-H. Lee, S.-T. Hwang, The transport of condensible vapors through a
  microporous {Vycor} glass membrane, Journal of colloid and interface science
  110~(2) (1986) 544--555.

\bibitem{jaguste1995}
D.~N. Jaguste, S.~K. Bhatia, Combined surface and viscous flow of condensable
  vapor in porous media, Chemical engineering science 50~(2) (1995) 167--182.

\bibitem{do2001}
H.~D. Do, D.~D. Do, A new diffusion and flow theory for activated carbon from
  low pressure to capillary condensation range, Chemical engineering journal
  84~(3) (2001) 295--308.

\bibitem{durlofsky1998}
L.~J. Durlofsky, Coarse scale models of two phase flow in heterogeneous
  reservoirs: volume averaged equations and their relationship to existing
  upscaling techniques, Computational Geosciences 2~(2) (1998) 73--92.

\bibitem{arbogast2000}
T.~Arbogast, Numerical subgrid upscaling of two-phase flow in porous media, in:
  Numerical treatment of multiphase flows in porous media, Springer, 2000, pp.
  35--49.

\bibitem{cushman2002}
J.~H. Cushman, L.~S. Bennethum, B.~X. Hu, A primer on upscaling tools for
  porous media, Advances in Water Resources 25~(8-12) (2002) 1043--1067.

\bibitem{wood2009}
B.~D. Wood, The role of scaling laws in upscaling, Advances in Water Resources
  32~(5) (2009) 723--736.

\bibitem{li2006}
L.~Li, C.~A. Peters, M.~A. Celia, Upscaling geochemical reaction rates using
  pore-scale network modeling, Advances in water resources 29~(9) (2006)
  1351--1370.

\bibitem{lenormand1983}
R.~Lenormand, C.~Zarcone, A.~Sarr, Mechanisms of the displacement of one fluid
  by another in a network of capillary ducts, J. Fluid Mech. 135 (1983)
  337--353.

\bibitem{pak2015}
T.~Pak, I.~B. Butler, S.~Geiger, M.~I. van Dijke, K.~S. Sorbie, Droplet
  fragmentation: 3d imaging of a previously unidentified pore-scale process
  during multiphase flow in porous media, Proceedings of the National Academy
  of Sciences 112~(7) (2015) 1947--1952.

\bibitem{holtzman2015}
R.~Holtzman, E.~Segre, Wettability stabilizes fluid invasion into porous media
  via nonlocal, cooperative pore filling, Physical review letters 115~(16)
  (2015) 164501.

\bibitem{zhao2016}
B.~Zhao, C.~W. MacMinn, R.~Juanes, Wettability control on multiphase flow in
  patterned microfluidics, Proceedings of the National Academy of Sciences
  113~(37) (2016) 10251--10256.

\bibitem{brooks1964}
R.~H. Brooks, A.~T. Corey, Hydraulic properties of porous media and their
  relation to drainage design, T. ASAE 7~(1) (1964) 26--0028.

\bibitem{van_genuchten1980}
M.~T. van Genuchten, A closed-form equation for predicting the hydraulic
  conductivity of unsaturated soils, Soil Sci. Soc. Am. J. 44~(5) (1980)
  892--898.

\bibitem{corey1954}
A.~T. Corey, The interrelation between gas and oil relative permeabilities,
  Produc. Mon. 19~(1) (1954) 38--41.

\bibitem{irmay1954}
S.~Irmay, On the hydraulic conductivity of unsaturated soils, Eos, Transactions
  American Geophysical Union 35~(3) (1954) 463--467.

\bibitem{pinder2008}
G.~F. Pinder, W.~G. Gray, Essentials of {Multiphase} {Flow} in {Porous}
  {Media}, John Wiley \& Sons, 2008.

\bibitem{thomson1872}
W.~Thomson, 4. {On} the equilibrium of vapour at a curved surface of liquid,
  Proceedings of the Royal Society of Edinburgh 7 (1872) 63--68.

\bibitem{washburn1921}
E.~W. Washburn, Note on a method of determining the distribution of pore sizes
  in a porous material, P. Natl. Acad. Sci. USA (1921) 115--116.

\bibitem{burdine1953}
N.~T. Burdine, Relative permeability calculations from pore size distribution
  data, J. Petrol. Technol. 5~(03) (1953) 71--78.

\bibitem{mualem1976}
Y.~Mualem, A new model for predicting the hydraulic conductivity of unsaturated
  porous media, Water Resour. Res. 12~(3) (1976) 513--522.

\bibitem{van_brakel1975}
J.~van Brakel, Pore space models for transport phenomena in porous media review
  and evaluation with special emphasis on capillary liquid transport, Powder
  Technol. 11~(3) (1975) 205--236.
\newblock \href {http://dx.doi.org/10.1016/0032-5910(75)80049-0}
  {\path{doi:10.1016/0032-5910(75)80049-0}}.

\bibitem{quiblier1984}
J.~A. Quiblier, A new three-dimensional modeling technique for studying porous
  media, J. Colloid Interf. Sci. 98~(1) (1984) 84--102.
\newblock \href {http://dx.doi.org/10.1016/0021-9797(84)90481-8}
  {\path{doi:10.1016/0021-9797(84)90481-8}}.

\bibitem{barenblatt1971}
G.~I. Barenblatt, Filtration of two nonmixing fluids in a homogeneous porous
  medium, Fluid Dynamics 6~(5) (1971) 857--864.

\bibitem{luckner1989}
L.~Luckner, M.~T. Van~Genuchten, D.~R. Nielsen, A consistent set of parametric
  models for the two-phase flow of immiscible fluids in the subsurface, Water
  Resources Research 25~(10) (1989) 2187--2193.

\bibitem{hassanizadeh1990}
S.~M. Hassanizadeh, W.~G. Gray, Mechanics and thermodynamics of multiphase flow
  in porous media including interphase boundaries, Advances in water resources
  13~(4) (1990) 169--186.

\bibitem{barenblatt2003}
G.~I. Barenblatt, T.~W. Patzek, D.~B. Silin, The mathematical model of
  nonequilibrium effects in water-oil displacement, SPE journal 8~(04) (2003)
  409--416.

\bibitem{juanes2008}
R.~Juanes, Nonequilibrium effects in models of three-phase flow in porous
  media, Advances in Water Resources 31~(4) (2008) 661--673.

\bibitem{kalaydjian1987}
F.~Kalaydjian, A macroscopic description of multiphase flow in porous media
  involving spacetime evolution of fluid/fluid interface, Transport in Porous
  Media 2~(6) (1987) 537--552.

\bibitem{hassanizadeh1993}
S.~M. Hassanizadeh, W.~G. Gray, Thermodynamic basis of capillary pressure in
  porous media, Water resources research 29~(10) (1993) 3389--3405.

\bibitem{beliaev2001}
A.~Y. Beliaev, S.~M. Hassanizadeh, A theoretical model of hysteresis and
  dynamic effects in the capillary relation for two-phase flow in porous media,
  Transport in Porous media 43~(3) (2001) 487--510.

\bibitem{hassanizadeh2002}
S.~M. Hassanizadeh, M.~A. Celia, H.~K. Dahle, Dynamic effect in the capillary
  pressure–saturation relationship and its impacts on unsaturated flow,
  Vadose Zone Journal 1~(1) (2002) 38--57.

\bibitem{cheng2004}
J.-T. Cheng, L.~J. Pyrak-Nolte, D.~D. Nolte, N.~J. Giordano, Linking pressure
  and saturation through interfacial areas in porous media, Geophys. Res. Lett.
  31~(8).

\bibitem{chen2007}
D.~Chen, L.~J. Pyrak-Nolte, J.~Griffin, N.~J. Giordano, Measurement of
  interfacial area per volume for drainage and imbibition, Water Resour. Res.
  43~(12).

\bibitem{pyrak-nolte2008}
L.~J. Pyrak-Nolte, D.~D. Nolte, D.~Chen, N.~J. Giordano, Relating capillary
  pressure to interfacial areas, Water Resour. Res. 44~(6).

\bibitem{karadimitriou2014}
N.~K. Karadimitriou, S.~M. Hassanizadeh, V.~Joekar-Niasar, P.~J. Kleingeld,
  Micromodel study of two-phase flow under transient conditions: {Quantifying}
  effects of specific interfacial area, Water Resour. Res. 50~(10) (2014)
  8125--8140.

\bibitem{porter2009}
M.~L. Porter, M.~G. Schaap, D.~Wildenschild, Lattice-{Boltzmann} simulations of
  the capillary pressure-saturation-interfacial area relationship for porous
  media, Adv. Water Resour. 32~(11) (2009) 1632--1640.

\bibitem{reeves1996}
P.~C. Reeves, M.~A. Celia, A {Functional} {Relationship} {Between} {Capillary}
  {Pressure}, {Saturation}, and {Interfacial} {Area} as {Revealed} by a
  {Pore}-{Scale} {Network} {Model}, Water Resour. Res. 32~(8) (1996)
  2345--2358.

\bibitem{held2001}
R.~J. Held, M.~A. Celia, Modeling support of functional relationships between
  capillary pressure, saturation, interfacial area and common lines, Adv. Water
  Resour. 24~(3) (2001) 325--343.

\bibitem{joekar-niasar2008}
V.~Joekar-Niasar, S.~M. Hassanizadeh, A.~Leijnse, Insights into the
  relationships among capillary pressure, saturation, interfacial area and
  relative permeability using pore-network modeling, Transport Porous Med.
  74~(2) (2008) 201--219.

\bibitem{joekar-niasar2012-1}
V.~Joekar-Niasar, S.~M. Hassanizadeh, Uniqueness of specific interfacial area -
  capillary pressure-saturation relationship under non-equilibrium conditions
  in two-phase porous media flow, Transport Porous Med. 94~(2) (2012) 465--486.

\bibitem{helland2007}
J.~O. Helland, S.~M. Skjaeveland, Relationship between capillary pressure,
  saturation, and interfacial area from a model of mixed-wet triangular tubes,
  Water Resour. Res. 43~(12).

\bibitem{hilfer1998}
R.~Hilfer, Macroscopic equations of motion for two-phase flow in porous media,
  Physical Review E 58~(2) (1998) 2090.

\bibitem{hilfer2006}
R.~Hilfer, Macroscopic capillarity and hysteresis for flow in porous media,
  Physical Review E 73~(1) (2006) 016307.

\bibitem{doster2010}
F.~Doster, P.~A. Zegeling, R.~Hilfer, Numerical solutions of a generalized
  theory for macroscopic capillarity, Physical Review E 81~(3) (2010) 036307.

\bibitem{meakin2009}
P.~Meakin, A.~M. Tartakovsky, Modeling and simulation of pore-scale multiphase
  fluid flow and reactive transport in fractured and porous media, Reviews of
  Geophysics 47~(3).

\bibitem{essaid1993}
H.~I. Essaid, W.~N. Herkelrath, K.~M. Hess, Simulation of fluid distributions
  observed at a crude oil spill site incorporating hysteresis, oil entrapment,
  and spatial variability of hydraulic properties, Water Resources Research
  29~(6) (1993) 1753--1770.

\bibitem{diamond2000}
S.~Diamond, Mercury porosimetry: an inappropriate method for the measurement of
  pore size distributions in cement-based materials, Cement Concrete Res.
  30~(10) (2000) 1517--1525.
\newblock \href {http://dx.doi.org/10.1016/S0008-8846(00)00370-7}
  {\path{doi:10.1016/S0008-8846(00)00370-7}}.

\bibitem{hunt2013}
A.~G. Hunt, R.~P. Ewing, R.~Horton, What’s wrong with soil physics?, Soil
  Sci. Soc. Am. J. 77~(6) (2013) 1877--1887.
\newblock \href {http://dx.doi.org/10.2136/sssaj2013.01.0020}
  {\path{doi:10.2136/sssaj2013.01.0020}}.

\bibitem{epstein1989}
N.~Epstein, On tortuosity and the tortuosity factor in flow and diffusion
  through porous media, Chem. Eng. Sci. 44~(3) (1989) 777--779.
\newblock \href {http://dx.doi.org/10.1016/0009-2509(89)85053-5}
  {\path{doi:10.1016/0009-2509(89)85053-5}}.

\bibitem{cooper2016}
S.~J. Cooper, A.~Bertei, P.~R. Shearing, J.~A. Kilner, N.~P. Brandon,
  {TauFactor}: {An} open-source application for calculating tortuosity factors
  from tomographic data, SoftwareX 5 (2016) 203--210.

\bibitem{deen2011}
W.~M. Deen, Analysis of {Transport} {Phenomena}, second edition Edition, Oxford
  University Press, New York, 2011.

\bibitem{hilfer1996}
R.~Hilfer, Transport and relaxation phenomena in porous media, Adv. Chem. Phys.
  92 (1996) 299--424.

\bibitem{wilkinson1983}
D.~Wilkinson, D.~F. Willemsen, Invasion percolation: a new form of percolation
  theory, J. Phys. A-Math. Gen. 16~(14) (1983) 3365.
\newblock \href {http://dx.doi.org/10.1088/0305-4470/16/14/028}
  {\path{doi:10.1088/0305-4470/16/14/028}}.

\bibitem{abell1999}
A.~B. Abell, K.~L. Willis, D.~A. Lange, Mercury intrusion porosimetry and image
  analysis of cement-based materials, J. Colloid Interf. Sci. 211~(1) (1999)
  39--44.
\newblock \href {http://dx.doi.org/10.1006/jcis.1998.5986}
  {\path{doi:10.1006/jcis.1998.5986}}.

\bibitem{lowell2012}
S.~Lowell, J.~E. Shields, M.~A. Thomas, M.~Thommes, Characterization of
  {Porous} {Solids} and {Powders}: {Surface} {Area}, {Pore} {Size} and
  {Density}, Springer Science \& Business Media, 2012.

\bibitem{fatt1956}
I.~Fatt, Network model of porous media, J. Petrol. Technol. 8~(7) (1956)
  144--177.

\bibitem{celia1995}
M.~A. Celia, P.~C. Reeves, L.~A. Ferrand, Recent advances in pore scale models
  for multiphase flow in porous media, Rev. Geophys. 33~(S2) (1995) 1049--1057.
\newblock \href {http://dx.doi.org/10.1029/95RG00248}
  {\path{doi:10.1029/95RG00248}}.

\bibitem{bakke1997}
S.~Bakke, P.-E. Øren, 3-{D} pore-scale modelling of sandstones and flow
  simulations in the pore networks, SPE J. 2~(02) (1997) 136--149.
\newblock \href {http://dx.doi.org/10.2118/35479-PA}
  {\path{doi:10.2118/35479-PA}}.

\bibitem{blunt2001}
M.~J. Blunt, Flow in porous media - pore-network models and multiphase flow,
  Curr. Opin. Colloid In. 6~(3) (2001) 197--207.
\newblock \href {http://dx.doi.org/10.1016/S1359-0294(01)00084-X}
  {\path{doi:10.1016/S1359-0294(01)00084-X}}.

\bibitem{dullien1991}
F.~A.~L. Dullien, Porous media: fluid transport and pore structure, Academic
  press, 1991.

\bibitem{joekar-niasar2012}
V.~Joekar-Niasar, S.~M. Hassanizadeh, Analysis of fundamentals of two-phase
  flow in porous media using dynamic pore-network models: a review, Cr. Rev.
  Env. Sci. Techol. 42~(18) (2012) 1895--1976.
\newblock \href {http://dx.doi.org/10.1080/10643389.2011.574101}
  {\path{doi:10.1080/10643389.2011.574101}}.

\bibitem{sahimi1994}
M.~Sahimi, Applications of percolation theory, CRC Press, 1994.

\bibitem{stauffer2014}
D.~Stauffer, A.~Aharony, Introduction to percolation theory: revised second
  edition, CRC press, 2014.

\bibitem{selyakov2013}
V.~I. Selyakov, V.~Kadet, Percolation models for transport in porous media:
  with applications to reservoir engineering, Vol.~9, Springer Science \&
  Business Media, 2013.

\bibitem{larson1981}
R.~G. Larson, N.~R. Morrow, Effects of sample size on capillary pressures in
  porous media, Powder Technol. 30~(2) (1981) 123--138.
\newblock \href {http://dx.doi.org/10.1016/0032-5910(81)80005-8}
  {\path{doi:10.1016/0032-5910(81)80005-8}}.

\bibitem{mason1982}
G.~Mason, The effect of pore space connectivity on the hysteresis of capillary
  condensation in adsorption—desorption isotherms, Journal of Colloid and
  Interface Science 88~(1) (1982) 36--46.

\bibitem{mason1983}
G.~Mason,
  \href{http://www.sciencedirect.com/science/article/pii/0021979783901005}{The
  effect of pore lattice structure on the pore size distributions calculated
  from sorption isotherms using percolation theory}, Journal of Colloid and
  Interface Science 95~(1) (1983) 277--278.
\newblock \href {http://dx.doi.org/10.1016/0021-9797(83)90100-5}
  {\path{doi:10.1016/0021-9797(83)90100-5}}.
\newline\urlprefix\url{http://www.sciencedirect.com/science/article/pii/0021979783901005}

\bibitem{parlar1988}
M.~Parlar, Y.~C. Yortsos, Percolation theory of vapor adsorption—desorption
  processes in porous materials, Journal of colloid and interface science
  124~(1) (1988) 162--176.

\bibitem{seaton1991}
N.~A. Seaton, Determination of the connectivity of porous solids from nitrogen
  sorption measurements, Chemical Engineering Science 46~(8) (1991) 1895--1909.

\bibitem{kadet2013}
V.~V. Kadet, A.~M. Galechyan, Percolation model of relative permeability
  hysteresis, J. Appl. Mech. Tech. Phys. 54~(3) (2013) 423--432.
\newblock \href {http://dx.doi.org/10.1134/S002189441}
  {\path{doi:10.1134/S002189441}}.

\bibitem{pinson2014}
M.~B. Pinson, T.~Zhao, H.~M. Jennings, M.~Z. Bazant,
  \href{http://arxiv.org/abs/1402.3377}{Inferring {Pore} {Size} and {Network}
  {Structure} from {Sorption} {Hysteresis}}, arXiv:1402.3377 [cond-mat]ArXiv:
  1402.3377.
\newline\urlprefix\url{http://arxiv.org/abs/1402.3377}

\bibitem{masoero2015}
E.~Masoero, M.~B. Pinson, P.~A. Bonnaud, H.~Manzano, Q.~Ji, S.~Yip, J.~J.
  Thomas, M.~Z. Bazant, K.~Van~Vliet, H.~M. Jennings, Modelling {Hysteresis} in
  the {Water} {Sorption} and {Drying} {Shrinkage} of {Cement} {Paste}, in:
  {CONCREEP} 10, 2015, pp. 306--312.

\bibitem{pinson2015}
M.~B. Pinson, E.~Masoero, P.~A. Bonnaud, H.~Manzano, Q.~Ji, S.~Yip, J.~J.
  Thomas, M.~Z. Bazant, K.~J. Van~Vliet, H.~M. Jennings, Hysteresis from
  multiscale porosity: modeling water sorption and shrinkage in cement paste,
  Physical Review Applied 3~(6) (2015) 064009.

\bibitem{salmas2001}
C.~Salmas, G.~Androutsopoulos, Mercury porosimetry: contact angle hysteresis of
  materials with controlled pore structure, J. Colloid Interf. Sci. 239~(1)
  (2001) 178--189.
\newblock \href {http://dx.doi.org/10.1006/jcis.2001.7531}
  {\path{doi:10.1006/jcis.2001.7531}}.

\bibitem{giesche2006}
H.~Giesche, Mercury {Porosimetry}: {A} {General} ({Practical}) {Overview},
  Part. Part. Syst. Char. 23~(1) (2006) 9--19.

\bibitem{gostick2008}
J.~T. Gostick, M.~A. Ioannidis, M.~W. Fowler, M.~D. Pritzker, Direct
  measurement of the capillary pressure characteristics of water–air–gas
  diffusion layer systems for {PEM} fuel cells, Electrochemistry Communications
  10~(10) (2008) 1520--1523.

\bibitem{forner-cuenca2016}
A.~Forner-Cuenca, J.~Biesdorf, A.~Lamibrac, V.~Manzi-Orezzoli, F.~N. Büchi,
  L.~Gubler, T.~J. Schmidt, P.~Boillat, Advanced {Water} {Management} in
  {PEFCs}: {Diffusion} {Layers} with {Patterned} {Wettability} {II}.
  {Measurement} of {Capillary} {Pressure} {Characteristic} with {Neutron} and
  {Synchrotron} {Imaging}, Journal of The Electrochemical Society 163~(9)
  (2016) F1038--F1048.

\bibitem{lamibrac2016}
A.~Lamibrac, J.~Roth, M.~Toulec, F.~Marone, M.~Stampanoni, F.~N. Büchi,
  Characterization of liquid water saturation in gas diffusion layers by x-ray
  tomographic microscopy, Journal of The Electrochemical Society 163~(3) (2016)
  F202--F209.

\bibitem{sabharwal2018}
M.~Sabharwal, J.~T. Gostick, M.~Secanell, Virtual {Liquid} {Water} {Intrusion}
  in {Fuel} {Cell} {Gas} {Diffusion} {Media}, Journal of The Electrochemical
  Society 165~(7) (2018) F553--F563.

\bibitem{tranter2018}
T.~G. Tranter, J.~T. Gostick, A.~D. Burns, W.~F. Gale, Capillary {Hysteresis}
  in {Neutrally} {Wettable} {Fibrous} {Media}: {A} {Pore} {Network} {Study} of
  a {Fuel} {Cell} {Electrode}, Transport in Porous Media 121~(3) (2018)
  597--620.

\bibitem{hill1960}
R.~D. Hill, A study of pore size distribution of fired clay bodies, Trans.
  Brit. Ceram. Soc 59~(6) (1960) 189--212.

\bibitem{kozeny1927}
J.~Kozeny, Über kapillare leitung des wassers im boden:(aufstieg, versickerung
  und anwendung auf die bewässerung), Hölder-Pichler-Tempsky, 1927.

\bibitem{carman1937}
P.~C. Carman, Fluid flow through granular beds, T. Inst. Chem. Eng. 15 (1937)
  150--166.
\newblock \href {http://dx.doi.org/10.1016/S0263-8762(97)80003-2}
  {\path{doi:10.1016/S0263-8762(97)80003-2}}.

\bibitem{scheidegger1953}
A.~E. Scheidegger, Theoretical models of porous matter, Produc. Mon. 17~(10)
  (1953) 17--23.

\end{thebibliography}
\end{document}